\documentclass[%
 reprint,
superscriptaddress,
nofootinbib,
 amsmath,amssymb,
 prd,
]{revtex4-2}

\usepackage[margin=5pt,caption=false]{subfig}
\usepackage{slantsc}
\usepackage{comment}
\usepackage{threeparttable}
\usepackage{booktabs}
\usepackage[figuresright]{rotating}
\usepackage{textcase}
\usepackage{multirow}
\newcommand{\be}{\begin{equation}}
\newcommand{\ee}{\end{equation}}
\newcommand{\bea}{\begin{eqnarray}}
\newcommand{\eea}{\end{eqnarray}}
\newcommand{\hunit}{$\rm{km \ s^{-1} \ Mpc^{-1}}$}
\newcommand{\lcdm}{$\Lambda$CDM}
\newcommand{\pcdm}{$\phi$CDM}
\usepackage{array}

\newcommand{\hiig}{H\,\textsc{ii}G}
\newcommand{\hii}{H\,\textsc{ii}}
\newcommand{\Om}{\Omega_{m0}}
\newcommand{\Ok}{\Omega_{k0}}
\newcommand{\om}{$\Omega_{m0}$}
\newcommand{\ok}{$\Omega_{k0}$}
\newcommand{\wx}{$w_{\rm X}$}
\newcommand{\wX}{w_{\rm X}}

\newcommand{\mii}{Mg\,\textsc{ii}}

\newcommand{\civ}{C\,\textsc{iv}}

\newcommand{\obh}{\Omega_{b}h^2}
\newcommand{\och}{\Omega_{c}h^2}
\newcommand{\onh}{\Omega_{\nu}h^2}
\newcommand{\obhs}{$\Omega_{b}h^2$}
\newcommand{\ochs}{$\Omega_{c}h^2$}

\usepackage[T1]{fontenc}
\usepackage{scalerel}
\usepackage{tikz}
\usetikzlibrary{svg.path}

\definecolor{orcidlogocol}{HTML}{A6CE39}
\tikzset{
  orcidlogo/.pic={
    \fill[orcidlogocol] svg{M256,128c0,70.7-57.3,128-128,128C57.3,256,0,198.7,0,128C0,57.3,57.3,0,128,0C198.7,0,256,57.3,256,128z};
    \fill[white] svg{M86.3,186.2H70.9V79.1h15.4v48.4V186.2z}
                 svg{M108.9,79.1h41.6c39.6,0,57,28.3,57,53.6c0,27.5-21.5,53.6-56.8,53.6h-41.8V79.1z M124.3,172.4h24.5c34.9,0,42.9-26.5,42.9-39.7c0-21.5-13.7-39.7-43.7-39.7h-23.7V172.4z}
                 svg{M88.7,56.8c0,5.5-4.5,10.1-10.1,10.1c-5.6,0-10.1-4.6-10.1-10.1c0-5.6,4.5-10.1,10.1-10.1C84.2,46.7,88.7,51.3,88.7,56.8z};
  }
}

\newcommand\orcidicon[1]{\href{https://orcid.org/#1}{\mbox{\scalerel*{
\begin{tikzpicture}[yscale=-1,transform shape]
\pic{orcidlogo};
\end{tikzpicture}
}{|}}}}

\usepackage{amsmath,bm}	
\usepackage{amssymb}	
\usepackage{fnpct}

\usepackage[colorlinks=true,allcolors=blue]{hyperref}
\begin{document}

\preprint{APS/123-QED}

\title{Low- and high-redshift \hii\ starburst galaxies obey different luminosity-velocity dispersion relations}

\author{Shulei Cao$^{\orcidicon{0000-0003-2421-7071}}$}
 \email{shuleicao@boisestate.edu}
\affiliation{Department of Physics, Kansas State University, 116 Cardwell Hall, Manhattan, KS 66506, USA}%
\affiliation{Department of Physics, Boise State University, 1910 University Drive, Boise, ID 83725, USA}%
\author{Bharat Ratra$^{\orcidicon{0000-0002-7307-0726}}$}%
 \email{ratra@phys.ksu.edu}
\affiliation{Department of Physics, Kansas State University, 116 Cardwell Hall, Manhattan, KS 66506, USA}%

\date{\today}

\begin{abstract}
To determine whether or not \hii\ starburst galaxies (\hiig) are standardizable candles, we study the correlation between the H$\beta$ luminosity ($L$) and the velocity dispersion ($\sigma$) of the ionized gas from \hiig\ measurements by simultaneously constraining the $L-\sigma$ relation parameters and the cosmological model parameters. We investigate six flat and nonflat relativistic dark energy cosmological models. We find that low-redshift and high-redshift \hiig\ data subsets are standardizable but obey  different $L-\sigma$ relations. Current \hiig\ data are too sparse and too non-uniformly distributed in redshift to allow for a determination of why the samples follow different relations, but it could be caused by the high-redshift sample containing relatively fewer intrinsically dimmer sources (Malmquist bias) or it could be a consequence of \hiig\ evolution. Until this issue is better understood, \hiig\ data cosmological constraints must be treated with caution.
\end{abstract}

\maketitle

\section{Introduction}

Current better-established cosmological data probe either the lower redshift universe, at $z < 2.3$ \cite{Yuetal2018, eBOSS_2020, Brout:2022vxf}, or the universe at $z \sim 1100$, \cite{planck2018b}. Among potential cosmological probes of the intermediate redshift range are bright \hii\ starburst galaxies (\hiig) that currently reach to $z\sim2.5$, see e.g. Refs.\ \cite{Melnick_2000, Siegel_2005, Plionis_2009, Plionis_2010, Plionis_2011, Mania_2012, Chavez_2012, Chavez_2014, Chavez_2016, Terlevich_2015, G-M_2019, GM2021, CaoRyanRatra2020, CaoRyanRatra2021, CaoRyanRatra2022, Johnsonetal2022, Mehrabietal2022}. These galaxies, with their prominent \hii\ regions illuminated by Balmer emission lines, obey an $L-\sigma$ correlation between their H$\beta$ luminosity ($L$) and the velocity dispersion ($\sigma$) of the ionized gas within the \hiig\ starburst galaxy. If valid, with time-independent intercept and slope, this $L-\sigma$ relation allows one to use observed H$\beta$ fluxes and observed ionized gas velocity dispersions of \hiig\ galaxies to constrain cosmological parameters.

Other similar emerging cosmological probes include reverberation-measured \mii\ and \civ\ quasar (QSO) measurements that reach to $z\sim3.4$, \citep{Czernyetal2021, Zajaceketal2021, Yuetal2021, Khadkaetal_2021a, Khadkaetal2021c, Khadka:2022ooh, Cao:2022pdv, Czerny:2022xfj, Caoetal2024}, and gamma-ray burst (GRB) data that reach to $z\sim8.2$, \citep{Wang_2016, Dirirsa2019, KhadkaRatra2020c, Dainottetal2020, Huetal_2021, Daietal_2021, Demianskietal_2021, Khadkaetal_2021b, CaoDainottiRatra2022b, DainottiNielson2022}.\footnote{Of which only 118 Amati-correlated (A118) GRBs, with lower intrinsic dispersion, are suitable for cosmological purposes, \citep{Khadkaetal_2021b, LuongoMuccino2021, CaoKhadkaRatra2021, CaoDainottiRatra2022, Liuetal2022}.} Both these probes are based on correlations similar to the $L-\sigma$ relation for \hiig, and we have analyzed these data by simultaneously constraining correlation parameters and cosmological parameters in a number of different cosmological models. This is required to avoid the circularity problem, and also to determine whether or not the correlation is independent of cosmological model and thus standardizable \citep{KhadkaRatra2020c, Caoetal_2021}. In papers cited above, we have determined that these QSO and GRB data have relevant correlations that are independent of the assumed cosmological model and so these data are standardizable and suitable for cosmological purposes.

QSO X-ray and UV flux observations that reach to $z \sim 7.5$ have been also been studied as a potential cosmological probe, based on a similar correlation, \citep{RisalitiLusso2015, RisalitiLusso2019, KhadkaRatra2020a, Yangetal2020, KhadkaRatra2020b, Lussoetal2020, KhadkaRatra2021, KhadkaRatra2022, Rezaeietal2022, Luongoetal2021, DainottiBardiacchi2022}, however, these QSOs are not standardizable as the correlation is neither cosmological model independent nor redshift independent for the latest QSO flux compilation, \cite{Lussoetal2020}, and so these data cannot be used for cosmology, \citep{KhadkaRatra2021, KhadkaRatra2022, Petrosian:2022tlp, Khadka:2022aeg, Zajaceketal2024}.

All previous cosmological analyses of \hiig\ data have assumed a fixed $L-\sigma$ relation with slope and intercept parameter values assumed to be independent of cosmological model and redshift, and taken from the analyses of Ref.\ \cite{GM2021}, as discussed in more detail below in Sec.\ \ref{sec:data}. In this paper we examine for the first time whether or not \hiig\ data obey the $L-\sigma$ relation with cosmological model and time independent intercept and slope. To do this we use \hiig\ flux and velocity dispersion measurements to simultaneously constrain $L-\sigma$ relation parameters and cosmological parameters in a number of different cosmological models. In this paper we also examine for the first time cosmological constraints from just \hiig\ data --- in particular, we do not use Giant Extragalactic \hii\ Regions (GH\,\textsc{ii}R) sources that have been calibrated with primary distance indicators --- as we want to determine \hiig\ data cosmological constraints that are independent of local calibration data. Our technique of simultaneously constraining $L-\sigma$ relation parameters and cosmological parameters in a number of different cosmological models allows us compute \hiig\ data alone cosmological constraints.

In this paper we consider six flat and nonflat relativistic dark energy cosmological models, including the standard \lcdm\ model and models with dynamical dark energy. In the spatially-flat \lcdm\ model, \cite{peeb84}, dark energy is a cosmological constant, $\Lambda$, dark matter is cold dark matter (CDM), and spatially-flat hypersurfaces are considered. While it is the simplest model and fits reasonably well with most observations, there are some potential discrepancies, see, e.g., Refs.\ \cite{PerivolaropoulosSkara2021,Morescoetal2022,Abdallaetal2022,Hu:2023jqc}. These potential discrepancies motivate the consideration of alternate cosmological models with spatially nonflat hypersurfaces or dynamical dark energy. By considering a variety of cosmological models, we can test whether \hiig\ data are standardizable through the $L-\sigma$ relation. 

We also analyze low-redshift and high-redshift subsets of these \hiig\ data, to determine whether or not the $L-\sigma$ relation intercept and slope parameters are independent of redshift, and for reasons described next. Initially, we analyzed the entire sample of 181 \hiig\ measurements \citep{GM2021} by assuming that they obey the same $L-\sigma$ relation, but we found that their cosmological parameter constraints were not consistent with those from better-established Hubble parameter and baryon acoustic oscillation [$H(z)$ + BAO] data. To better understand these results, we performed separate analyses of the 107 low-$z$ \hiig\ measurements and of the 74 high-$z$ \hiig\ measurements.

Analyses of the low-$z$ and high-$z$ \hiig\ data sets showed that both low-$z$ and high-$z$ \hii\ starburst galaxies can be standardized, but obey very different $L-\sigma$ relations, contradicting the assumption used in all previous analyses of these data. It is possible that all we have discovered is that the high-$z$ data set suffers from Malmquist bias and is missing intrinsically dimmer sources, or that \hii\ starburst galaxies evolve, but more data, more uniformly distributed in redshift space, and especially more intrinsically-dimmer high-$z$ sources, are needed to properly determine the cause of the effect we have found.

The derived low-$z$ and high-$z$ data cosmological constraints are weak, and are not consistent in two of the cosmological models we study. In the other four cosmological models we jointly analyze low-$z$ + high-$z$ data, assuming independent slope and intercept parameters for the low-$z$ and high-$z$ data $L-\sigma$ relations, and find that the low-$z$ + high-$z$ \hiig\ sources are also standardizable. However, prior to using \hiig\ data for cosmological purposes, it is essential to understand the cause of the different low-$z$ and high-$z$ data $L-\sigma$ relations.

This paper is organized as follows. In Section \ref{sec:model} we briefly describe the cosmological models used in our analyses. We introduce the data sets used in Section \ref{sec:data} and summarize the analysis methodology in Section \ref{sec:analysis}. We then present our main results in Section \ref{sec:results}. Finally, we draw conclusions in Section \ref{sec:conclusion}.

\section{Cosmological models}
\label{sec:model}

The Hubble parameter, $H(z)$, a function of redshift $z$ and cosmological parameters, is fundamental to each cosmological model we study. $H(z)$ is defined by the first Friedmann equation, derived from the Friedmann-Lema\^{i}tre-Robertson-Walker metric in the framework of general relativity.

We consider one massive and two massless neutrino species in our models. With an effective number of relativistic neutrino species $N_{\rm eff} = 3.046$ and a total neutrino mass $\sum m_{\nu}=0.06$ eV, we compute the current value of the non-relativistic neutrino physical energy density parameter as $\onh=\sum m_{\nu}/(93.14\ \rm eV)$, where $h$ is the Hubble constant ($H_0$) in units of 100 \hunit. Consequently, the present value of the non-relativistic matter density parameter is $\Om = (\onh + \obh + \och)/{h^2}$, where \obhs\ and \ochs\ are the present values of the baryonic and cold dark matter physical energy density parameters, respectively. As our analysis focuses on late-time measurements, we ignore the contribution from photons to the cosmological energy budget. 

In the current study we explore the \lcdm\ models and the XCDM parametrizations, which serve as an extension to \lcdm. These frameworks handle the dark energy equation of state parameter, $w_{\rm DE}=p_{\rm DE}/\rho_{\rm DE}$, the ratio of the dark energy fluid pressure and energy density, differently. The \lcdm\ models fix $w_{\rm DE}$ at $-1$, while the XCDM parametrizations allow for its variability. The governing Friedmann equation can be expressed as
\be
\label{eq:HzLX}
\resizebox{0.475\textwidth}{!}{%
    $H(z) = H_0\sqrt{\Om\left(1 + z\right)^3 + \Ok\left(1 + z\right)^2 + \Omega_{\rm DE}\left(1+z\right)^{1+w_{\rm DE}}},$%
    }
\ee
where \ok\ is the present spatial curvature energy density parameter\footnote{For recent observational constraints on spatial curvature see Refs.\ \cite{Oobaetal2018b, ParkRatra2019b, DiValentinoetal2021a, ArjonaNesseris2021, Dhawanetal2021, Renzietal2022, Gengetal2022, MukherjeeBanerjee2022, Glanvilleetal2022, Wuetal2023, deCruzPerezetal2023, DahiyaJain2022, Stevensetal2023, Favaleetal2023, Qietal2023, deCruzPerezetal2024} and references therein.} and $\Omega_{\rm DE} = 1 - \Om - \Ok$ is the present dark energy density parameter. In \lcdm, dark energy is the cosmological constant $\Lambda$, making $\Omega_{\rm DE} =\Omega_{\Lambda}$. In XCDM, dark energy is treated as an X-fluid, with a dynamical dark energy equation of state parameter, making $\Omega_{\rm DE} =\Omega_{\rm X0}$. In the analysis involving \hiig\ data, we set $H_0=70$ \hunit\ and $\Omega_{b}=0.05$ because these parameters cannot be constrained by \hiig\ data. Therefore, the free cosmological parameters being constrained are $\{\Omega_c, \Ok\}$ for \lcdm\ and $\{\Omega_c, \wX, \Ok\}$ for XCDM. When analyzing $H(z)$ + BAO data, the constrained cosmological parameters are $\{H_0, \obh\!, \och\!, \Ok\}$ for \lcdm\ and $\{H_0, \obh\!, \och\!, \wX, \Ok\}$ for XCDM. In spatially-flat models $\Ok$ is not a free parameter and is set to $0$.

In addition, we also investigate the \pcdm\ models (see e.g. Ref.\ \cite{peebrat88,ratpeeb88,pavlov13})\footnote{For recent cosmological constraints on the \pcdm\ models, see Refs.\ \cite{ooba_etal_2018b, ooba_etal_2019, park_ratra_2018, park_ratra_2019b, park_ratra_2020, Singhetal2019, UrenaLopezRoy2020, SinhaBanerjee2021, deCruzetal2021, Xuetal2022, Jesusetal2022, Adiletal2023, Dongetal2023, VanRaamsdonkWaddell2023, CaoRatra2024} and references therein.} where dark energy is a dynamical scalar field $\phi$ governed by an inverse power-law potential energy density
\be
\label{PE}
V(\phi)=\frac{1}{2}\kappa m_p^2\phi^{-\alpha}.
\ee
Here $m_p$ stands for the Planck mass and $\alpha$ is a positive constant; when $\alpha=0$ \pcdm\ reduces to \lcdm. The constant $\kappa$ is determined using the shooting method in the Cosmic Linear Anisotropy Solving System (\textsc{class}) code, \cite{class}. The Friedmann equation is
\be
\label{eq:Hzp}
    H(z) = H_0\sqrt{\Om\left(1 + z\right)^3 + \Ok\left(1 + z\right)^2 + \Omega_{\phi}(z,\alpha)},
\ee
where the scalar field dynamical dark energy density parameter
\be
\label{Op}
\Omega_{\phi}(z,\alpha)=\frac{1}{6H_0^2}\bigg[\frac{1}{2}\dot{\phi}^2+V(\phi)\bigg]
\ee
is computed by numerically solving the Friedmann equation \eqref{eq:Hzp} and the equation of motion of the scalar field
\be
\label{em}
\ddot{\phi}+3H\dot{\phi}+V'(\phi)=0.
\ee 
In these equations, an overdot and a prime denote derivatives with respect to time and $\phi$, respectively. For \hiig\ data, the free cosmological parameters being constrained are $\{\Omega_c, \alpha, \Ok\}$, while for $H(z)$ + BAO data, the constrained cosmological parameters are $\{H_0, \obh\!, \och\!, \alpha, \Ok\}$, with $\Ok=0$ in flat \pcdm.

\section{Data}
\label{sec:data}

In this paper we test whether \hiig\ data obey the $L-\sigma$ correlation in a model- and redshift-independent manner, by simultaneously constraining the $L-\sigma$ correlation parameters and cosmological parameters for both low- and high-redshift \hiig\ data. These \hiig\ data, as well as $H(z)$ + BAO data used for comparison purposes, are summarized below.

\begin{itemize}

\item[]{\it H\,\textsl{\textsc{ii}}G data.} We use 181 \hiig\ measurements listed in table A3 of Ref.\ \cite{GM2021}, with 107 low-$z$ ones from Ref.\ \cite{Chavez_2014} recalibrated in Ref.\ \cite{G-M_2019}, spanning the redshift range $0.0088 \leq z \leq 0.16417$, and 74 high-$z$ ones spanning the redshift range $0.63427 \leq z \leq 2.545$. (In what follows we refer to these data sets as low-$z$ and high-$z$ data.) It is believed that these sources follow the $L-\sigma$ correlation represented by the equation $\log L = \beta \log \sigma + \gamma$, where $\log = \log_{10}$, and $L$ and $\sigma$ are in units of $\text{erg s}^{-1}$ and $\text{km s}^{-1}$, respectively.\footnote{We do not consider three-parameter generalizations of this $L-\sigma$ relation \citep{BordaloTelles2011, Chavez_2012, Chavez_2014}.} In Ref.\ \cite{G-M_2019}, 107 low-$z$ \hiig\ and 36 GH\,\textsc{ii}R data are used to determine the $L-\sigma$ relation slope and intercept parameters, $\beta$ and $\gamma$, which are found to be $5.022 \pm 0.058$ and $33.268 \pm 0.083$, respectively. To infer distances to the GH\,\textsc{ii}R sources, the authors relied on primary indicators such as Cepheids, TRGB, and theoretical model calibrations, \cite{FernandezArenas}. In Ref.\ \cite{GM2021}, the authors also simultaneously constrained the cosmological parameters and $L-\sigma$ correlation parameters, but with different data (including 36 GH\,\textsc{ii}R sources) and likelihood function (without intrinsic scatter and Gaussian likelihood coefficient). In order to test the correlation, here we consider $\beta$ and $\gamma$ as free parameters to be simultaneously constrained with the cosmological parameters. The observed distance modulus of an \hiig\ can be computed as $\mu_{\rm obs} = 2.5\log L - 2.5\log f - 100.2$, where $f(z)$ is the measured flux in units of $\text{erg s}^{-1}\text{cm}^{-2}$ at redshift $z$ corrected for extinction using the Gordon, \cite{Gordon_2003}, extinction law and $L$ is obtained from the $L-\sigma$ correlation. The theoretical distance modulus in a given cosmological model is $\mu_{\rm th}(z) = 5\log D_{L}(z) + 25$, where $D_L(z)$ is the luminosity distance
\begin{equation}
  \label{eq:DL}
\resizebox{0.475\textwidth}{!}{%
    $D_L(z) = 
    \begin{cases}
    \frac{c(1+z)}{H_0\sqrt{\Ok}}\sinh\left[\frac{\sqrt{\Ok}H_0}{c}D_C(z)\right] & \text{if}\ \Ok > 0, \\
    \vspace{1mm}
    (1+z)D_C(z) & \text{if}\ \Ok = 0,\\
    \vspace{1mm}
    \frac{c(1+z)}{H_0\sqrt{|\Ok|}}\sin\left[\frac{H_0\sqrt{|\Ok|}}{c}D_C(z)\right] & \text{if}\ \Ok < 0,
    \end{cases}$%
    }
\end{equation}
where $D_C(z)$ is the comoving distance
\begin{equation}
\label{eq:gz}
   D_C(z) = c\int^z_0 \frac{dz'}{H(z')},
\end{equation}
and $c$ is the speed of light.

\item[]{$H(z)\ +\ BAO\ data$.} Here we use 32 $H(z)$ and 12 BAO measurements listed in Tables 1 and 2 of Ref.\ \cite{CaoRatra2023}, spanning the redshift ranges $0.07 \leq z \leq 1.965$ and $0.122 \leq z \leq 2.334$, respectively.

\end{itemize}

\section{Data Analysis Methodology}
\label{sec:analysis}

The natural log of the \hiig\ data likelihood function is
\begin{equation}
\label{eq:LF_s1}
    \ln\mathcal{L}_{\rm\hiig}= -\frac{1}{2}\Bigg[\chi^2_{\rm\hiig}+\sum^{N}_{i=1}\ln\left(2\pi\epsilon^2_{\mathrm{tot},i}\right)\Bigg],
\end{equation}
where
\begin{equation}
\label{eq:chi2_s1}
    \chi^2_{\rm\hiig} = \sum^{N}_{i=1}\bigg[\frac{(\mathbf{\mu}_{\mathrm{obs},i} - \mathbf{\mu}_{\mathrm{th},i})^2}{\epsilon^2_{\mathrm{tot},i}}\bigg]
\end{equation}
with total uncertainty
\begin{equation}
\label{eq:sigma_s2}
\epsilon^2_{\mathrm{tot},i}=\sigma_{\mathrm{int}}^2+\epsilon_{{\mathbf{\mu}_{\mathrm{obs},i}}}^2+\epsilon_{{\mathbf{\mu}_{\mathrm{th},i}}}^2,
\end{equation}
where $\sigma_{\mathrm{int}}$ is the intrinsic scatter parameter for \hiig\ data, which also accounts for unknown systematic uncertainties. In the cases of low-$z$ and high-$z$ data, we denote $\beta$ and $\gamma$, and $\sigma_{\mathrm{int}}$ by adding subscripts of low and high, respectively. 

In this study, we compute the likelihoods associated with $H(z)$ and BAO data by following the methods of Ref.\ \cite{CaoRatra2023}. 

The parameters we constrain, which are subject to flat priors, are presented in Table \ref{tab:priors}. We apply the {\footnotesize MontePython} Markov chain Monte Carlo code, \cite{Audrenetal2013,Brinckmann2019}, to perform likelihood analysis, targeting both cosmological and $L-\sigma$ correlation parameters. For subsequent statistical analysis and visualization, we make use of the {\footnotesize GetDist} \textsc{python} package, \cite{Lewis_2019}. For definitions and details of Information Criteria (IC) like AIC, BIC, and DIC, we refer readers to our previous work (see, e.g. Ref.\ \cite{CaoDainottiRatra2022}). We assess model performance using $\Delta \mathrm{IC}$ comparing each alternate dark energy model against the flat \lcdm\ baseline. A $\Delta \mathrm{IC}$ value, whether positive or negative, signifies how well (worse or better) the model aligns with the data set relative to this baseline. In evaluating the models, we categorize the strength of evidence against them based on $\Delta \mathrm{IC}$ values relative to the model with the minimum IC: weak $(0, 2]$, positive $(2, 6]$, strong $(6, 10]$, and very strong $>10$.

\begin{table}
\centering
\resizebox{\columnwidth}{!}{%
\begin{threeparttable}
\caption{Flat priors of the constrained parameters.}
\label{tab:priors}
\begin{tabular}{lcc}
\toprule\toprule
Parameter & & Prior\\
\midrule
 & Cosmological Parameters & \\
\midrule
$H_0$\,\tnote{a} &  & [None, None]\\
\obhs\,\tnote{b} &  & [0, 1]\\
\ochs\,\tnote{c} &  & [0, 1]\\
\ok &  & [$-2$, 2]\\
$\alpha$ &  & [0, 10]\\
\wx &  & [$-5$, 0.33]\\
\\
 & $L-\sigma$ Correlation Parameters & \\
$\beta$ &  & [$-5$, 15]\\
$\gamma$ &  & [20, 60]\\
$\sigma_{\mathrm{int}}$ &  & [0, 5]\\
\bottomrule\bottomrule
\end{tabular}
\begin{tablenotes}
\item [a] \hunit. In the \hiig\ cases, $H_0=70$ \hunit.
\item [b] In the \hiig\ cases, $\Omega_{b}=0.05$.
\item [c] In the \hiig\ cases, $\Om\in[0,1]$ is ensured.
\end{tablenotes}
\end{threeparttable}%
}
\end{table}

\section{Results}
\label{sec:results}

\begin{figure*}
\centering
 \subfloat[Flat \lcdm]{%
    \includegraphics[width=0.45\textwidth,height=0.35\textwidth]{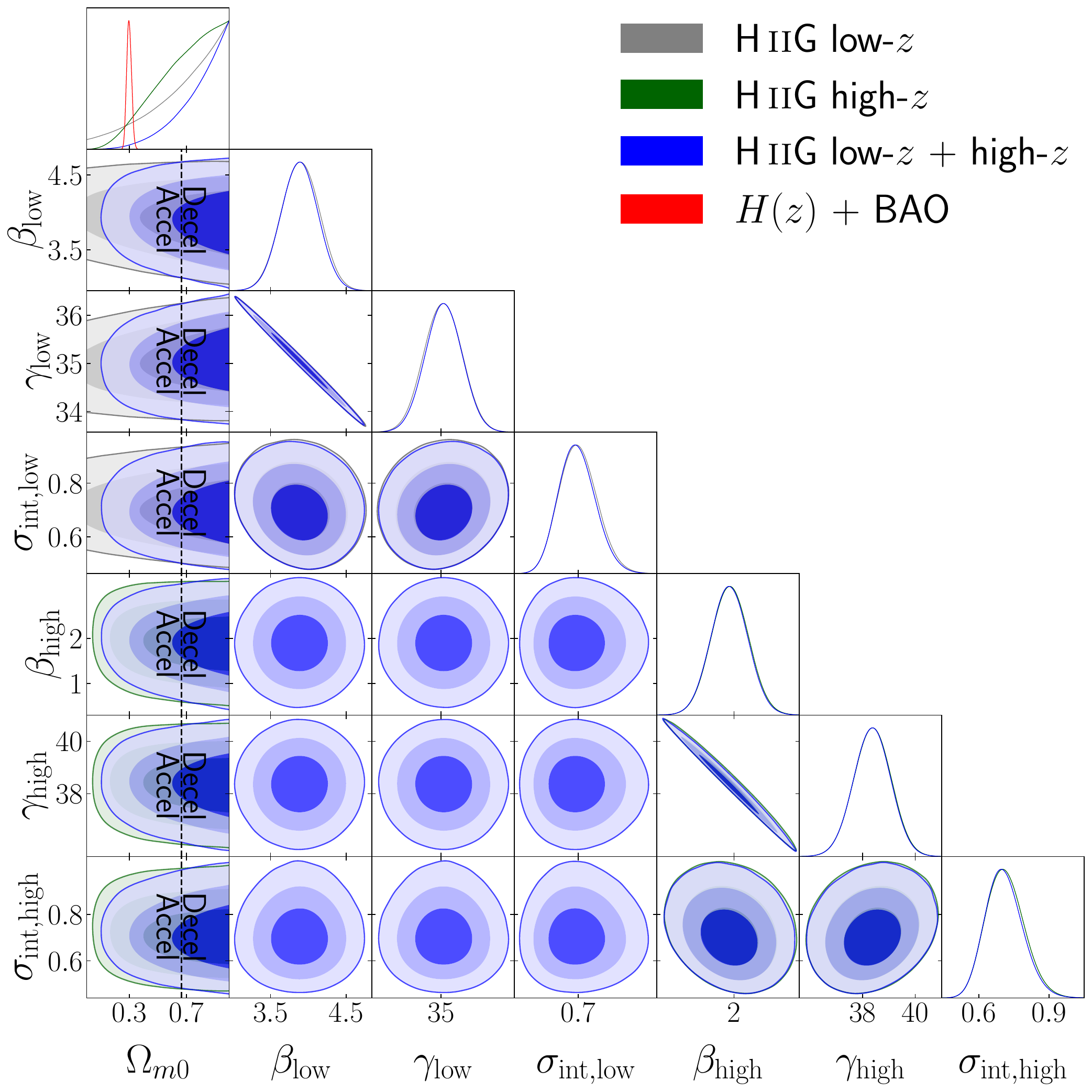}}
 \subfloat[Nonflat \lcdm]{%
    \includegraphics[width=0.45\textwidth,height=0.35\textwidth]{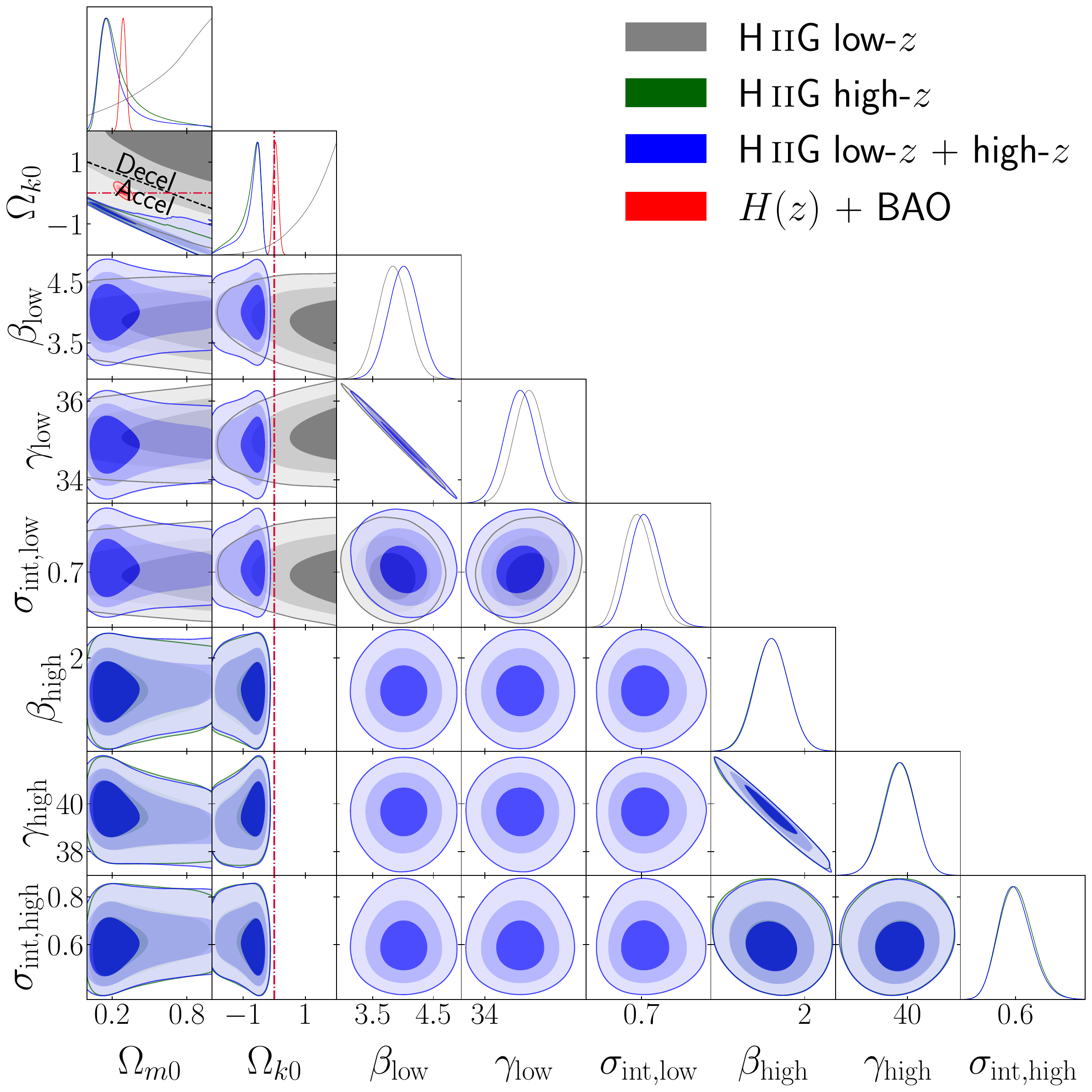}}\\
 \subfloat[Flat XCDM]{%
    \includegraphics[width=0.45\textwidth,height=0.35\textwidth]{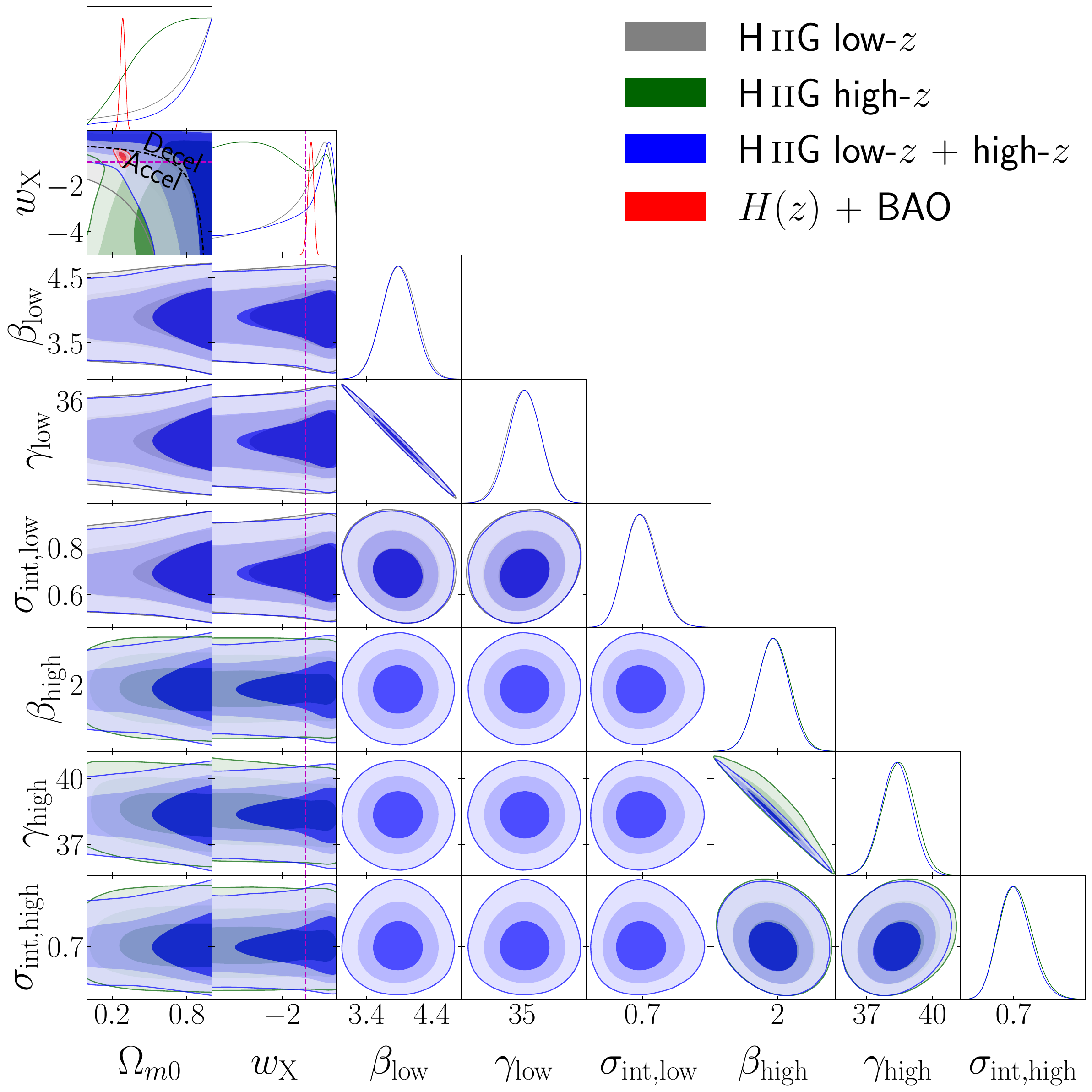}}
 \subfloat[Nonflat XCDM]{%
    \includegraphics[width=0.45\textwidth,height=0.35\textwidth]{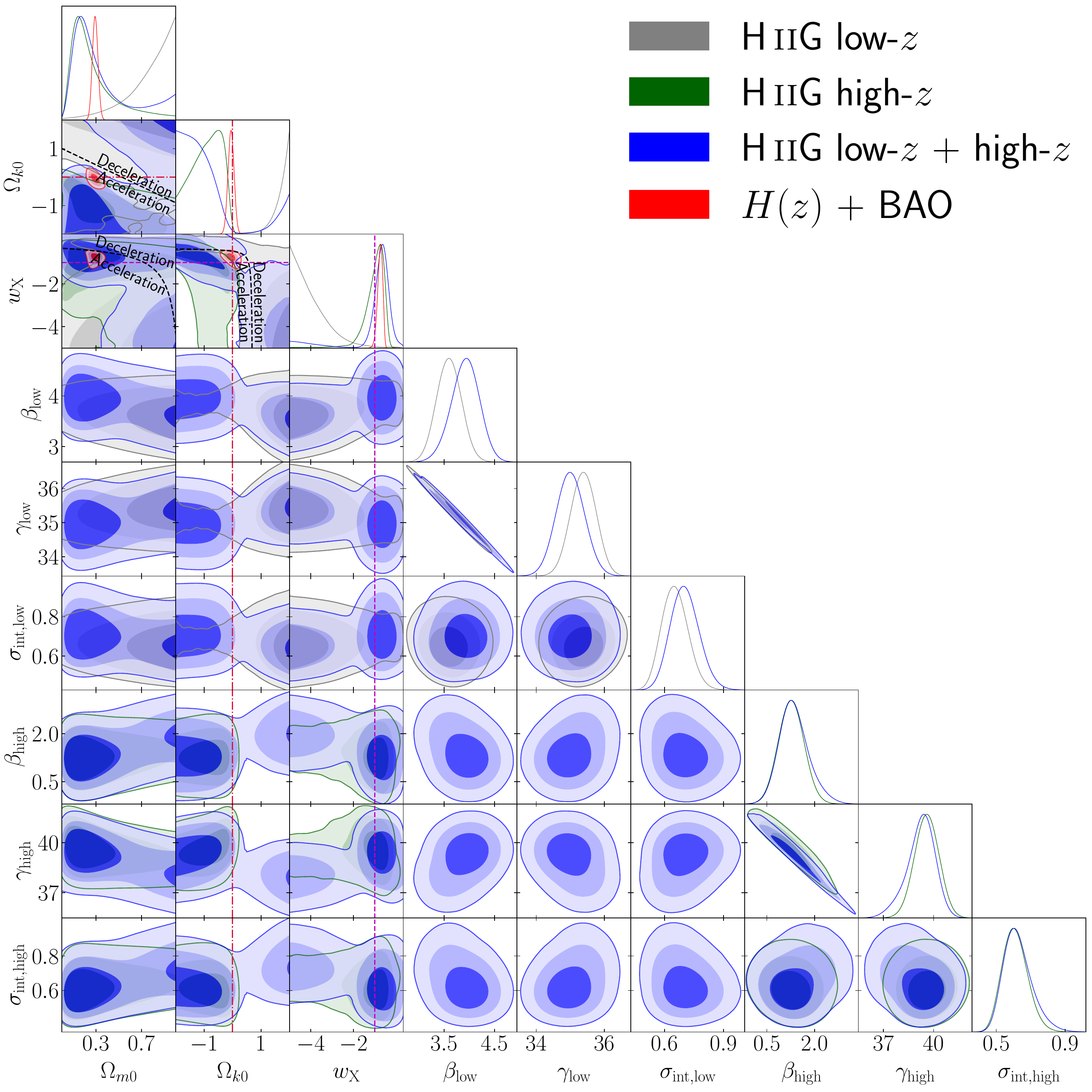}}\\
 \subfloat[Flat \pcdm]{%
    \includegraphics[width=0.45\textwidth,height=0.35\textwidth]{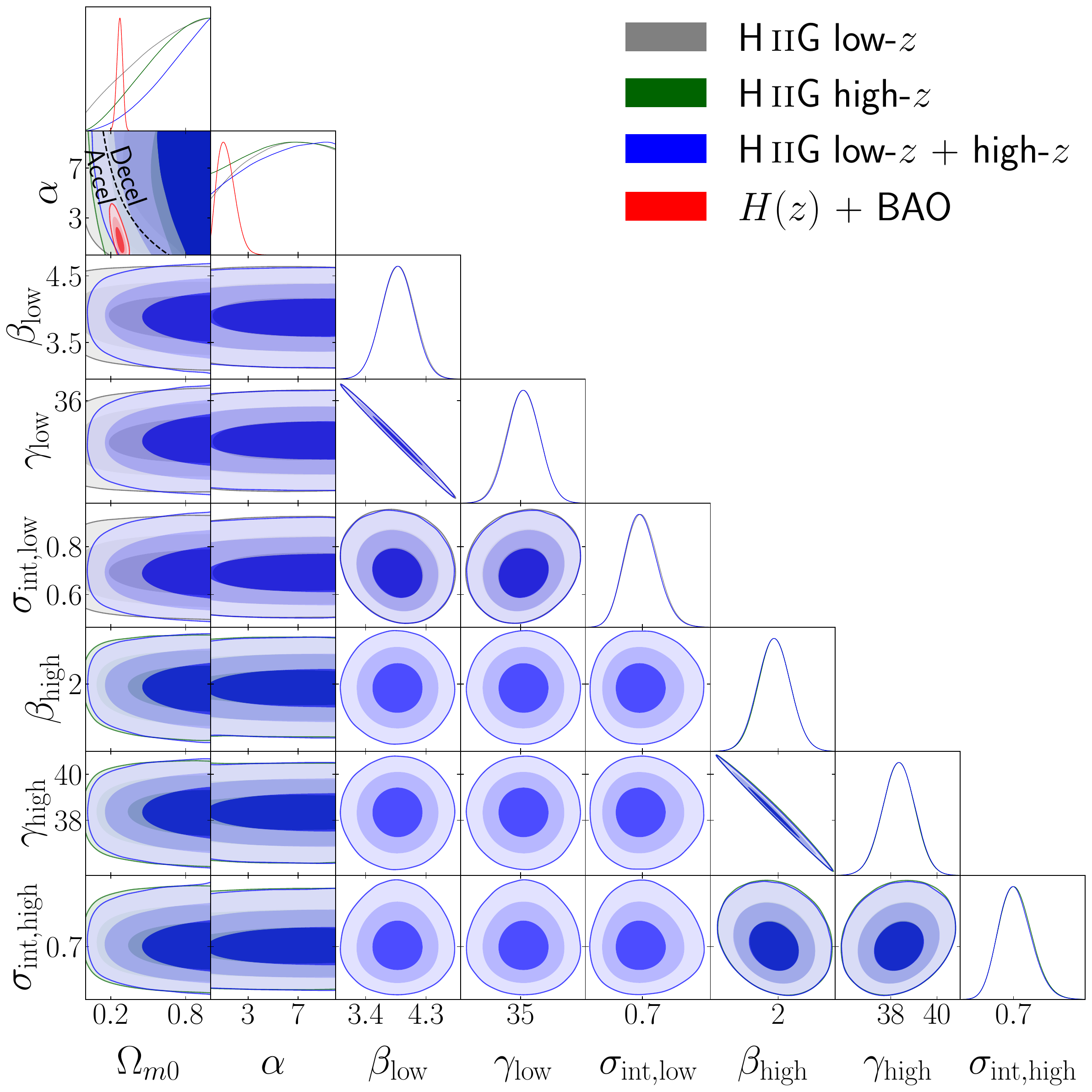}}
 \subfloat[Nonflat \pcdm]{%
    \includegraphics[width=0.45\textwidth,height=0.35\textwidth]{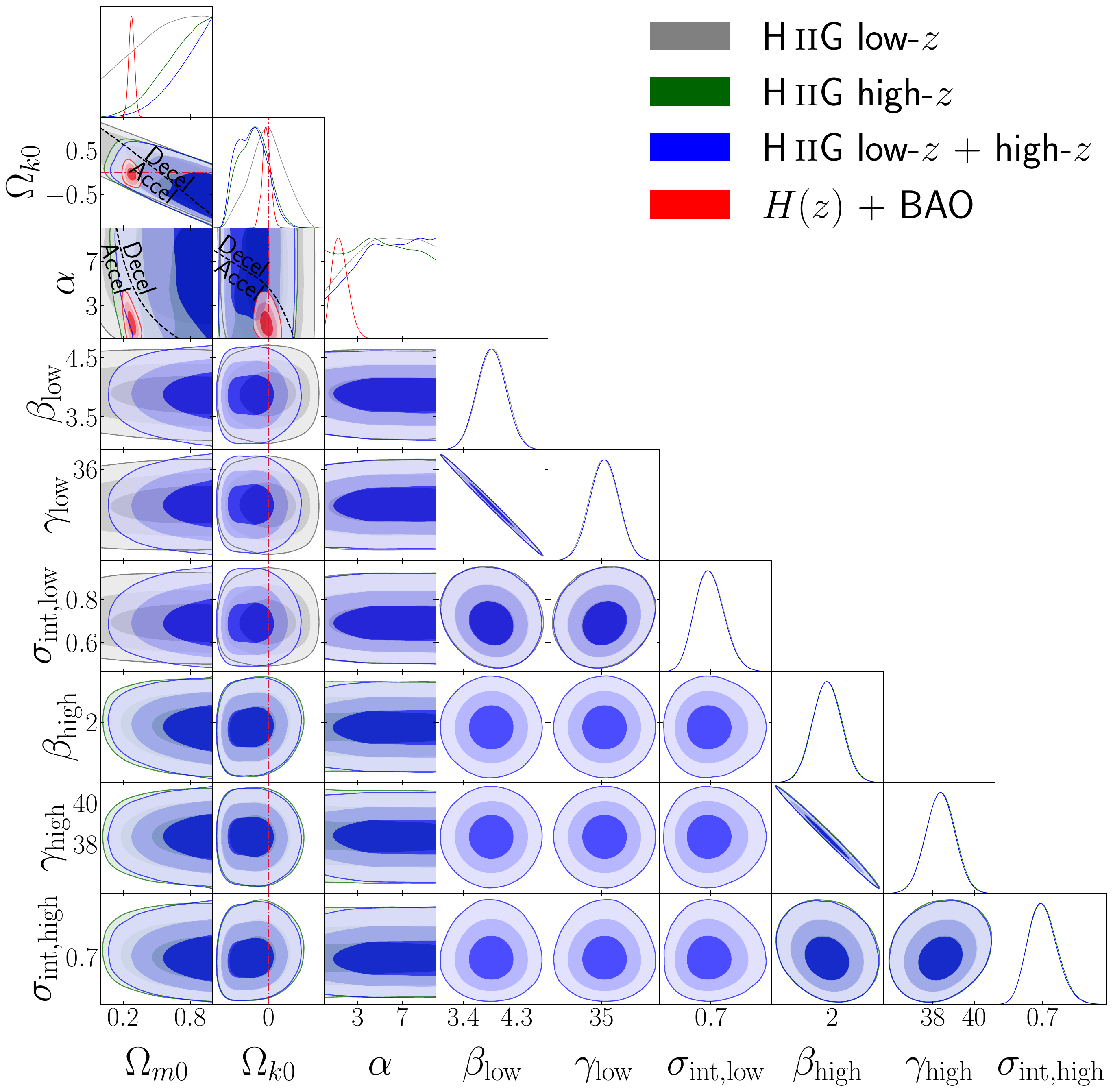}}\\
\caption{One-dimensional likelihoods and 1$\sigma$, 2$\sigma$, and 3$\sigma$ two-dimensional likelihood confidence contours from \hiig\ low-$z$ (gray), \hiig\ high-$z$ (green), \hiig\ low-$z$ + high-$z$ (blue), and $H(z)$ + BAO (red) data for six different models, with \lcdm, XCDM, and \pcdm\ in the top, middle, and bottom rows, and flat (nonflat) models in the left (right) column. The black dashed zero-acceleration lines, computed for the third cosmological parameter set to the $H(z)$ + BAO data best-fitting values listed in Table \ref{tab:BFP} in panels (d) and (f), divide the parameter space into regions associated with currently-accelerating (below or below left) and currently-decelerating (above or above right) cosmological expansion. The crimson dash-dot lines represent flat hypersurfaces, with closed spatial hypersurfaces either below or to the left. The magenta lines represent $w_{\rm X}=-1$, i.e.\ flat or nonflat \lcdm\ models. The $\alpha = 0$ axes correspond to flat and nonflat \lcdm\ models in panels (e) and (f), respectively.}
\label{fig1}
\vspace{-50pt}
\end{figure*}

\begin{figure*}
\centering
 \subfloat[Flat \lcdm]{%
    \includegraphics[width=0.4\textwidth,height=0.35\textwidth]{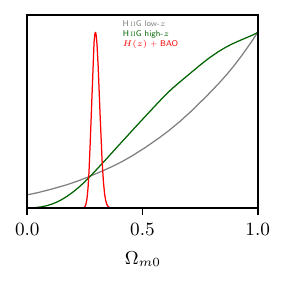}}
 \subfloat[]{%
    \includegraphics[width=0.4\textwidth,height=0.35\textwidth]{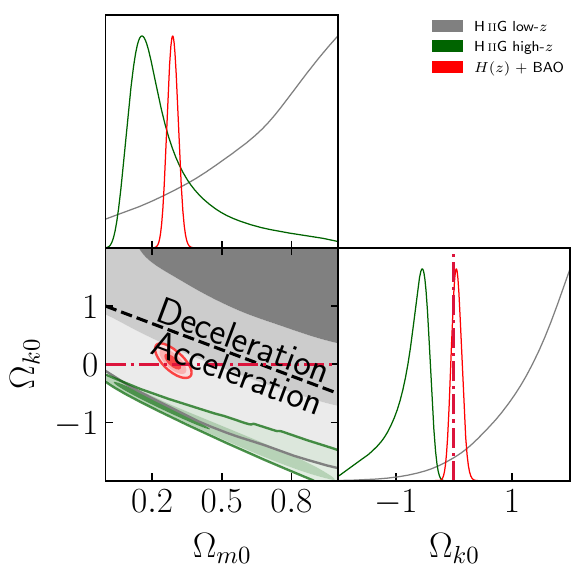}}\\
 \subfloat[]{%
    \includegraphics[width=0.4\textwidth,height=0.35\textwidth]{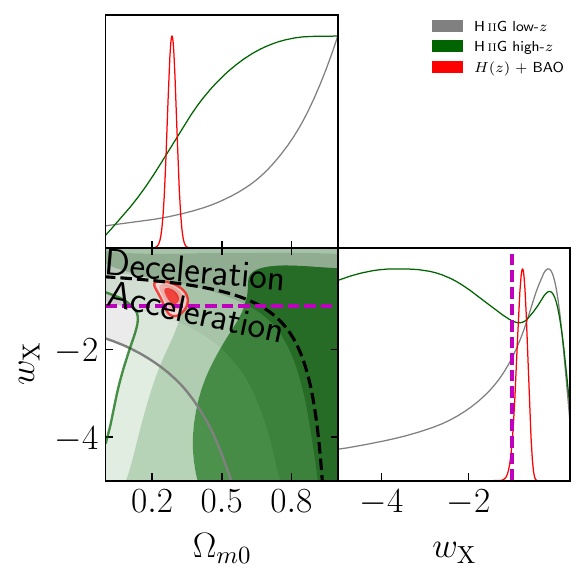}}
 \subfloat[]{%
    \includegraphics[width=0.4\textwidth,height=0.35\textwidth]{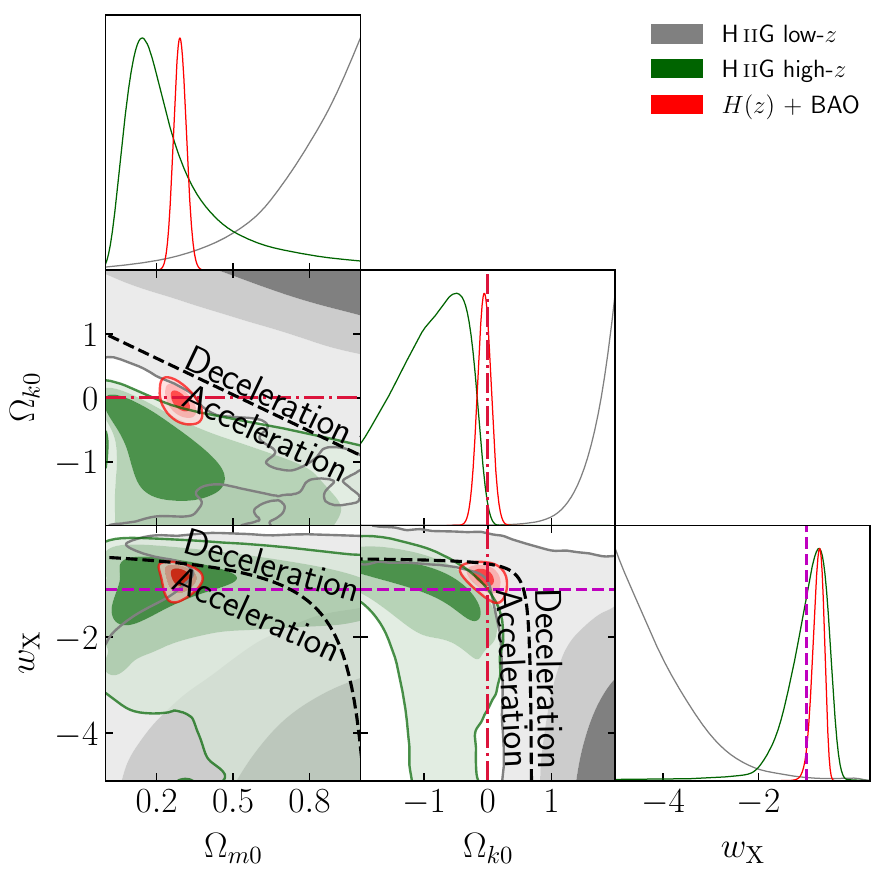}}\\
 \subfloat[]{%
    \includegraphics[width=0.4\textwidth,height=0.35\textwidth]{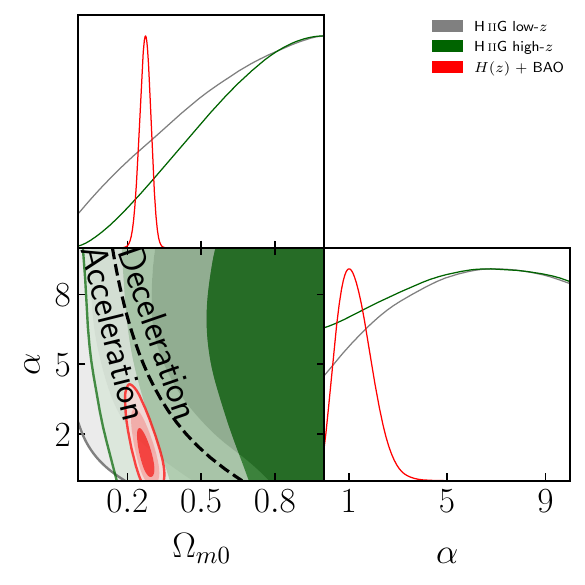}}
 \subfloat[]{%
    \includegraphics[width=0.4\textwidth,height=0.35\textwidth]{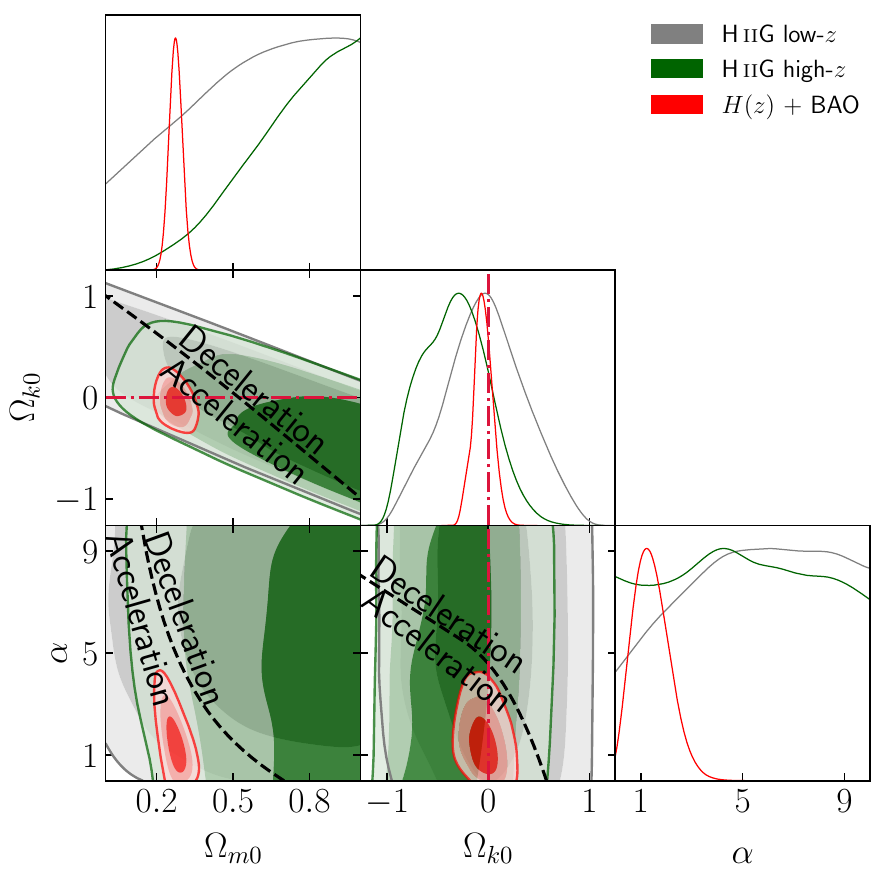}}\\
\caption{Same as Fig. \ref{fig1}, but excluding \hiig\ low-$z$ + high-$z$ data results, and for cosmological parameters only.}
\label{fig2}
\end{figure*}

In Fig.\ \ref{fig1} we present all-parameter triangle plots for all models; the corresponding only cosmological parameter triangle plots are in Fig.\ \ref{fig2}. These are for \hiig\ low-$z$, high-$z$, low-$z$ + high-$z$ (not in Fig.\ \ref{fig2}), and $H(z)$ + BAO data.

\begin{turnpage}
\begin{table*}
\centering
\resizebox*{2.7\columnwidth}{1.8\columnwidth}{%
\begin{threeparttable}
\caption{Unmarginalized best-fitting parameter values for all models from \hiig\ data.}\label{tab:BFP}
\begin{tabular}{lcccccccccccccccccccc}
\toprule\toprule
Model & Data set & $\Omega_{b}h^2$ & $\Omega_{c}h^2$ & $\Omega_{m0}$ & $\Omega_{k0}$ & $w_{\mathrm{X}}$/$\alpha$\tnote{a} & $H_0$\tnote{b} & $\beta_{\mathrm{low}}$ & $\gamma_{\mathrm{low}}$ & $\sigma_{\mathrm{int,low}}$ & $\beta_{\mathrm{high}}$ & $\gamma_{\mathrm{high}}$ & $\sigma_{\mathrm{int,high}}$ & $-2\ln\mathcal{L}_{\mathrm{max}}$ & AIC & BIC & DIC & $\Delta \mathrm{AIC}$ & $\Delta \mathrm{BIC}$ & $\Delta \mathrm{DIC}$ \\
\midrule
Flat \lcdm & $H(z)$ + BAO & 0.0254 & 0.1200 & 0.297 & $\cdots$ & $\cdots$ & 70.12 & $\cdots$ & $\cdots$ & $\cdots$ & $\cdots$ & $\cdots$ & $\cdots$ & 30.56 & 36.56 & 41.91 & 37.32 & 0.00 & 0.00 & 0.00\\
 & \hiig\ low-$z$\tnote{c} & $\cdots$ & $\cdots$ & 0.999 & $\cdots$ & $\cdots$ & $\cdots$ & 3.868 & 35.07 & 0.676 & $\cdots$ & $\cdots$ & $\cdots$ & 251.03 & 259.03 & 271.83 & 261.21 & 0.00 & 0.00 & 0.00\\
 & \hiig\ high-$z$\tnote{c} & $\cdots$ & $\cdots$ & 1.000 & $\cdots$ & $\cdots$ & $\cdots$ & $\cdots$ & $\cdots$ & $\cdots$ & 1.853 & 38.37 & 0.680 & 169.97 & 177.97 & 190.76 & 178.79 & 0.00 & 0.00 & 0.00\\
 & \hiig\ low-$z$ + high-$z$\tnote{c} & $\cdots$ & $\cdots$ & 0.997 & $\cdots$ & $\cdots$ & $\cdots$ & 3.832 & 35.13 & 0.674 & 1.945 & 38.21 & 0.686 & 421.14 & 435.14 & 457.53 & 437.38 & 0.00 & 0.00 & 0.00\\[6pt]
Nonflat \lcdm & $H(z)$ + BAO & 0.0269 & 0.1128 & 0.289 & 0.041 & $\cdots$ & 69.61 & $\cdots$ & $\cdots$ & $\cdots$ & $\cdots$ & $\cdots$ & $\cdots$ & 30.34 & 38.34 & 45.48 & 38.80 & 1.78 & 3.56 & 1.48\\
 & \hiig\ low-$z$\tnote{c} & $\cdots$ & $\cdots$ & 0.994 & 1.997 & $\cdots$ & $\cdots$ & 3.752 & 35.21 & 0.649 & $\cdots$ & $\cdots$ & $\cdots$ & 245.74 & 255.74 & 271.73 & 259.41 & $-3.29$ & $-0.09$ & $-1.80$\\
 & \hiig\ high-$z$\tnote{c} & $\cdots$ & $\cdots$ & 0.114 & $-0.453$ & $\cdots$ & $\cdots$ & $\cdots$ & $\cdots$ & $\cdots$ & 1.193 & 39.92 & 0.555 & 139.81 & 149.81 & 165.80 & 158.91 & $-28.16$ & $-24.96$ & $-19.89$\\
 & \hiig\ low-$z$ + high-$z$\tnote{c} & $\cdots$ & $\cdots$ & 0.116 & $-0.459$ & $\cdots$ & $\cdots$ & 3.965 & 34.96 & 0.694 & 1.150 & 39.99 & 0.568 & 397.86 & 413.86 & 439.45 & 422.28 & $-21.27$ & $-18.08$ & $-15.10$\\[6pt]
Flat XCDM & $H(z)$ + BAO & 0.0318 & 0.0938 & 0.283 & $\cdots$ & $-0.734$ & 66.67 & $\cdots$ & $\cdots$ & $\cdots$ & $\cdots$ & $\cdots$ & $\cdots$ & 26.58 & 34.58 & 41.71 & 34.83 & $-1.98$ & $-0.20$ & $-2.49$\\
 & \hiig\ low-$z$\tnote{c} & $\cdots$ & $\cdots$ & 0.017 & $\cdots$ & 0.134 & $\cdots$ & 3.873 & 35.06 & 0.673 & $\cdots$ & $\cdots$ & $\cdots$ & 250.32 & 260.32 & 276.31 & 261.27 & 1.28 & 4.48 & 0.06\\
 & \hiig\ high-$z$\tnote{c} & $\cdots$ & $\cdots$ & 0.039 & $\cdots$ & 0.137 & $\cdots$ & $\cdots$ & $\cdots$ & $\cdots$ & 1.855 & 38.29 & 0.672 & 168.35 & 178.35 & 194.34 & 180.30 & 0.38 & 3.58 & 1.51\\
 & \hiig\ low-$z$ + high-$z$\tnote{c} & $\cdots$ & $\cdots$ & 0.050 & $\cdots$ & 0.142 & $\cdots$ & 3.907 & 35.01 & 0.671 & 1.896 & 38.20 & 0.685 & 418.83 & 434.83 & 460.42 & 438.91 & $-0.31$ & 2.89 & 1.53\\[6pt]
Nonflat XCDM & $H(z)$ + BAO & 0.0305 & 0.0998 & 0.293 & $-0.084$ & $-0.703$ & 66.79 & $\cdots$ & $\cdots$ & $\cdots$ & $\cdots$ & $\cdots$ & $\cdots$ & 26.00 & 36.00 & 44.92 & 36.11 & $-0.56$ & 3.01 & $-1.21$\\
 & \hiig\ low-$z$\tnote{c} & $\cdots$ & $\cdots$ & 0.996 & 1.990 & $-4.971$ & $\cdots$ & 3.518 & 35.46 & 0.613 & $\cdots$ & $\cdots$ & $\cdots$ & 233.25 & 245.25 & 264.44 & 250.63 & $-13.78$ & $-7.38$ & $-10.58$\\
 & \hiig\ high-$z$\tnote{c} & $\cdots$ & $\cdots$ & 0.068 & $-0.225$ & $-1.350$ & $\cdots$ & $\cdots$ & $\cdots$ & $\cdots$ & 1.224 & 40.15 & 0.550 & 138.50 & 150.50 & 169.69 & 162.58 & $-27.47$ & $-21.07$ & $-16.21$\\
 & \hiig\ low-$z$ + high-$z$\tnote{c} & $\cdots$ & $\cdots$ & 0.097 & $-0.373$ & $-1.078$ & $\cdots$ & 3.959 & 34.98 & 0.727 & 1.127 & 40.11 & 0.556 & 397.98 & 415.98 & 444.76 & 426.34 & $-19.16$ & $-12.76$ & $-11.04$\\[6pt]
Flat \pcdm & $H(z)$ + BAO & 0.0336 & 0.0866 & 0.271 & $\cdots$ & 1.157 & 66.80 & $\cdots$ & $\cdots$ & $\cdots$ & $\cdots$ & $\cdots$ & $\cdots$ & 26.50 & 34.50 & 41.64 & 34.15 & $-2.05$ & $-0.27$ & $-3.17$\\
 & \hiig\ low-$z$\tnote{c} & $\cdots$ & $\cdots$ & 0.999 & $\cdots$ & 2.334 & $\cdots$ & 3.864 & 35.08 & 0.673 & $\cdots$ & $\cdots$ & $\cdots$ & 251.03 & 261.03 & 277.02 & 259.51 & 2.00 & 5.20 & $-1.69$\\
 & \hiig\ high-$z$\tnote{c} & $\cdots$ & $\cdots$ & 1.000 & $\cdots$ & 2.263 & $\cdots$ & $\cdots$ & $\cdots$ & $\cdots$ & 1.856 & 38.37 & 0.676 & 169.97 & 179.97 & 195.96 & 178.59 & 2.00 & 5.20 & $-0.20$\\
 & \hiig\ low-$z$ + high-$z$\tnote{c} & $\cdots$ & $\cdots$ & 0.984 & $\cdots$ & 9.923 & $\cdots$ & 3.870 & 35.07 & 0.665 & 1.899 & 38.29 & 0.672 & 421.08 & 437.08 & 462.67 & 436.70 & 1.95 & 5.15 & $-0.68$\\[6pt]
Nonflat \pcdm & $H(z)$ + BAO & 0.0337 & 0.0894 & 0.275 & $-0.074$ & 1.393 & 67.16 & $\cdots$ & $\cdots$ & $\cdots$ & $\cdots$ & $\cdots$ & $\cdots$ & 25.92 & 35.92 & 44.84 & 35.29 & $-0.64$ & 2.93 & $-2.03$\\
 & \hiig\ low-$z$\tnote{c} & $\cdots$ & $\cdots$ & 0.979 & $-0.958$ & 9.937 & $\cdots$ & 3.807 & 35.16 & 0.663 & $\cdots$ & $\cdots$ & $\cdots$ & 249.91 & 261.91 & 281.10 & 260.41 & 2.88 & 9.27 & $-0.80$\\
 & \hiig\ high-$z$\tnote{c} & $\cdots$ & $\cdots$ & 0.999 & $-0.999$ & 0.027 & $\cdots$ & $\cdots$ & $\cdots$ & $\cdots$ & 1.897 & 38.30 & 0.664 & 164.89 & 176.89 & 196.08 & 180.71 & $-1.08$ & 5.32 & 1.92\\
 & \hiig\ low-$z$ + high-$z$\tnote{c} & $\cdots$ & $\cdots$ & 0.969 & $-0.966$ & 8.392 & $\cdots$ & 3.795 & 35.19 & 0.674 & 1.881 & 38.19 & 0.653 & 417.01 & 435.01 & 463.80 & 436.93 & $-0.12$ & 6.27 & $-0.45$\\
\bottomrule\bottomrule
\end{tabular}
\begin{tablenotes}
\item [a] \wx\ corresponds to flat/nonflat XCDM and $\alpha$ corresponds to flat/nonflat \pcdm.
\item [b] \hunit.
\item [c] $\Omega_b=0.05$ and $H_0=70$ \hunit\ with prior range of $\Omega_{m0}\in[0,1]$.
\end{tablenotes}
\end{threeparttable}%
}
\end{table*}
\end{turnpage}

\begin{turnpage}
\begin{table*}
\centering
\resizebox*{2.7\columnwidth}{1.8\columnwidth}{%
\begin{threeparttable}
\caption{One-dimensional marginalized posterior mean values and uncertainties ($\pm 1\sigma$ error bars or $2\sigma$ limits) of the parameters for all models from various combinations of data.}\label{tab:1d_BFP}
\begin{tabular}{lccccccccccccc}
\toprule\toprule
Model & Data set & $\Omega_{b}h^2$ & $\Omega_{c}h^2$ & $\Omega_{m0}$ & $\Omega_{k0}$ & $w_{\mathrm{X}}$/$\alpha$\tnote{a} & $H_0$\tnote{b} & $\beta_{\mathrm{low}}$ & $\gamma_{\mathrm{low}}$ & $\sigma_{\mathrm{int,low}}$ & $\beta_{\mathrm{high}}$ & $\gamma_{\mathrm{high}}$ & $\sigma_{\mathrm{int,high}}$ \\
\midrule
Flat \lcdm & $H(z)$ + BAO & $0.0260\pm0.0040$ & $0.1212^{+0.0091}_{-0.0101}$ & $0.297^{+0.015}_{-0.017}$ & $\cdots$ & $\cdots$ & $70.49\pm2.74$ & $\cdots$ & $\cdots$ & $\cdots$ & $\cdots$ & $\cdots$ & $\cdots$ \\
 & \hiig\ low-$z$\tnote{c} & $\cdots$ & $\cdots$ & $>0.177$ & $\cdots$ & $\cdots$ & $\cdots$ & $3.896\pm0.247$ & $35.04\pm0.39$ & $0.701^{+0.062}_{-0.074}$ & $\cdots$ & $\cdots$ & $\cdots$ \\%
 & \hiig\ high-$z$\tnote{c} & $\cdots$ & $\cdots$ & $>0.310$ & $\cdots$ & $\cdots$ & $\cdots$ & $\cdots$ & $\cdots$ & $\cdots$ & $1.912\pm0.401$ & $38.38\pm0.69$ & $0.713^{+0.071}_{-0.087}$ \\%
 & \hiig\ low-$z$ + high-$z$\tnote{c} & $\cdots$ & $\cdots$ & $>0.446$ & $\cdots$ & $\cdots$ & $\cdots$ & $3.885\pm0.238$ & $35.05\pm0.37$ & $0.697^{+0.060}_{-0.072}$ & $1.897\pm0.396$ & $38.36\pm0.68$ & $0.708^{+0.068}_{-0.085}$ \\[6pt]%
Nonflat \lcdm & $H(z)$ + BAO & $0.0275^{+0.0046}_{-0.0051}$ & $0.1131^{+0.0180}_{-0.0181}$ & $0.289\pm0.023$ & $0.047^{+0.082}_{-0.089}$ & $\cdots$ & $69.81\pm2.80$ & $\cdots$ & $\cdots$ & $\cdots$ & $\cdots$ & $\cdots$ & $\cdots$ \\
 & \hiig\ low-$z$\tnote{c} & $\cdots$ & $\cdots$ & $>0.548$\tnote{d} & $>-0.148$ & $\cdots$ & $\cdots$ & $3.829\pm0.241$ & $35.12\pm0.37$ & $0.687^{+0.061}_{-0.073}$ & $\cdots$ & $\cdots$ & $\cdots$ \\%
 & \hiig\ high-$z$\tnote{c} & $\cdots$ & $\cdots$ & $0.295^{+0.034}_{-0.208}$ & $-0.797^{+0.402}_{-0.100}$ & $\cdots$ & $\cdots$ & $\cdots$ & $\cdots$ & $\cdots$ & $1.245\pm0.375$ & $39.64\pm0.66$ & $0.604^{+0.060}_{-0.075}$ \\
 & \hiig\ low-$z$ + high-$z$\tnote{c} & $\cdots$ & $\cdots$ & $0.274^{+0.009}_{-0.189}$ & $-0.746^{+0.359}_{-0.061}$ & $\cdots$ & $\cdots$ & $4.010\pm0.246$ & $34.89\pm0.38$ & $0.718^{+0.062}_{-0.074}$ & $1.253\pm0.369$ & $39.65\pm0.65$ & $0.599^{+0.058}_{-0.075}$  \\[6pt]%
Flat XCDM & $H(z)$ + BAO & $0.0308^{+0.0053}_{-0.0046}$ & $0.0978^{+0.0184}_{-0.0164}$ & $0.285\pm0.019$ & $\cdots$ & $-0.776^{+0.130}_{-0.103}$ & $67.18\pm3.18$ & $\cdots$ & $\cdots$ & $\cdots$ & $\cdots$ & $\cdots$ & $\cdots$ \\
 & \hiig\ low-$z$\tnote{c} & $\cdots$ & $\cdots$ & $>0.128$ & $\cdots$ & $-1.442^{+1.601}_{-0.465}$ & $\cdots$ & $3.895\pm0.249$ & $35.04\pm0.39$ & $0.699^{+0.063}_{-0.074}$ & $\cdots$ & $\cdots$ & $\cdots$ \\%
 & \hiig\ high-$z$\tnote{c} & $\cdots$ & $\cdots$ & $>0.182$ & $\cdots$ & $-2.477^{+2.619}_{-2.516}$ & $\cdots$ & $\cdots$ & $\cdots$ & $\cdots$ & $1.901\pm0.409$ & $38.46\pm0.72$ & $0.713^{+0.071}_{-0.088}$ \\%
 & \hiig\ low-$z$ + high-$z$\tnote{c} & $\cdots$ & $\cdots$ & $>0.160$ & $\cdots$ & $-1.491^{+1.655}_{-0.704}$ & $\cdots$ & $3.882\pm0.240$ & $35.06\pm0.37$ & $0.697^{+0.061}_{-0.073}$ & $1.883\pm0.395$ & $38.36\pm0.68$ & $0.707^{+0.068}_{-0.085}$ \\[6pt]%
Nonflat XCDM & $H(z)$ + BAO & $0.0303^{+0.0054}_{-0.0048}$ & $0.1021\pm0.0193$ & $0.292\pm0.024$ & $-0.054\pm0.103$ & $-0.757^{+0.135}_{-0.093}$ & $67.33\pm2.96$ & $\cdots$ & $\cdots$ & $\cdots$ & $\cdots$ & $\cdots$ & $\cdots$ \\
 & \hiig\ low-$z$\tnote{c} & $\cdots$ & $\cdots$ & $>0.331$ & $>1.079$ & $<-2.269$ & $\cdots$ & $3.599\pm0.240$ & $35.39\pm0.37$ & $0.653^{+0.060}_{-0.070}$ & $\cdots$ & $\cdots$ & $\cdots$ \\%
 & \hiig\ high-$z$\tnote{c} & $\cdots$ & $\cdots$ & $0.276^{+0.041}_{-0.219}$ & $-0.927^{+0.718}_{-0.336}$ & $-1.040^{+0.540}_{-0.140}$ & $\cdots$ & $\cdots$ & $\cdots$ & $\cdots$ & $1.264\pm0.392$ & $39.59\pm0.71$ & $0.610^{+0.063}_{-0.079}$ \\
 & \hiig\ low-$z$ + high-$z$\tnote{c} & $\cdots$ & $\cdots$ & $0.340^{+0.042}_{-0.276}$ & $<-0.888$\tnote{d} & $<-0.386$\tnote{e} & $\cdots$ & $3.920^{+0.285}_{-0.256}$ & $35.01\pm0.41$ & $0.704^{+0.065}_{-0.076}$ & $1.359^{+0.376}_{-0.504}$ & $39.22^{+0.95}_{-0.66}$ & $0.627^{+0.060}_{-0.094}$ \\[6pt]%
Flat \pcdm & $H(z)$ + BAO & $0.0326^{+0.0061}_{-0.0030}$ & $0.0866^{+0.0197}_{-0.0180}$ & $0.272^{+0.024}_{-0.022}$ & $\cdots$ & $1.271^{+0.507}_{-0.836}$ & $66.19^{+2.89}_{-2.88}$ & $\cdots$ & $\cdots$ & $\cdots$ & $\cdots$ & $\cdots$ & $\cdots$ \\
 & \hiig\ low-$z$\tnote{c} & $\cdots$ & $\cdots$ & $>0.480$\tnote{d} & $\cdots$ & $>3.833$\tnote{d} & $\cdots$ & $3.883\pm0.240$ & $35.06\pm0.38$ & $0.698^{+0.061}_{-0.072}$ & $\cdots$ & $\cdots$ & $\cdots$ \\%
 & \hiig\ high-$z$\tnote{c} & $\cdots$ & $\cdots$ & $>0.243$ & $\cdots$ & $\cdots$ & $\cdots$ & $\cdots$ & $\cdots$ & $\cdots$ & $1.909\pm0.391$ & $38.34\pm0.68$ & $0.712^{+0.069}_{-0.086}$ \\%
 & \hiig\ low-$z$ + high-$z$\tnote{c} & $\cdots$ & $\cdots$ & $>0.317$ & $\cdots$ & $>3.863$\tnote{d} & $\cdots$ & $3.875\pm0.236$ & $35.07\pm0.37$ & $0.695^{+0.060}_{-0.071}$ & $1.899\pm0.397$ & $38.34\pm0.68$ & $0.709^{+0.068}_{-0.085}$ \\[6pt]%
Nonflat \pcdm & $H(z)$ + BAO & $0.0325^{+0.0064}_{-0.0029}$ & $0.0881^{+0.0199}_{-0.0201}$ & $0.275\pm0.025$ & $-0.052^{+0.093}_{-0.087}$ & $1.427^{+0.572}_{-0.830}$ & $66.24\pm2.88$ & $\cdots$ & $\cdots$ & $\cdots$ & $\cdots$ & $\cdots$ & $\cdots$ \\
 & \hiig\ low-$z$\tnote{c} & $\cdots$ & $\cdots$ & $>0.430$\tnote{d} & $-0.027^{+0.403}_{-0.393}$ & $5.402^{+4.376}_{-1.747}$ & $\cdots$ & $3.886\pm0.240$ & $35.05\pm0.37$ & $0.696^{+0.061}_{-0.072}$ & $\cdots$ & $\cdots$ & $\cdots$ \\%
 & \hiig\ high-$z$\tnote{c} & $\cdots$ & $\cdots$ & $>0.337$ & $-0.331^{+0.329}_{-0.354}$ & $\cdots$ & $\cdots$ & $\cdots$ & $\cdots$ & $\cdots$ & $1.863\pm0.404$ & $38.37\pm0.69$ & $0.703^{+0.069}_{-0.086}$ \\%
 & \hiig\ low-$z$ + high-$z$\tnote{c} & $\cdots$ & $\cdots$ & $>0.422$ & $-0.385^{+0.273}_{-0.405}$ & $\cdots$ & $\cdots$ & $3.870\pm0.237$ & $35.07\pm0.37$ & $0.694^{+0.060}_{-0.072}$ & $1.858\pm0.394$ & $38.36\pm0.68$ & $0.700^{+0.068}_{-0.085}$ \\%
\bottomrule\bottomrule
\end{tabular}
\begin{tablenotes}
\item [a] \wx\ corresponds to flat/nonflat XCDM and $\alpha$ corresponds to flat/nonflat \pcdm.
\item [b] \hunit.
\item [c] $\Omega_b=0.05$ and $H_0=70$ \hunit\ with prior range of $\Omega_{m0}\in[0,1]$.
\item [d] This is the 1$\sigma$ limit. The 2$\sigma$ limit is set by the prior and not shown here.
\item [e] The 2$\sigma$ limit is shown here because the posterior mean value is lower than 1$\sigma$ lower limit.
\end{tablenotes}
\end{threeparttable}%
}
\end{table*}
\end{turnpage}

\begin{table*}
\centering
\begin{threeparttable}
\caption{The largest differences between models from \hiig\ data, where $1\sigma$ means the quadrature sum of the two corresponding $1\sigma$ error bars.}\label{tab:diff}
\setlength{\tabcolsep}{16pt}
\begin{tabular}{lcccccc}
\toprule\toprule
 Data set & $\Delta\beta_{\mathrm{low}}$ & $\Delta\gamma_{\mathrm{low}}$ & $\Delta\sigma_{\mathrm{int,low}}$ & $\Delta\beta_{\mathrm{high}}$ & $\Delta\gamma_{\mathrm{high}}$ & $\Delta\sigma_{\mathrm{int,high}}$ \\
\midrule
\hiig\ low-$z$ & $0.86\sigma$ & $0.65\sigma$ & $0.50\sigma$ & $\cdots$ & $\cdots$ & $\cdots$\\
\hiig\ high-$z$ & $\cdots$ & $\cdots$ & $\cdots$ & $1.21\sigma$ & $1.37\sigma$ & $1.03\sigma$\\
\hiig\ low-$z$ + high-$z$ & $0.41\sigma$ & $0.33\sigma$ & $0.25\sigma$ & $1.19\sigma$ & $1.39\sigma$ & $1.07\sigma$\\
\bottomrule\bottomrule
\end{tabular}
\end{threeparttable}%
\end{table*}

\begin{table}
\begin{threeparttable}
\caption{Low-$z$ and high-$z$ \hiig\ data $L-\sigma$ correlation parameters differences, where $1\sigma$ means the quadrature sum of the two corresponding $1\sigma$ error bars.}
\label{tab:comp}
\setlength{\tabcolsep}{22pt}
\begin{tabular}{lcc}
\toprule\toprule
Model & $\Delta\beta$ & $\Delta\gamma$ \\
\midrule
Flat \lcdm\  & $4.21\sigma$ & $-4.21\sigma$ \\
Nonflat \lcdm\ & $5.80\sigma$ & $-5.97\sigma$ \\
Flat XCDM  & $4.16\sigma$ & $-4.18\sigma$ \\
Nonflat XCDM  & $5.08\sigma$ & $-5.25\sigma$ \\
Flat $\phi$CDM  & $4.30\sigma$ & $-4.21\sigma$ \\
Nonflat $\phi$CDM & $4.31\sigma$ & $-4.24\sigma$ \\
\bottomrule\bottomrule
\end{tabular}
\end{threeparttable}
\end{table}

We list unmarginalized best-fitting parameter values in Table \ref{tab:BFP}, along with corresponding maximum likelihood $\mathcal{L}_{\rm max}$, AIC, BIC, DIC, $\Delta \mathrm{AIC}$, $\Delta \mathrm{BIC}$, and $\Delta \mathrm{DIC}$ values, for all models and data sets. In Table \ref{tab:1d_BFP} we list the one dimensional marginalized posterior mean parameter values and their uncertainties ($\pm 1\sigma$ error bars and 1 or 2$\sigma$ limits) for all models and data sets.

Table \ref{tab:diff} lists the largest differences between $L-\sigma$ relation (and $\sigma_{\rm int}$) parameter values measured in the different cosmological models, for \hiig\ low-$z$, high-$z$, and low-$z$ + high-$z$ data. Table \ref{tab:comp} lists the differences in the $L-\sigma$ intercept and slope parameters between low-$z$ and high-$z$ \hiig\ data, for all six cosmological models. 

From the small $\Delta\beta$ and $\Delta\gamma$ values for \hiig\ low-$z$ and \hiig\ high-$z$ data in Table \ref{tab:diff} we conclude that both low-$z$ \hiig\ and high-$z$ \hiig\ data are standardizable candles.

Panels (b) and (d) of Fig.\ \ref{fig2}, for the nonflat \lcdm\ and nonflat XCDM models, show that low-$z$ and high-$z$ \hiig\ data favor different regions of cosmological parameters space, with significant amount of 2$\sigma$ contours not overlapping. This is consistent with the different $\Omega_{m0}$, $\Omega_{k0}$, and $w_{\mathrm{X}}$ (for nonflat XCDM) low-$z$ and high-$z$ \hiig\ data limits in Table \ref{tab:1d_BFP} for these two models. Assuming that the low-$z$ and high-$z$ \hiig\ data sets are correct, this inconsistency between low-$z$ and high-$z$ data constraints means that these data rule out the nonflat \lcdm\ and nonflat XCDM models. 

Parenthetically we note that excluding the nonflat \lcdm\ and nonflat XCDM models, the largest differences between $L-\sigma$ relation parameter values measured in the remaining four different cosmological models are $\Delta\beta_{\mathrm{low}} = 0.038\sigma$ and $\Delta\gamma_{\mathrm{low}} = 0.037\sigma$ for \hiig\ low-$z$ data and $\Delta\beta_{\mathrm{high}} = 0.086\sigma$ and  $\Delta\gamma_{\mathrm{high}} = 0.121\sigma$ for high-$z$ data. These are significantly smaller than the corresponding six cosmological model values listed in Table \ref{tab:diff}, indicating that in models where the low-$z$ and high-$z$ cosmological constraints are not mutually inconsistent both low-$z$ \hiig\ and high-$z$ \hiig\ data are good standardizable candles.

More significantly, when we compare the differences in the $L-\sigma$ intercept and slope parameters between low-$z$ and high-$z$ \hiig\ data in each of the six cosmological models, we find significant differences at the $\sim 4-5\sigma$ level, as shown in Table \ref{tab:comp}. 

\begin{figure*}
\centering
 \subfloat[Flat \lcdm]{%
    \includegraphics[width=0.45\textwidth,height=0.35\textwidth]{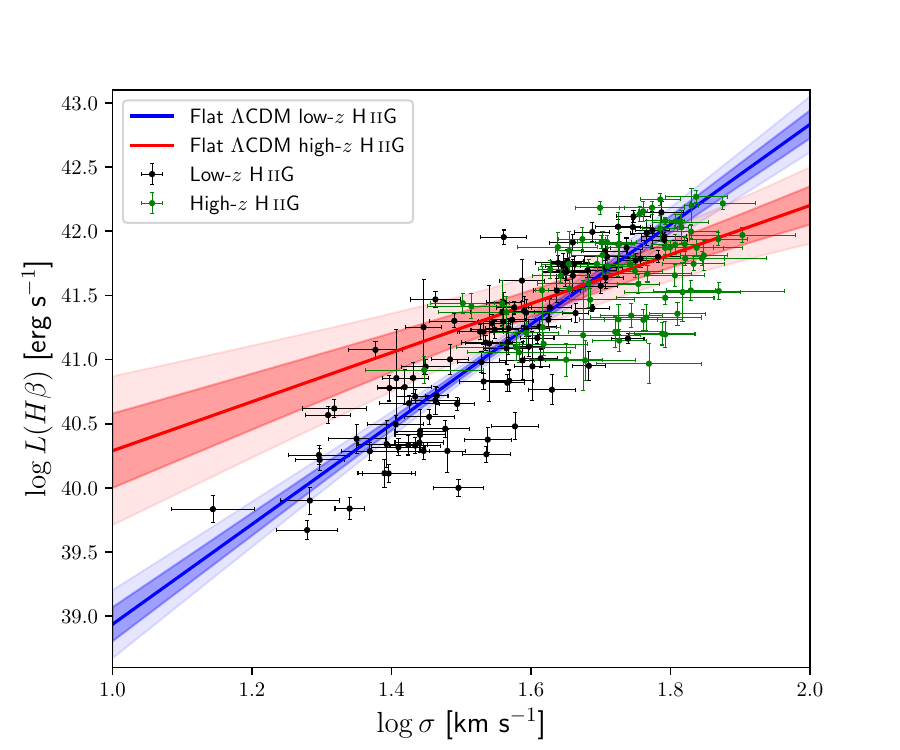}}
 \subfloat[Nonflat \lcdm]{%
    \includegraphics[width=0.45\textwidth,height=0.35\textwidth]{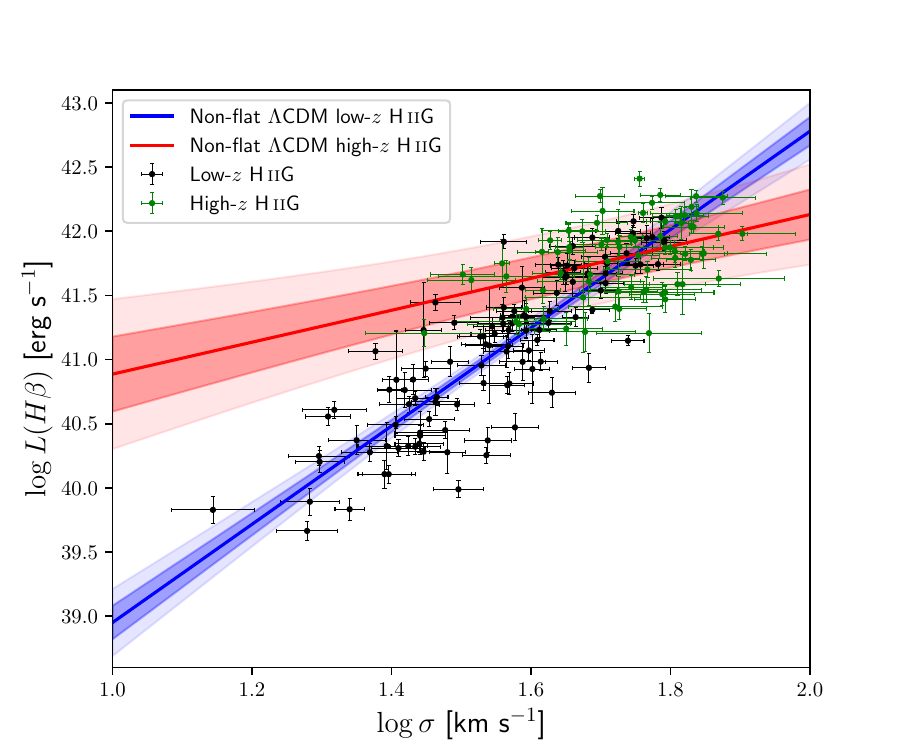}}\\
 \subfloat[Flat XCDM]{%
    \includegraphics[width=0.45\textwidth,height=0.35\textwidth]{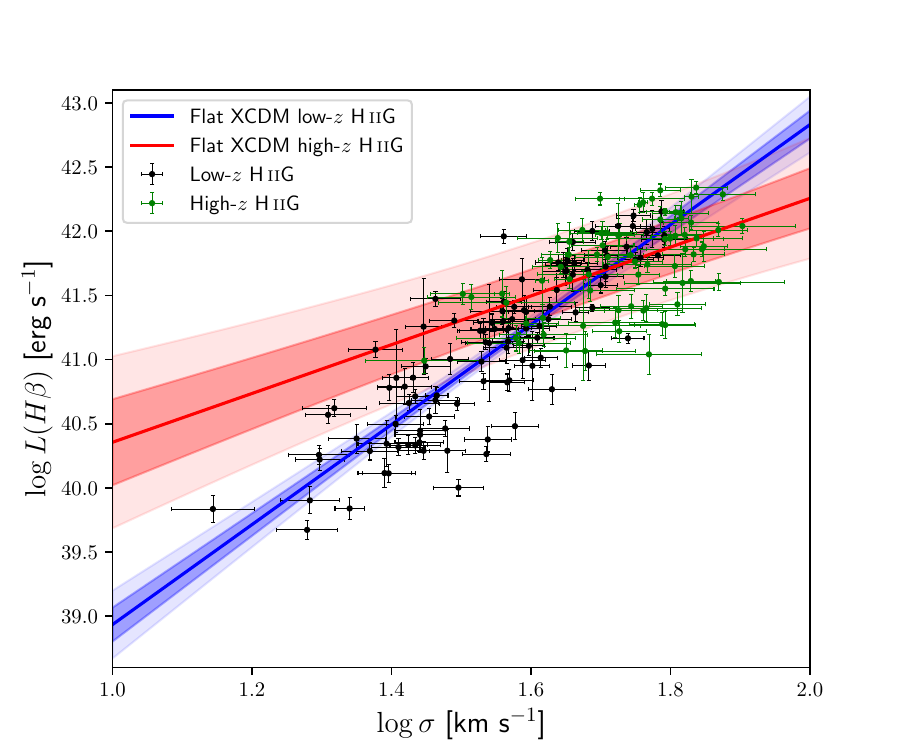}}
 \subfloat[Nonflat XCDM]{%
    \includegraphics[width=0.45\textwidth,height=0.35\textwidth]{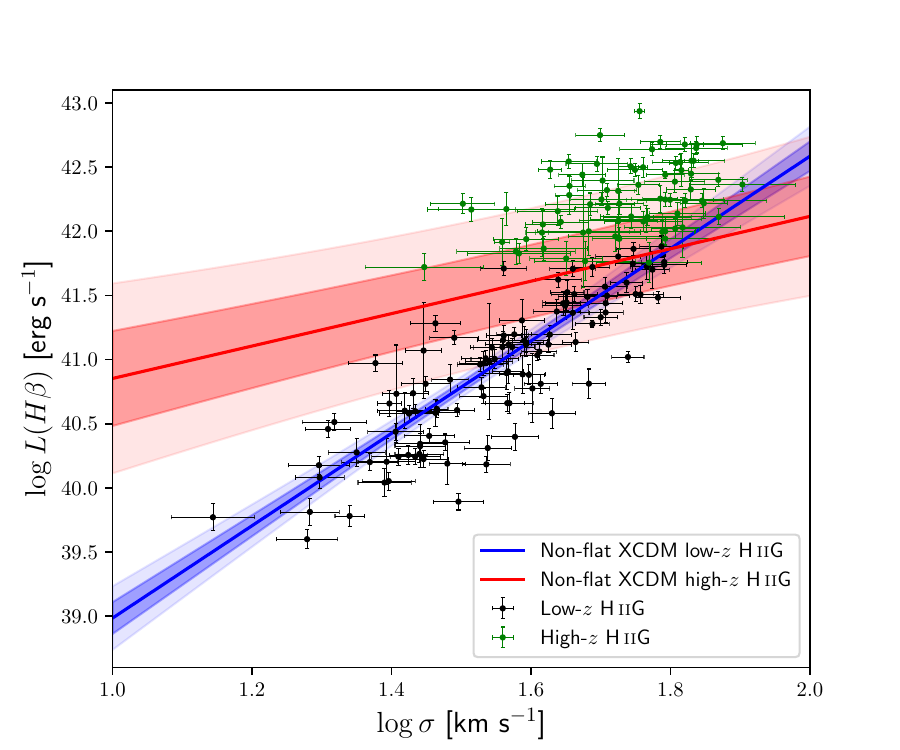}}\\
 \subfloat[Flat \pcdm]{%
    \includegraphics[width=0.45\textwidth,height=0.35\textwidth]{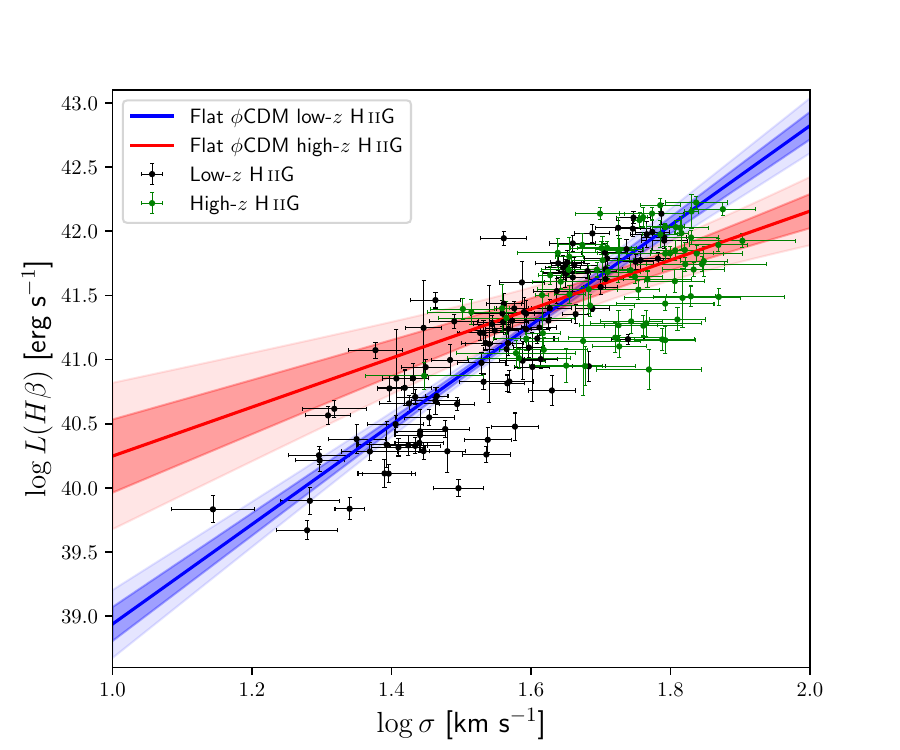}}
 \subfloat[Nonflat \pcdm]{%
    \includegraphics[width=0.45\textwidth,height=0.35\textwidth]{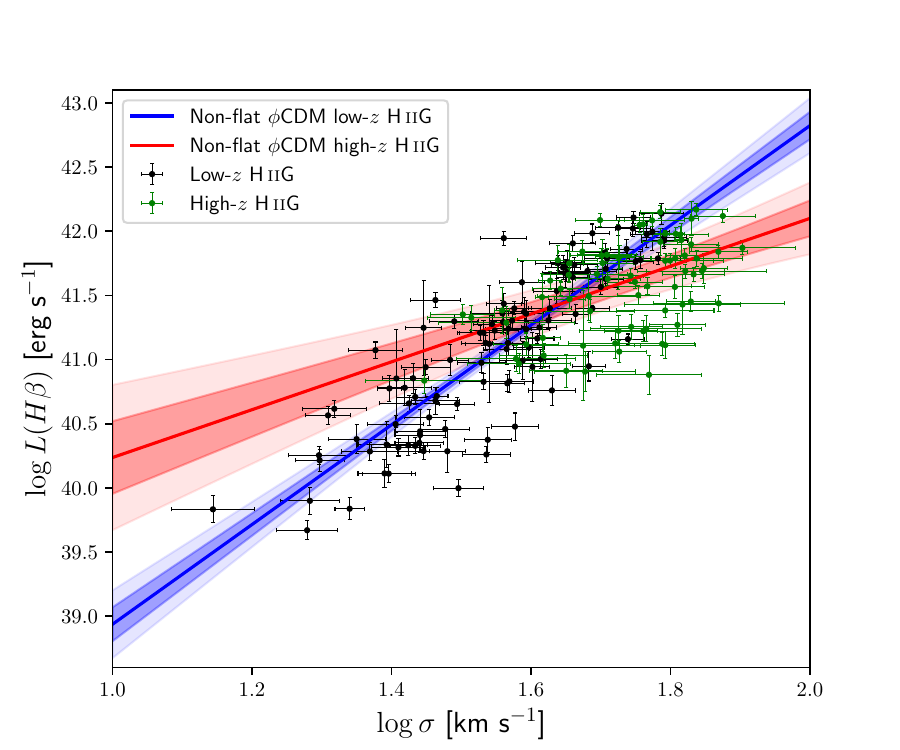}}\\
\caption{$\log L(H\beta)$--$\log \sigma$ relations for six different models, with \lcdm, XCDM, and \pcdm\ in the top, middle, and bottom rows, and flat (nonflat) models in the left (right) column. Here data point $D_L$ values are computed using the corresponding posterior mean or best-fitting (for the nonflat XCDM parameterization due to instability around the mean) values listed in Tables \ref{tab:1d_BFP} or \ref{tab:BFP} and the $\log L(H\beta)$--$\log \sigma$ relations are derived from Monte Carlo simulations with given posterior mean values and covariance matrices of the intercept and slope parameters.}
\label{fig3}
\end{figure*}

Supporting these results, in Fig.\ \ref{fig3} we display the $L-\sigma$ relations determined in each of the six cosmological models for both  low-$z$ and high-$z$ \hiig\ data, as well as the data points. The data point $D_L$ values are computed using the posterior mean or best-fitting (for the nonflat XCDM parameterization due to instability around the mean) cosmological parameter values from Table \ref{tab:1d_BFP} or \ref{tab:BFP}. The $L-\sigma$ relations are derived from Monte Carlo simulations constructed using the posterior mean values of the slope and intercept parameters listed in Table \ref{tab:1d_BFP} along with their corresponding covariance matrices. While, for each of the two data sets, each model has different $L-\sigma$ relation and data predictions, the high-$z$ data results appear to exhibit more variability between models, and, for the reason mentioned above, the high-$z$ $L-\sigma$ relation (red line and bands) for the nonflat XCDM parameterization does not align well with their (green) data points. In contrast, the (black) low-$z$ data points seem less variable between models and seem to align well with their (blue) $L-\sigma$ relation.

The significantly different low-$z$ data $L-\sigma$ relations (blue lines and bands) and high-$z$ data $L-\sigma$ relations (red lines and bands) in each panel of Fig.\ \ref{fig3} support the numerical results of Table \ref{tab:comp} that show significant differences at the $\sim 4-5\sigma$ level in the $L-\sigma$ intercept and slope parameters between low-$z$ and high-$z$ \hiig\ data. This means that it is incorrect to assume that low-$z$ and high-$z$ \hiig\ data obey the same $L-\sigma$ relations when jointly analyzing low-$z$ + high-$z$ data, as has been assumed in all previous analyses of these data. However, since the results of Table \ref{tab:diff} indicate that both low-$z$ \hiig\ and high-$z$ \hiig\ data are standardizable candles, it is still possible to jointly analyze low-$z$ and high-$z$ data but it is necessary to use different $\beta$, $\gamma$, and $\sigma_{\mathrm{int}}$ parameters for low-$z$ and high-$z$ data, and this is what we do in our joint low-$z$ + high-$z$ data analyses. However, this doubles the number of nuisance parameters and so reduces the effectiveness of \hiig\ data (as far as cosmology is concerned). On the other hand, if the difference between the low-$z$ and high-$z$ $L-\sigma$ relations is real, then we have discovered something new, and possibly interesting, about H\,\textsc{ii}Gs.

From the small $\Delta\beta$ and $\Delta\gamma$ values from the six cosmological models joint \hiig\ low-$z$ + high-$z$ data analyses in the last line of Table \ref{tab:diff} we conclude that low-$z$ + high-$z$ data are standardizable candles. Excluding the nonflat \lcdm\ and nonflat XCDM models, the largest differences between $L-\sigma$ relation parameter values measured in the remaining four different cosmological models joint \hiig\ low-$z$ + high-$z$ data analyses are $\Delta\beta_{\mathrm{low}} = 0.045\sigma$, $\Delta\gamma_{\mathrm{low}} = 0.038\sigma$, $\Delta\beta_{\mathrm{high}} = 0.073\sigma$, and  $\Delta\gamma_{\mathrm{high}} = 0.021\sigma$. These are significantly smaller than the corresponding six cosmological model values listed Table \ref{tab:diff}, indicating that in models where the low-$z$ and high-$z$ cosmological constraints are not mutually inconsistent joint low-$z$ + high-$z$ \hiig\ data are good standardizable candles.

Regarding cosmological constraints, even though the inconsistency between the low-$z$ and high-$z$ cosmological constraints rule out the nonflat \lcdm\ model and the noflat XCDM parameterization, we note that for individual low-$z$ and high-$z$ \hiig\ samples, the nonflat \lcdm\ and nonflat XCDM $\Om$--$\Ok$ constraints also exhibit tensions of $>2\sigma$ with $H(z)$ + BAO $\Om$--$\Ok$ constraints. 

In flat \lcdm, although individual \hiig\ low-$z$ and high-$z$ data result in \om\ $2\sigma$ constraints of $>0.177$ and $>0.310$, respectively, consistent with $H(z)$ + BAO ($0.297^{+0.015}_{-0.017}$), the joint low-$z$ + high-$z$ data yield constraints of $>0.446$ ($2\sigma$), which is inconsistent with $H(z)$ + BAO which favor lower \om\ values.

In flat XCDM, \hiig\ low-$z$, high-$z$, and low-$z$ + high-$z$ data provide \om\ $2\sigma$ constraints of $>0.128$, $>0.182$, and $>0.160$, respectively, consistent with $H(z)$ + BAO ($0.285\pm0.019$). Weak \wx\ constraints are also obtained, with values of $-1.442^{+1.601}_{-0.465}$, $-2.477^{+2.619}_{-2.516}$, $-1.491^{+1.655}_{-0.704}$, respectively, consistent with $H(z)$ + BAO ($-0.776^{+0.130}_{-0.103}$). 

In flat \pcdm, \hiig\ low-$z$, high-$z$, and low-$z$ + high-$z$ data yield \om\ constraints of $>0.480$ ($1\sigma$), $>0.243$ ($2\sigma$), and $>0.317$ ($2\sigma$), respectively, consistent with $H(z)$ + BAO ($0.272^{+0.024}_{-0.022}$). Weak $\alpha$ $1\sigma$ constraints are also obtained, with values of $>3.833$, none, and $>3.863$, respectively, consistent (at the $2\sigma$ level) with $H(z)$ + BAO ($1.271^{+0.507}_{-0.836}$).

In nonflat \pcdm, \hiig\ low-$z$, high-$z$, and low-$z$ + high-$z$ data yield \om\ constraints of $>0.430$ ($1\sigma$), $>0.337$ ($2\sigma$), and $>0.422$ ($2\sigma$), respectively. The last two are inconsistent with $H(z)$ + BAO ($0.275\pm0.025$). Only \hiig\ low-$z$ data provide weak $\alpha$ $1\sigma$ constraints of $5.402^{+4.376}_{-1.747}$, consistent (at the $2\sigma$ level) with $H(z)$ + BAO ($1.427^{+0.572}_{-0.830}$). Slightly improved \ok\ constraints are also obtained, with values of $-0.027^{+0.403}_{-0.393}$, $-0.331^{+0.329}_{-0.354}$, and $-0.385^{+0.273}_{-0.405}$, respectively, consistent (at the $1\sigma$ level) with $H(z)$ + BAO ($-0.052^{+0.093}_{-0.087}$), and indicating a mild preference for closed hypersurfaces.

\begin{figure*}
\centering
 \subfloat[Flat \lcdm]{%
    \includegraphics[width=0.4\textwidth,height=0.35\textwidth]{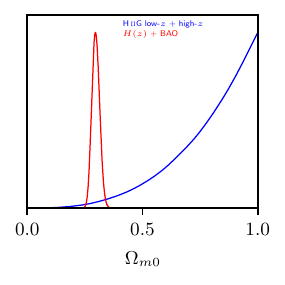}}
 \subfloat[Nonflat \lcdm]{%
    \includegraphics[width=0.4\textwidth,height=0.35\textwidth]{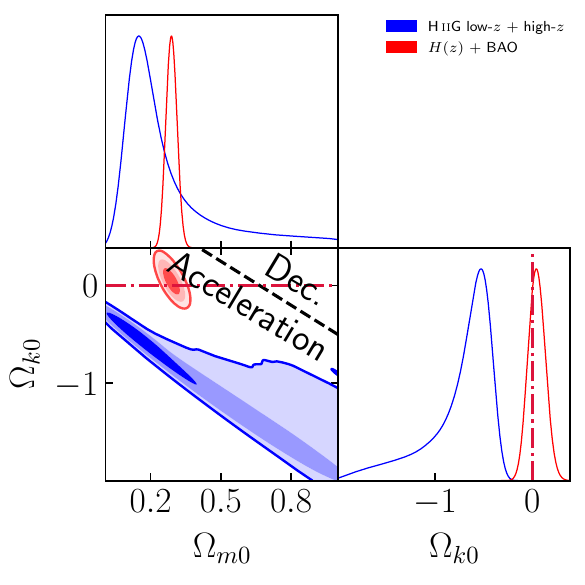}}\\
 \subfloat[Flat XCDM]{%
    \includegraphics[width=0.4\textwidth,height=0.35\textwidth]{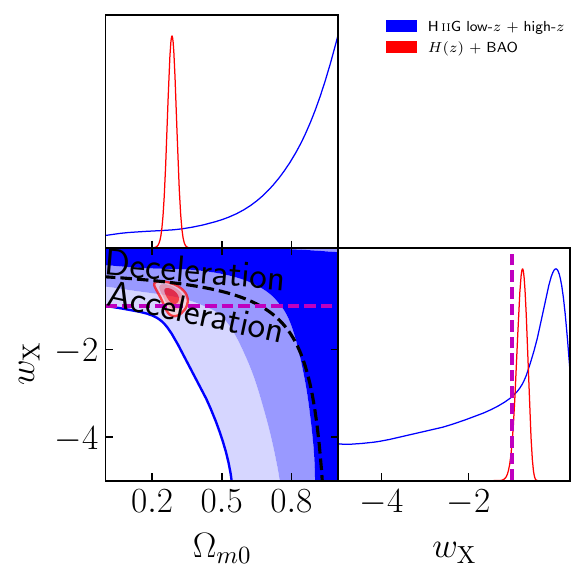}}
 \subfloat[Nonflat XCDM]{%
    \includegraphics[width=0.4\textwidth,height=0.35\textwidth]{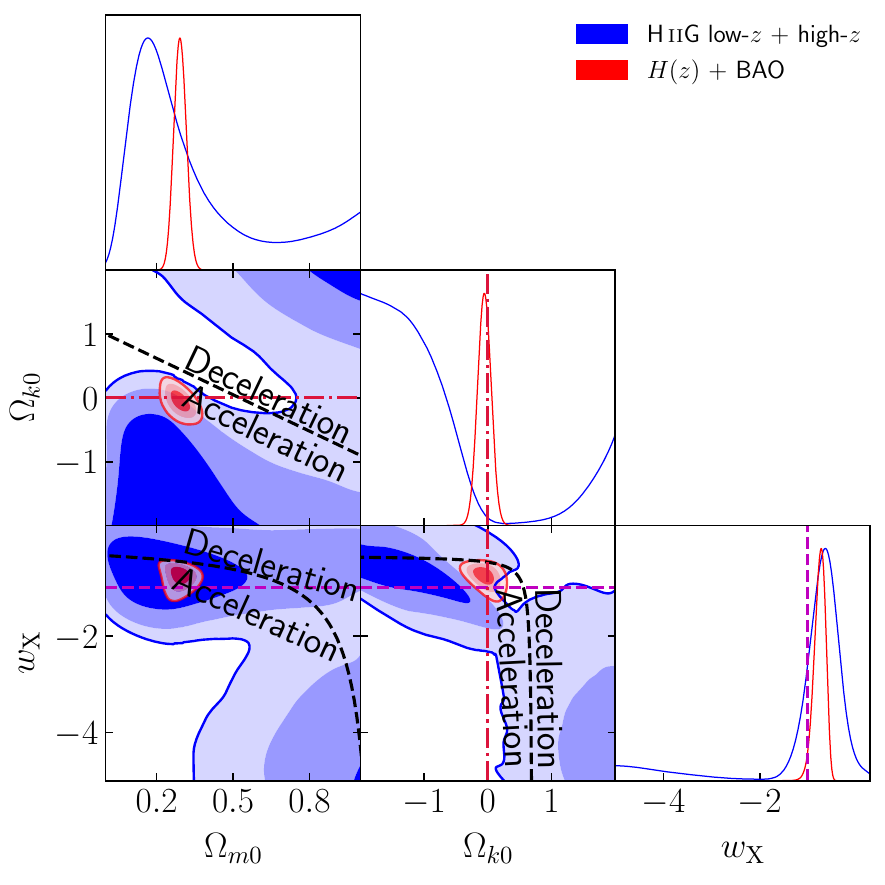}}\\
 \subfloat[Flat \pcdm]{%
    \includegraphics[width=0.4\textwidth,height=0.35\textwidth]{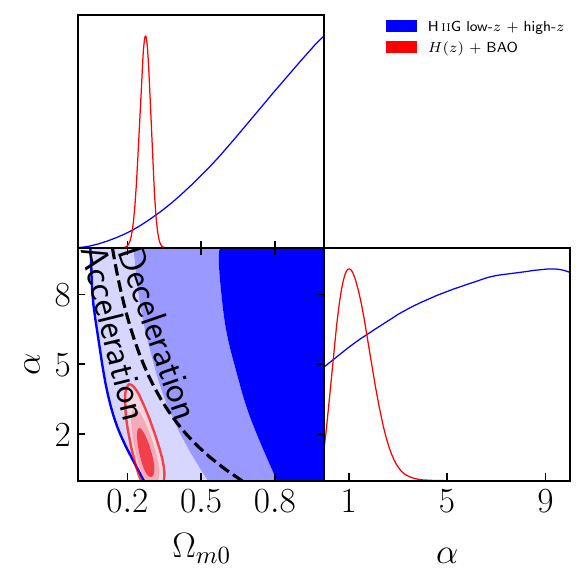}}
 \subfloat[Nonflat \pcdm]{%
    \includegraphics[width=0.4\textwidth,height=0.35\textwidth]{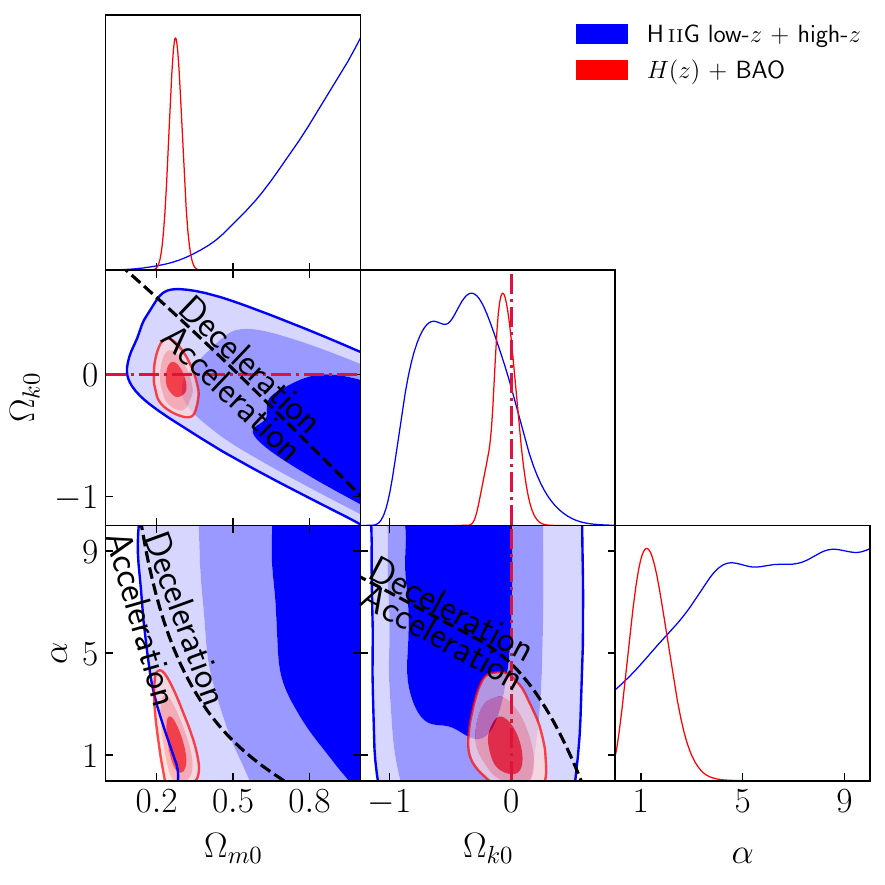}}\\
\caption{Same as Fig. \ref{fig1}, but excluding low-$z$ and high-$z$ \hiig\ contours, and for cosmological parameters only.}
\label{fig4}
\end{figure*}

Figure \ref{fig4} shows the cosmological parameter contours from joint low-$z$ + high-$z$ \hiig\ data and from joint $H(z)$ + BAO data. In the four cosmological models with not inconsistent low-$z$ and high-$z$ \hiig\ data cosmological contours, joint low-$z$ + high-$z$ \hiig\ data contours, unlike joint $H(z)$ + BAO data contours, tend to favor currently decelerating cosmological expansion. Flat XCDM is the only case where joint low-$z$ + high-$z$ cosmological constraints and $H(z)$ + BAO cosmological constraints are consistent enough to warrant a joint low-$z$ + high-$z$ + $H(z)$ + BAO data analysis. We do not do this here as we feel it is important to first more carefully examine the difference between the low-$z$ and high-$z$ $L-\sigma$ relations.

We note in passing that the large $\sigma_{\rm int} = 0.6-0.7$ values in Table \ref{tab:1d_BFP} indicate that current \hiig\ data are not that constraining, as we have seen. For comparison, reverberation measured quasars have $\sigma_{\rm int} = 0.24-0.31$, \cite{Khadkaetal_2021a, Khadkaetal2021c, Khadka:2022ooh, Cao:2022pdv, Khadka:2022aeg, Caoetal2024}, while gamma-ray bursts have $\sigma_{\rm int} = 0.35-0.42$, \cite{KhadkaRatra2020c, Khadkaetal_2021b, CaoKhadkaRatra2021, CaoDainottiRatra2022}.

When considering the AIC, BIC, and DIC values, it becomes evident that nonflat \lcdm\ and nonflat XCDM, although ruled out, emerge as the most preferred models. This observation raises questions about the ability of these criteria to effectively distinguish between observationally viable and unviable models.

After excluding these two models, the AIC indicates that in the \hiig\ low-$z$, high-$z$, and low-$z$ + high-$z$ cases, flat \lcdm, nonflat \pcdm, and flat XCDM are the most favored models, respectively. The evidence against other models is either weak or positive according to this criterion. Conversely, the BIC strongly favors flat \lcdm\ as the top model, with the evidence against other models ranging from positive to strong.

When relying on the relatively more reliable DIC, flat \pcdm\ emerges as the most favored model, while the evidence against other models remains weak or positive.

\section{Conclusion}
\label{sec:conclusion}

In our previous works that used \hiig\ data to constrain cosmological parameters, \cite{CaoRyanRatra2020, CaoKhadkaRatra2021, CaoRyanRatra2021, CaoRyanRatra2022, CaoRatra2022, CaoRatra2023}, we followed the procedure of Refs.\ \cite{G-M_2019,GM2021} where the \hiig\ $L-\sigma$ intercept and slope parameters were determined by 107 low-$z$ \hiig\ and 36 GH\,\textsc{ii}R data. To infer distances to the GH\,\textsc{ii}R sources, the authors relied on primary indicators such as Cepheids, TRGB, and theoretical model calibrations, \cite{FernandezArenas}. Although in Ref.\ \cite{GM2021}, the authors also simultaneously constrained the cosmological parameters and $L-\sigma$ correlation parameters using $\chi^2$ technique. Here we explore whether the apparent magnitude measurements of 181 \hii\ starburst galaxies conform to the $L-\sigma$ relation by simultaneously constraining the $L-\sigma$ relation and cosmological parameters. Our simultaneous constraining approach allows us to not only determine, for the first time, whether these \hiig\ data obey the $L-\sigma$ relation, but also, for the first time, to determine \hiig\ data cosmological constraints that are independent of other data sets. However, as we are also simultaneously constraining the $L-\sigma$ relation parameters, our approach offers no constraining power on cosmological parameters \obhs\ and $H_0$.\footnote{When also varying \obhs\ and $H_0$ we found that the resulting \hiig\ data cosmological constraints are largely not physical. For example, in the flat \lcdm\ model the derived \om\ can spread up to $\sim 20$ with $H_0$ peaked very low, $\sim 0$. We hence decided to set $\Omega_b=0.05$ and $H_0 = 70$ \hunit\ in our analyses here. This differs from the choice made in Refs.\ \cite{G-M_2019, GM2021} using a likelihood function that is marginalized over $H_0$. We emphasize that we treat $\beta$ and $\gamma$ as two additional (compared to the cosmological) free parameters to be determined from the data set being used in the analysis. We also emphasize that the conclusions are not sensitive to a reasonably chosen $H_0$ prior value.}

Our analysis shows that the 107 low-$z$ and 74 high-$z$ \hiig\ sources obey very different $L-\sigma$ relations.\footnote{This could be the consequence of \hiig\ starburst galaxy evolution, but we cannot tell from these data whether this is the case, and if it is the case whether the evolution is a somewhat continuous function of redshift. This is because the low-$z$ sample spans a small redshift range, $0.0088 \leq z \leq 0.16417$, that is very distinct from the redshift range of the 74 high-$z$ sources, $0.63427 \leq z \leq 2.545$, and there are too few high-$z$ sources to study evolution in the high-$z$ range. More \hiig\ data in the intermediate and high redshift ranges will allow for a test of \hiig\ starburst galaxy evolution, and if continual (in $z$) evolution is found to be responsible for the different low-$z$ and high-$z$ $L-\sigma$ relations this will likely make it impossible to cosmologically use \hiig\ data.} \footnote{There could be other explanations (besides \hiig\ evolution) of the effect we found. We thank the referee for noting that the high-$z$ sample suffers from Malmquist bias and suggesting that this could be significant enough to also be a contributing cause. (Figure \ref{fig3} suggests that Malmquist bias is significant, with the black low-$z$ data points reaching to significantly lower absolute luminosities than do the green high-$z$ data points; note that at the higher absolute luminosity end there is much less difference between the high-$z$ and low-$z$ data points.) To superficially study the significance of this known Malmquist bias we computed absolute luminosities of the \hiig\ sources in a flat $\Lambda$CDM model with $\Omega_{m0} =0.29$ and then compared a truncated low-$z$ sample of just 57 (of 107) sources whose absolute luminosity range was restricted to match that of the high-$z$ sample (by discarding 50 intrinsically-dimmer low-$z$ sources) and found that the $L-\sigma$ relation of the truncated low-$z$ sample was consistent, within the error bars, with that of the high-$z$ sample. While this is consistent with Malmquist bias of the high-$z$ sample being a significant contributor to the cause of the effect we have found, this is not definitive and certainly an analysis of the effect based on the assumption of a specific cosmological model cannot be used to correctly study this. Ideally it is much more desirable to correct for Malmquist bias instead of discarding a significant fraction of low-$z$ data to deal with this problem with current \hiig\ data.}  Both however are standardizable candles. The low-$z$ and high-$z$ cosmological constraints are mutually inconsistent in the nonflat \lcdm\ and nonflat XCDM models; they are however consistent in the other four cosmological models and so in these four models it is possible to do a joint analysis of low-$z$ + high-$z$ \hiig\ data. However, the joint low-$z$ + high-$z$ \hiig\ data and joint (better-established) $H(z)$ + BAO data cosmological constraints are not inconsistent in only the flat XCDM parametrization. 

Given our results, we believe that prior to using \hiig\ data for cosmological purposes it is necessary to more carefully examine and understand the difference we have discovered between the (currently available) low-$z$ and high-$z$ \hiig\ data $L-\sigma$ relations.

\begin{acknowledgments}
The computations for this project were performed on the Beocat Research Cluster at Kansas State University, which is funded in part by NSF grants CNS-1006860, EPS-1006860, EPS-0919443, ACI-1440548, CHE-1726332, and NIH P20GM113109.
\end{acknowledgments}




\bibliography{apssamp}

\newcommand{\apjl}{Astrophys. J. Lett.}\newcommand{\apjs}{Astrophys. J. Suppl.}\newcommand{\mnras}{Mon. Not. R. Astron. Soc.}\newcommand{\jcap}{J. Cosmol. Astropart. Phys.}\newcommand{\aap}{Astron. Astrophys.}\newcommand{\rmxaa}{Revista Mexicana de Astronom{\'i}a y Astrof{\'i}sica}\newcommand{\phiv}{$\phi$}\newcommand{\apss}{Astrophys. Space Sci.}\newcommand{\pasp}{PASP}\newcommand{\nar}{New Astronomy Reviews}\newcommand{\pasj}{PASJ}
\begin{thebibliography}{107}%
\makeatletter
\providecommand \@ifxundefined [1]{%
 \@ifx{#1\undefined}
}%
\providecommand \@ifnum [1]{%
 \ifnum #1\expandafter \@firstoftwo
 \else \expandafter \@secondoftwo
 \fi
}%
\providecommand \@ifx [1]{%
 \ifx #1\expandafter \@firstoftwo
 \else \expandafter \@secondoftwo
 \fi
}%
\providecommand \natexlab [1]{#1}%
\providecommand \enquote  [1]{``#1''}%
\providecommand \bibnamefont  [1]{#1}%
\providecommand \bibfnamefont [1]{#1}%
\providecommand \citenamefont [1]{#1}%
\providecommand \href@noop [0]{\@secondoftwo}%
\providecommand \href [0]{\begingroup \@sanitize@url \@href}%
\providecommand \@href[1]{\@@startlink{#1}\@@href}%
\providecommand \@@href[1]{\endgroup#1\@@endlink}%
\providecommand \@sanitize@url [0]{\catcode `\\12\catcode `\$12\catcode `\&12\catcode `\#12\catcode `\^12\catcode `\_12\catcode `\%12\relax}%
\providecommand \@@startlink[1]{}%
\providecommand \@@endlink[0]{}%
\providecommand \url  [0]{\begingroup\@sanitize@url \@url }%
\providecommand \@url [1]{\endgroup\@href {#1}{\urlprefix }}%
\providecommand \urlprefix  [0]{URL }%
\providecommand \Eprint [0]{\href }%
\providecommand \doibase [0]{https://doi.org/}%
\providecommand \selectlanguage [0]{\@gobble}%
\providecommand \bibinfo  [0]{\@secondoftwo}%
\providecommand \bibfield  [0]{\@secondoftwo}%
\providecommand \translation [1]{[#1]}%
\providecommand \BibitemOpen [0]{}%
\providecommand \bibitemStop [0]{}%
\providecommand \bibitemNoStop [0]{.\EOS\space}%
\providecommand \EOS [0]{\spacefactor3000\relax}%
\providecommand \BibitemShut  [1]{\csname bibitem#1\endcsname}%
\let\auto@bib@innerbib\@empty
\bibitem [{\citenamefont {{Yu}}\ \emph {et~al.}(2018)\citenamefont {{Yu}}, \citenamefont {{Ratra}},\ and\ \citenamefont {{Wang}}}]{Yuetal2018}%
  \BibitemOpen
  \bibfield  {author} {\bibinfo {author} {\bibfnamefont {H.}~\bibnamefont {{Yu}}}, \bibinfo {author} {\bibfnamefont {B.}~\bibnamefont {{Ratra}}},\ and\ \bibinfo {author} {\bibfnamefont {F.-Y.}\ \bibnamefont {{Wang}}},\ }\href {https://doi.org/10.3847/1538-4357/aab0a2} {\bibfield  {journal} {\bibinfo  {journal} {\apj}\ }\textbf {\bibinfo {volume} {856}},\ \bibinfo {eid} {3} (\bibinfo {year} {2018})},\ \Eprint {https://arxiv.org/abs/1711.03437} {arXiv:1711.03437 [astro-ph.CO]} \BibitemShut {NoStop}%
\bibitem [{\citenamefont {{\MakeLowercase{E}BOSS Collaboration}}(2021)}]{eBOSS_2020}%
  \BibitemOpen
  \bibfield  {author} {\bibinfo {author} {\bibnamefont {{\MakeLowercase{E}BOSS Collaboration}}},\ }\href {https://doi.org/10.1103/PhysRevD.103.083533} {\bibfield  {journal} {\bibinfo  {journal} {Phys. Rev. D}\ }\textbf {\bibinfo {volume} {103}},\ \bibinfo {eid} {083533} (\bibinfo {year} {2021})},\ \Eprint {https://arxiv.org/abs/2007.08991} {arXiv:2007.08991 [astro-ph.CO]} \BibitemShut {NoStop}%
\bibitem [{\citenamefont {{Brout}}\ \emph {et~al.}(2022)\citenamefont {{Brout}}, \citenamefont {{Scolnic}}, \citenamefont {{Popovic}}, \citenamefont {{Riess}}, \citenamefont {{Carr}}, \citenamefont {{Zuntz}}, \citenamefont {{Kessler}}, \citenamefont {{Davis}}, \citenamefont {{Hinton}}, \citenamefont {{Jones}} \emph {et~al.}}]{Brout:2022vxf}%
  \BibitemOpen
  \bibfield  {author} {\bibinfo {author} {\bibfnamefont {D.}~\bibnamefont {{Brout}}}, \bibinfo {author} {\bibfnamefont {D.}~\bibnamefont {{Scolnic}}}, \bibinfo {author} {\bibfnamefont {B.}~\bibnamefont {{Popovic}}}, \bibinfo {author} {\bibfnamefont {A.~G.}\ \bibnamefont {{Riess}}}, \bibinfo {author} {\bibfnamefont {A.}~\bibnamefont {{Carr}}}, \bibinfo {author} {\bibfnamefont {J.}~\bibnamefont {{Zuntz}}}, \bibinfo {author} {\bibfnamefont {R.}~\bibnamefont {{Kessler}}}, \bibinfo {author} {\bibfnamefont {T.~M.}\ \bibnamefont {{Davis}}}, \bibinfo {author} {\bibfnamefont {S.}~\bibnamefont {{Hinton}}}, \bibinfo {author} {\bibfnamefont {D.}~\bibnamefont {{Jones}}}, \emph {et~al.},\ }\href {https://doi.org/10.3847/1538-4357/ac8e04} {\bibfield  {journal} {\bibinfo  {journal} {Astrophys. J.}\ }\textbf {\bibinfo {volume} {938}},\ \bibinfo {eid} {110} (\bibinfo {year} {2022})},\ \Eprint {https://arxiv.org/abs/2202.04077} {arXiv:2202.04077 [astro-ph.CO]} \BibitemShut {NoStop}%
\bibitem [{\citenamefont {{Planck Collaboration}}(2020)}]{planck2018b}%
  \BibitemOpen
  \bibfield  {author} {\bibinfo {author} {\bibnamefont {{Planck Collaboration}}},\ }\href {https://doi.org/10.1051/0004-6361/201833910} {\bibfield  {journal} {\bibinfo  {journal} {Astron. Astrophys.}\ }\textbf {\bibinfo {volume} {641}},\ \bibinfo {eid} {A6} (\bibinfo {year} {2020})},\ \Eprint {https://arxiv.org/abs/1807.06209} {arXiv:1807.06209 [astro-ph.CO]} \BibitemShut {NoStop}%
\bibitem [{\citenamefont {{Melnick}}\ \emph {et~al.}(2000)\citenamefont {{Melnick}}, \citenamefont {{Terlevich}},\ and\ \citenamefont {{Terlevich}}}]{Melnick_2000}%
  \BibitemOpen
  \bibfield  {author} {\bibinfo {author} {\bibfnamefont {J.}~\bibnamefont {{Melnick}}}, \bibinfo {author} {\bibfnamefont {R.}~\bibnamefont {{Terlevich}}},\ and\ \bibinfo {author} {\bibfnamefont {E.}~\bibnamefont {{Terlevich}}},\ }\href {https://doi.org/10.1046/j.1365-8711.2000.03112.x} {\bibfield  {journal} {\bibinfo  {journal} {Mon. Not. R. Astron. Soc.}\ }\textbf {\bibinfo {volume} {311}},\ \bibinfo {pages} {629} (\bibinfo {year} {2000})},\ \Eprint {https://arxiv.org/abs/astro-ph/9908346} {arXiv:astro-ph/9908346 [astro-ph]} \BibitemShut {NoStop}%
\bibitem [{\citenamefont {{Siegel}}\ \emph {et~al.}(2005)\citenamefont {{Siegel}}, \citenamefont {{Guzm{\'a}n}}, \citenamefont {{Gallego}}, \citenamefont {{Ordu{\~n}a L{\'o}pez}},\ and\ \citenamefont {{Rodr{\'\i}guez Hidalgo}}}]{Siegel_2005}%
  \BibitemOpen
  \bibfield  {author} {\bibinfo {author} {\bibfnamefont {E.~R.}\ \bibnamefont {{Siegel}}}, \bibinfo {author} {\bibfnamefont {R.}~\bibnamefont {{Guzm{\'a}n}}}, \bibinfo {author} {\bibfnamefont {J.~P.}\ \bibnamefont {{Gallego}}}, \bibinfo {author} {\bibfnamefont {M.}~\bibnamefont {{Ordu{\~n}a L{\'o}pez}}},\ and\ \bibinfo {author} {\bibfnamefont {P.}~\bibnamefont {{Rodr{\'\i}guez Hidalgo}}},\ }\href {https://doi.org/10.1111/j.1365-2966.2004.08539.x} {\bibfield  {journal} {\bibinfo  {journal} {Mon. Not. R. Astron. Soc.}\ }\textbf {\bibinfo {volume} {356}},\ \bibinfo {pages} {1117} (\bibinfo {year} {2005})},\ \Eprint {https://arxiv.org/abs/astro-ph/0410612} {arXiv:astro-ph/0410612 [astro-ph]} \BibitemShut {NoStop}%
\bibitem [{\citenamefont {{Plionis}}\ \emph {et~al.}(2009)\citenamefont {{Plionis}}, \citenamefont {{Terlevich}}, \citenamefont {{Basilakos}}, \citenamefont {{Bresolin}}, \citenamefont {{Terlevich}}, \citenamefont {{Melnick}},\ and\ \citenamefont {{Georgantopoulos}}}]{Plionis_2009}%
  \BibitemOpen
  \bibfield  {author} {\bibinfo {author} {\bibfnamefont {M.}~\bibnamefont {{Plionis}}}, \bibinfo {author} {\bibfnamefont {R.}~\bibnamefont {{Terlevich}}}, \bibinfo {author} {\bibfnamefont {S.}~\bibnamefont {{Basilakos}}}, \bibinfo {author} {\bibfnamefont {F.}~\bibnamefont {{Bresolin}}}, \bibinfo {author} {\bibfnamefont {E.}~\bibnamefont {{Terlevich}}}, \bibinfo {author} {\bibfnamefont {J.}~\bibnamefont {{Melnick}}},\ and\ \bibinfo {author} {\bibfnamefont {I.}~\bibnamefont {{Georgantopoulos}}},\ }in\ \href {https://doi.org/10.1088/1742-6596/189/1/012032} {\emph {\bibinfo {booktitle} {Journal of Physics Conference Series}}},\ \bibinfo {series} {Journal of Physics Conference Series}, Vol.\ \bibinfo {volume} {189}\ (\bibinfo {year} {2009})\ p.\ \bibinfo {pages} {012032},\ \Eprint {https://arxiv.org/abs/0903.0131} {arXiv:0903.0131 [astro-ph.CO]} \BibitemShut {NoStop}%
\bibitem [{\citenamefont {{Plionis}}\ \emph {et~al.}(2010)\citenamefont {{Plionis}}, \citenamefont {{Terlevich}}, \citenamefont {{Basilakos}}, \citenamefont {{Bresolin}}, \citenamefont {{Terlevich}}, \citenamefont {{Melnick}},\ and\ \citenamefont {{Chavez}}}]{Plionis_2010}%
  \BibitemOpen
  \bibfield  {author} {\bibinfo {author} {\bibfnamefont {M.}~\bibnamefont {{Plionis}}}, \bibinfo {author} {\bibfnamefont {R.}~\bibnamefont {{Terlevich}}}, \bibinfo {author} {\bibfnamefont {S.}~\bibnamefont {{Basilakos}}}, \bibinfo {author} {\bibfnamefont {F.}~\bibnamefont {{Bresolin}}}, \bibinfo {author} {\bibfnamefont {E.}~\bibnamefont {{Terlevich}}}, \bibinfo {author} {\bibfnamefont {J.}~\bibnamefont {{Melnick}}},\ and\ \bibinfo {author} {\bibfnamefont {R.}~\bibnamefont {{Chavez}}},\ }in\ \href@noop {} {\emph {\bibinfo {booktitle} {American Institute of Physics Conference Series}}},\ \bibinfo {series} {American Institute of Physics Conference Series}, Vol.\ \bibinfo {volume} {1241},\ \bibinfo {editor} {edited by\ \bibinfo {editor} {\bibfnamefont {J.-M.}\ \bibnamefont {{Alimi}}}\ and\ \bibinfo {editor} {\bibfnamefont {A.}~\bibnamefont {{Fu{\"o}zfa}}}}\ (\bibinfo {year} {2010})\ pp.\ \bibinfo {pages} {267--276},\ \Eprint {https://arxiv.org/abs/0911.3198} {arXiv:0911.3198 [astro-ph.CO]} \BibitemShut {NoStop}%
\bibitem [{\citenamefont {{Plionis}}\ \emph {et~al.}(2011)\citenamefont {{Plionis}}, \citenamefont {{Terlevich}}, \citenamefont {{Basilakos}}, \citenamefont {{Bresolin}}, \citenamefont {{Terlevich}}, \citenamefont {{Melnick}},\ and\ \citenamefont {{Chavez}}}]{Plionis_2011}%
  \BibitemOpen
  \bibfield  {author} {\bibinfo {author} {\bibfnamefont {M.}~\bibnamefont {{Plionis}}}, \bibinfo {author} {\bibfnamefont {R.}~\bibnamefont {{Terlevich}}}, \bibinfo {author} {\bibfnamefont {S.}~\bibnamefont {{Basilakos}}}, \bibinfo {author} {\bibfnamefont {F.}~\bibnamefont {{Bresolin}}}, \bibinfo {author} {\bibfnamefont {E.}~\bibnamefont {{Terlevich}}}, \bibinfo {author} {\bibfnamefont {J.}~\bibnamefont {{Melnick}}},\ and\ \bibinfo {author} {\bibfnamefont {R.}~\bibnamefont {{Chavez}}},\ }\href {https://doi.org/10.1111/j.1365-2966.2011.19247.x} {\bibfield  {journal} {\bibinfo  {journal} {Mon. Not. R. Astron. Soc.}\ }\textbf {\bibinfo {volume} {416}},\ \bibinfo {pages} {2981} (\bibinfo {year} {2011})},\ \Eprint {https://arxiv.org/abs/1106.4558} {arXiv:1106.4558 [astro-ph.CO]} \BibitemShut {NoStop}%
\bibitem [{\citenamefont {{Mania}}\ and\ \citenamefont {{Ratra}}(2012)}]{Mania_2012}%
  \BibitemOpen
  \bibfield  {author} {\bibinfo {author} {\bibfnamefont {D.}~\bibnamefont {{Mania}}}\ and\ \bibinfo {author} {\bibfnamefont {B.}~\bibnamefont {{Ratra}}},\ }\href {https://doi.org/10.1016/j.physletb.2012.07.011} {\bibfield  {journal} {\bibinfo  {journal} {Physics Letters B}\ }\textbf {\bibinfo {volume} {715}},\ \bibinfo {pages} {9} (\bibinfo {year} {2012})},\ \Eprint {https://arxiv.org/abs/1110.5626} {arXiv:1110.5626 [astro-ph.CO]} \BibitemShut {NoStop}%
\bibitem [{\citenamefont {{Ch{\'a}vez}}\ \emph {et~al.}(2012)\citenamefont {{Ch{\'a}vez}}, \citenamefont {{Terlevich}}, \citenamefont {{Terlevich}}, \citenamefont {{Plionis}}, \citenamefont {{Bresolin}}, \citenamefont {{Basilakos}},\ and\ \citenamefont {{Melnick}}}]{Chavez_2012}%
  \BibitemOpen
  \bibfield  {author} {\bibinfo {author} {\bibfnamefont {R.}~\bibnamefont {{Ch{\'a}vez}}}, \bibinfo {author} {\bibfnamefont {E.}~\bibnamefont {{Terlevich}}}, \bibinfo {author} {\bibfnamefont {R.}~\bibnamefont {{Terlevich}}}, \bibinfo {author} {\bibfnamefont {M.}~\bibnamefont {{Plionis}}}, \bibinfo {author} {\bibfnamefont {F.}~\bibnamefont {{Bresolin}}}, \bibinfo {author} {\bibfnamefont {S.}~\bibnamefont {{Basilakos}}},\ and\ \bibinfo {author} {\bibfnamefont {J.}~\bibnamefont {{Melnick}}},\ }\href {https://doi.org/10.1111/j.1745-3933.2012.01299.x} {\bibfield  {journal} {\bibinfo  {journal} {Mon. Not. R. Astron. Soc.}\ }\textbf {\bibinfo {volume} {425}},\ \bibinfo {pages} {L56} (\bibinfo {year} {2012})},\ \Eprint {https://arxiv.org/abs/1203.6222} {arXiv:1203.6222 [astro-ph.CO]} \BibitemShut {NoStop}%
\bibitem [{\citenamefont {{Ch{\'a}vez}}\ \emph {et~al.}(2014)\citenamefont {{Ch{\'a}vez}}, \citenamefont {{Terlevich}}, \citenamefont {{Terlevich}}, \citenamefont {{Bresolin}}, \citenamefont {{Melnick}}, \citenamefont {{Plionis}},\ and\ \citenamefont {{Basilakos}}}]{Chavez_2014}%
  \BibitemOpen
  \bibfield  {author} {\bibinfo {author} {\bibfnamefont {R.}~\bibnamefont {{Ch{\'a}vez}}}, \bibinfo {author} {\bibfnamefont {R.}~\bibnamefont {{Terlevich}}}, \bibinfo {author} {\bibfnamefont {E.}~\bibnamefont {{Terlevich}}}, \bibinfo {author} {\bibfnamefont {F.}~\bibnamefont {{Bresolin}}}, \bibinfo {author} {\bibfnamefont {J.}~\bibnamefont {{Melnick}}}, \bibinfo {author} {\bibfnamefont {M.}~\bibnamefont {{Plionis}}},\ and\ \bibinfo {author} {\bibfnamefont {S.}~\bibnamefont {{Basilakos}}},\ }\href {https://doi.org/10.1093/mnras/stu987} {\bibfield  {journal} {\bibinfo  {journal} {Mon. Not. R. Astron. Soc.}\ }\textbf {\bibinfo {volume} {442}},\ \bibinfo {pages} {3565} (\bibinfo {year} {2014})},\ \Eprint {https://arxiv.org/abs/1405.4010} {arXiv:1405.4010 [astro-ph.GA]} \BibitemShut {NoStop}%
\bibitem [{\citenamefont {{Ch{\'a}vez}}\ \emph {et~al.}(2016)\citenamefont {{Ch{\'a}vez}}, \citenamefont {{Plionis}}, \citenamefont {{Basilakos}}, \citenamefont {{Terlevich}}, \citenamefont {{Terlevich}}, \citenamefont {{Melnick}}, \citenamefont {{Bresolin}},\ and\ \citenamefont {{Gonz{\'a}lez-Mor{\'a}n}}}]{Chavez_2016}%
  \BibitemOpen
  \bibfield  {author} {\bibinfo {author} {\bibfnamefont {R.}~\bibnamefont {{Ch{\'a}vez}}}, \bibinfo {author} {\bibfnamefont {M.}~\bibnamefont {{Plionis}}}, \bibinfo {author} {\bibfnamefont {S.}~\bibnamefont {{Basilakos}}}, \bibinfo {author} {\bibfnamefont {R.}~\bibnamefont {{Terlevich}}}, \bibinfo {author} {\bibfnamefont {E.}~\bibnamefont {{Terlevich}}}, \bibinfo {author} {\bibfnamefont {J.}~\bibnamefont {{Melnick}}}, \bibinfo {author} {\bibfnamefont {F.}~\bibnamefont {{Bresolin}}},\ and\ \bibinfo {author} {\bibfnamefont {A.~L.}\ \bibnamefont {{Gonz{\'a}lez-Mor{\'a}n}}},\ }\href {https://doi.org/10.1093/mnras/stw1813} {\bibfield  {journal} {\bibinfo  {journal} {Mon. Not. R. Astron. Soc.}\ }\textbf {\bibinfo {volume} {462}},\ \bibinfo {pages} {2431} (\bibinfo {year} {2016})},\ \Eprint {https://arxiv.org/abs/1607.06458} {arXiv:1607.06458 [astro-ph.CO]} \BibitemShut {NoStop}%
\bibitem [{\citenamefont {{Terlevich}}\ \emph {et~al.}(2015)\citenamefont {{Terlevich}}, \citenamefont {{Terlevich}}, \citenamefont {{Melnick}}, \citenamefont {{Ch{\'a}vez}}, \citenamefont {{Plionis}}, \citenamefont {{Bresolin}},\ and\ \citenamefont {{Basilakos}}}]{Terlevich_2015}%
  \BibitemOpen
  \bibfield  {author} {\bibinfo {author} {\bibfnamefont {R.}~\bibnamefont {{Terlevich}}}, \bibinfo {author} {\bibfnamefont {E.}~\bibnamefont {{Terlevich}}}, \bibinfo {author} {\bibfnamefont {J.}~\bibnamefont {{Melnick}}}, \bibinfo {author} {\bibfnamefont {R.}~\bibnamefont {{Ch{\'a}vez}}}, \bibinfo {author} {\bibfnamefont {M.}~\bibnamefont {{Plionis}}}, \bibinfo {author} {\bibfnamefont {F.}~\bibnamefont {{Bresolin}}},\ and\ \bibinfo {author} {\bibfnamefont {S.}~\bibnamefont {{Basilakos}}},\ }\href {https://doi.org/10.1093/mnras/stv1128} {\bibfield  {journal} {\bibinfo  {journal} {Mon. Not. R. Astron. Soc.}\ }\textbf {\bibinfo {volume} {451}},\ \bibinfo {pages} {3001} (\bibinfo {year} {2015})},\ \Eprint {https://arxiv.org/abs/1505.04376} {arXiv:1505.04376 [astro-ph.CO]} \BibitemShut {NoStop}%
\bibitem [{\citenamefont {{Gonz{\'a}lez-Mor{\'a}n}}\ \emph {et~al.}(2019)\citenamefont {{Gonz{\'a}lez-Mor{\'a}n}}, \citenamefont {{Ch{\'a}vez}}, \citenamefont {{Terlevich}}, \citenamefont {{Terlevich}}, \citenamefont {{Bresolin}}, \citenamefont {{Fern{\'a}ndez-Arenas}}, \citenamefont {{Plionis}}, \citenamefont {{Basilakos}}, \citenamefont {{Melnick}},\ and\ \citenamefont {{Telles}}}]{G-M_2019}%
  \BibitemOpen
  \bibfield  {author} {\bibinfo {author} {\bibfnamefont {A.~L.}\ \bibnamefont {{Gonz{\'a}lez-Mor{\'a}n}}}, \bibinfo {author} {\bibfnamefont {R.}~\bibnamefont {{Ch{\'a}vez}}}, \bibinfo {author} {\bibfnamefont {R.}~\bibnamefont {{Terlevich}}}, \bibinfo {author} {\bibfnamefont {E.}~\bibnamefont {{Terlevich}}}, \bibinfo {author} {\bibfnamefont {F.}~\bibnamefont {{Bresolin}}}, \bibinfo {author} {\bibfnamefont {D.}~\bibnamefont {{Fern{\'a}ndez-Arenas}}}, \bibinfo {author} {\bibfnamefont {M.}~\bibnamefont {{Plionis}}}, \bibinfo {author} {\bibfnamefont {S.}~\bibnamefont {{Basilakos}}}, \bibinfo {author} {\bibfnamefont {J.}~\bibnamefont {{Melnick}}},\ and\ \bibinfo {author} {\bibfnamefont {E.}~\bibnamefont {{Telles}}},\ }\href {https://doi.org/10.1093/mnras/stz1577} {\bibfield  {journal} {\bibinfo  {journal} {Mon. Not. R. Astron. Soc.}\ }\textbf {\bibinfo {volume} {487}},\ \bibinfo {pages} {4669} (\bibinfo {year} {2019})},\ \Eprint {https://arxiv.org/abs/1906.02195} {arXiv:1906.02195 [astro-ph.GA]} \BibitemShut
  {NoStop}%
\bibitem [{\citenamefont {{Gonz{\'a}lez-Mor{\'a}n}}\ \emph {et~al.}(2021)\citenamefont {{Gonz{\'a}lez-Mor{\'a}n}}, \citenamefont {{Ch{\'a}vez}}, \citenamefont {{Terlevich}}, \citenamefont {{Terlevich}}, \citenamefont {{Fern{\'a}ndez-Arenas}}, \citenamefont {{Bresolin}}, \citenamefont {{Plionis}}, \citenamefont {{Melnick}}, \citenamefont {{Basilakos}},\ and\ \citenamefont {{Telles}}}]{GM2021}%
  \BibitemOpen
  \bibfield  {author} {\bibinfo {author} {\bibfnamefont {A.~L.}\ \bibnamefont {{Gonz{\'a}lez-Mor{\'a}n}}}, \bibinfo {author} {\bibfnamefont {R.}~\bibnamefont {{Ch{\'a}vez}}}, \bibinfo {author} {\bibfnamefont {E.}~\bibnamefont {{Terlevich}}}, \bibinfo {author} {\bibfnamefont {R.}~\bibnamefont {{Terlevich}}}, \bibinfo {author} {\bibfnamefont {D.}~\bibnamefont {{Fern{\'a}ndez-Arenas}}}, \bibinfo {author} {\bibfnamefont {F.}~\bibnamefont {{Bresolin}}}, \bibinfo {author} {\bibfnamefont {M.}~\bibnamefont {{Plionis}}}, \bibinfo {author} {\bibfnamefont {J.}~\bibnamefont {{Melnick}}}, \bibinfo {author} {\bibfnamefont {S.}~\bibnamefont {{Basilakos}}},\ and\ \bibinfo {author} {\bibfnamefont {E.}~\bibnamefont {{Telles}}},\ }\href {https://doi.org/10.1093/mnras/stab1385} {\bibfield  {journal} {\bibinfo  {journal} {Mon. Not. R. Astron. Soc.}\ }\textbf {\bibinfo {volume} {505}},\ \bibinfo {pages} {1441} (\bibinfo {year} {2021})},\ \Eprint {https://arxiv.org/abs/2105.04025} {arXiv:2105.04025 [astro-ph.CO]} \BibitemShut
  {NoStop}%
\bibitem [{\citenamefont {{Cao}}\ \emph {et~al.}(2020)\citenamefont {{Cao}}, \citenamefont {{Ryan}},\ and\ \citenamefont {{Ratra}}}]{CaoRyanRatra2020}%
  \BibitemOpen
  \bibfield  {author} {\bibinfo {author} {\bibfnamefont {S.}~\bibnamefont {{Cao}}}, \bibinfo {author} {\bibfnamefont {J.}~\bibnamefont {{Ryan}}},\ and\ \bibinfo {author} {\bibfnamefont {B.}~\bibnamefont {{Ratra}}},\ }\href {https://doi.org/10.1093/mnras/staa2190} {\bibfield  {journal} {\bibinfo  {journal} {Mon. Not. R. Astron. Soc.}\ }\textbf {\bibinfo {volume} {497}},\ \bibinfo {pages} {3191} (\bibinfo {year} {2020})},\ \Eprint {https://arxiv.org/abs/2005.12617} {arXiv:2005.12617 [astro-ph.CO]} \BibitemShut {NoStop}%
\bibitem [{\citenamefont {{Cao}}\ \emph {et~al.}(2021{\natexlab{a}})\citenamefont {{Cao}}, \citenamefont {{Ryan}},\ and\ \citenamefont {{Ratra}}}]{CaoRyanRatra2021}%
  \BibitemOpen
  \bibfield  {author} {\bibinfo {author} {\bibfnamefont {S.}~\bibnamefont {{Cao}}}, \bibinfo {author} {\bibfnamefont {J.}~\bibnamefont {{Ryan}}},\ and\ \bibinfo {author} {\bibfnamefont {B.}~\bibnamefont {{Ratra}}},\ }\href {https://doi.org/10.1093/mnras/stab942} {\bibfield  {journal} {\bibinfo  {journal} {Mon. Not. R. Astron. Soc.}\ }\textbf {\bibinfo {volume} {504}},\ \bibinfo {pages} {300} (\bibinfo {year} {2021}{\natexlab{a}})},\ \Eprint {https://arxiv.org/abs/2101.08817} {arXiv:2101.08817 [astro-ph.CO]} \BibitemShut {NoStop}%
\bibitem [{\citenamefont {{Cao}}\ \emph {et~al.}(2022)\citenamefont {{Cao}}, \citenamefont {{Ryan}},\ and\ \citenamefont {{Ratra}}}]{CaoRyanRatra2022}%
  \BibitemOpen
  \bibfield  {author} {\bibinfo {author} {\bibfnamefont {S.}~\bibnamefont {{Cao}}}, \bibinfo {author} {\bibfnamefont {J.}~\bibnamefont {{Ryan}}},\ and\ \bibinfo {author} {\bibfnamefont {B.}~\bibnamefont {{Ratra}}},\ }\href {https://doi.org/10.1093/mnras/stab3304} {\bibfield  {journal} {\bibinfo  {journal} {Mon. Not. R. Astron. Soc.}\ }\textbf {\bibinfo {volume} {509}},\ \bibinfo {pages} {4745} (\bibinfo {year} {2022})},\ \Eprint {https://arxiv.org/abs/2109.01987} {arXiv:2109.01987 [astro-ph.CO]} \BibitemShut {NoStop}%
\bibitem [{\citenamefont {{Johnson}}\ \emph {et~al.}(2022)\citenamefont {{Johnson}}, \citenamefont {{Sangwan}},\ and\ \citenamefont {{Shankaranarayanan}}}]{Johnsonetal2022}%
  \BibitemOpen
  \bibfield  {author} {\bibinfo {author} {\bibfnamefont {J.~P.}\ \bibnamefont {{Johnson}}}, \bibinfo {author} {\bibfnamefont {A.}~\bibnamefont {{Sangwan}}},\ and\ \bibinfo {author} {\bibfnamefont {S.}~\bibnamefont {{Shankaranarayanan}}},\ }\href {https://doi.org/10.1088/1475-7516/2022/01/024} {\bibfield  {journal} {\bibinfo  {journal} {J. Cosmol. Astropart. Phys.}\ }\textbf {\bibinfo {volume} {2022}}\bibfield  {number} {\bibinfo  {number} { (1)},\ \bibinfo {eid} {024}},\ }\Eprint {https://arxiv.org/abs/2102.12367} {arXiv:2102.12367 [astro-ph.CO]} \BibitemShut {NoStop}%
\bibitem [{\citenamefont {{Mehrabi}}\ \emph {et~al.}(2022)\citenamefont {{Mehrabi}}, \citenamefont {{Basilakos}}, \citenamefont {{Tsiapi}}, \citenamefont {{Plionis}}, \citenamefont {{Terlevich}}, \citenamefont {{Terlevich}}, \citenamefont {{Gonzalez Moran}}, \citenamefont {{Chavez}}, \citenamefont {{Bresolin}}, \citenamefont {{Fernandez Arenas}},\ and\ \citenamefont {{Telles}}}]{Mehrabietal2022}%
  \BibitemOpen
  \bibfield  {author} {\bibinfo {author} {\bibfnamefont {A.}~\bibnamefont {{Mehrabi}}}, \bibinfo {author} {\bibfnamefont {S.}~\bibnamefont {{Basilakos}}}, \bibinfo {author} {\bibfnamefont {P.}~\bibnamefont {{Tsiapi}}}, \bibinfo {author} {\bibfnamefont {M.}~\bibnamefont {{Plionis}}}, \bibinfo {author} {\bibfnamefont {R.}~\bibnamefont {{Terlevich}}}, \bibinfo {author} {\bibfnamefont {E.}~\bibnamefont {{Terlevich}}}, \bibinfo {author} {\bibfnamefont {A.~L.}\ \bibnamefont {{Gonzalez Moran}}}, \bibinfo {author} {\bibfnamefont {R.}~\bibnamefont {{Chavez}}}, \bibinfo {author} {\bibfnamefont {F.}~\bibnamefont {{Bresolin}}}, \bibinfo {author} {\bibfnamefont {D.}~\bibnamefont {{Fernandez Arenas}}},\ and\ \bibinfo {author} {\bibfnamefont {E.}~\bibnamefont {{Telles}}},\ }\href {https://doi.org/10.1093/mnras/stab2915} {\bibfield  {journal} {\bibinfo  {journal} {Mon. Not. R. Astron. Soc.}\ }\textbf {\bibinfo {volume} {509}},\ \bibinfo {pages} {224} (\bibinfo {year} {2022})},\ \Eprint {https://arxiv.org/abs/2107.08820}
  {arXiv:2107.08820 [astro-ph.CO]} \BibitemShut {NoStop}%
\bibitem [{\citenamefont {{Czerny}}\ \emph {et~al.}(2021)\citenamefont {{Czerny}}, \citenamefont {{Mart{\'\i}nez-Aldama}}, \citenamefont {{Wojtkowska}}, \citenamefont {{Zaja{\v{c}}ek}}, \citenamefont {{Marziani}}, \citenamefont {{Dultzin}}, \citenamefont {{Naddaf}}, \citenamefont {{Panda}}, \citenamefont {{Prince}}, \citenamefont {{Przyluski}}, \citenamefont {{Ralowski}},\ and\ \citenamefont {{{\'S}niegowska}}}]{Czernyetal2021}%
  \BibitemOpen
  \bibfield  {author} {\bibinfo {author} {\bibfnamefont {B.}~\bibnamefont {{Czerny}}}, \bibinfo {author} {\bibfnamefont {M.~L.}\ \bibnamefont {{Mart{\'\i}nez-Aldama}}}, \bibinfo {author} {\bibfnamefont {G.}~\bibnamefont {{Wojtkowska}}}, \bibinfo {author} {\bibfnamefont {M.}~\bibnamefont {{Zaja{\v{c}}ek}}}, \bibinfo {author} {\bibfnamefont {P.}~\bibnamefont {{Marziani}}}, \bibinfo {author} {\bibfnamefont {D.}~\bibnamefont {{Dultzin}}}, \bibinfo {author} {\bibfnamefont {M.~H.}\ \bibnamefont {{Naddaf}}}, \bibinfo {author} {\bibfnamefont {S.}~\bibnamefont {{Panda}}}, \bibinfo {author} {\bibfnamefont {R.}~\bibnamefont {{Prince}}}, \bibinfo {author} {\bibfnamefont {R.}~\bibnamefont {{Przyluski}}}, \bibinfo {author} {\bibfnamefont {M.}~\bibnamefont {{Ralowski}}},\ and\ \bibinfo {author} {\bibfnamefont {M.}~\bibnamefont {{{\'S}niegowska}}},\ }\href {https://doi.org/10.12693/APhysPolA.139.389} {\bibfield  {journal} {\bibinfo  {journal} {Acta Physica Polonica A}\ }\textbf {\bibinfo {volume} {139}},\ \bibinfo {pages}
  {389} (\bibinfo {year} {2021})},\ \Eprint {https://arxiv.org/abs/2011.12375} {arXiv:2011.12375 [astro-ph.CO]} \BibitemShut {NoStop}%
\bibitem [{\citenamefont {{Zaja{\v{c}}ek}}\ \emph {et~al.}(2021)\citenamefont {{Zaja{\v{c}}ek}}, \citenamefont {{Czerny}}, \citenamefont {{Martinez-Aldama}}, \citenamefont {{Ra{\l}owski}}, \citenamefont {{Olejak}}, \citenamefont {{Przy{\l}uski}}, \citenamefont {{Panda}}, \citenamefont {{Hryniewicz}}, \citenamefont {{{\'S}niegowska}}, \citenamefont {{Naddaf}} \emph {et~al.}}]{Zajaceketal2021}%
  \BibitemOpen
  \bibfield  {author} {\bibinfo {author} {\bibfnamefont {M.}~\bibnamefont {{Zaja{\v{c}}ek}}}, \bibinfo {author} {\bibfnamefont {B.}~\bibnamefont {{Czerny}}}, \bibinfo {author} {\bibfnamefont {M.~L.}\ \bibnamefont {{Martinez-Aldama}}}, \bibinfo {author} {\bibfnamefont {M.}~\bibnamefont {{Ra{\l}owski}}}, \bibinfo {author} {\bibfnamefont {A.}~\bibnamefont {{Olejak}}}, \bibinfo {author} {\bibfnamefont {R.}~\bibnamefont {{Przy{\l}uski}}}, \bibinfo {author} {\bibfnamefont {S.}~\bibnamefont {{Panda}}}, \bibinfo {author} {\bibfnamefont {K.}~\bibnamefont {{Hryniewicz}}}, \bibinfo {author} {\bibfnamefont {M.}~\bibnamefont {{{\'S}niegowska}}}, \bibinfo {author} {\bibfnamefont {M.-H.}\ \bibnamefont {{Naddaf}}}, \emph {et~al.},\ }\href {https://doi.org/10.3847/1538-4357/abe9b2} {\bibfield  {journal} {\bibinfo  {journal} {Astrophys. J.}\ }\textbf {\bibinfo {volume} {912}},\ \bibinfo {eid} {10} (\bibinfo {year} {2021})},\ \Eprint {https://arxiv.org/abs/2012.12409} {arXiv:2012.12409 [astro-ph.GA]} \BibitemShut {NoStop}%
\bibitem [{\citenamefont {{Yu}}\ \emph {et~al.}(2021)\citenamefont {{Yu}}, \citenamefont {{Martini}}, \citenamefont {{Penton}}, \citenamefont {{Davis}}, \citenamefont {{Malik}}, \citenamefont {{Lidman}}, \citenamefont {{Tucker}}, \citenamefont {{Sharp}}, \citenamefont {{Kochanek}}, \citenamefont {{Peterson}} \emph {et~al.}}]{Yuetal2021}%
  \BibitemOpen
  \bibfield  {author} {\bibinfo {author} {\bibfnamefont {Z.}~\bibnamefont {{Yu}}}, \bibinfo {author} {\bibfnamefont {P.}~\bibnamefont {{Martini}}}, \bibinfo {author} {\bibfnamefont {A.}~\bibnamefont {{Penton}}}, \bibinfo {author} {\bibfnamefont {T.~M.}\ \bibnamefont {{Davis}}}, \bibinfo {author} {\bibfnamefont {U.}~\bibnamefont {{Malik}}}, \bibinfo {author} {\bibfnamefont {C.}~\bibnamefont {{Lidman}}}, \bibinfo {author} {\bibfnamefont {B.~E.}\ \bibnamefont {{Tucker}}}, \bibinfo {author} {\bibfnamefont {R.}~\bibnamefont {{Sharp}}}, \bibinfo {author} {\bibfnamefont {C.~S.}\ \bibnamefont {{Kochanek}}}, \bibinfo {author} {\bibfnamefont {B.~M.}\ \bibnamefont {{Peterson}}}, \emph {et~al.},\ }\href {https://doi.org/10.1093/mnras/stab2244} {\bibfield  {journal} {\bibinfo  {journal} {Mon. Not. R. Astron. Soc.}\ }\textbf {\bibinfo {volume} {507}},\ \bibinfo {pages} {3771} (\bibinfo {year} {2021})},\ \Eprint {https://arxiv.org/abs/2103.01973} {arXiv:2103.01973 [astro-ph.GA]} \BibitemShut {NoStop}%
\bibitem [{\citenamefont {{Khadka}}\ \emph {et~al.}(2021{\natexlab{a}})\citenamefont {{Khadka}}, \citenamefont {{Yu}}, \citenamefont {{Zaja{\v{c}}ek}}, \citenamefont {{Martinez-Aldama}}, \citenamefont {{Czerny}},\ and\ \citenamefont {{Ratra}}}]{Khadkaetal_2021a}%
  \BibitemOpen
  \bibfield  {author} {\bibinfo {author} {\bibfnamefont {N.}~\bibnamefont {{Khadka}}}, \bibinfo {author} {\bibfnamefont {Z.}~\bibnamefont {{Yu}}}, \bibinfo {author} {\bibfnamefont {M.}~\bibnamefont {{Zaja{\v{c}}ek}}}, \bibinfo {author} {\bibfnamefont {M.~L.}\ \bibnamefont {{Martinez-Aldama}}}, \bibinfo {author} {\bibfnamefont {B.}~\bibnamefont {{Czerny}}},\ and\ \bibinfo {author} {\bibfnamefont {B.}~\bibnamefont {{Ratra}}},\ }\href {https://doi.org/10.1093/mnras/stab2807} {\bibfield  {journal} {\bibinfo  {journal} {Mon. Not. R. Astron. Soc.}\ }\textbf {\bibinfo {volume} {508}},\ \bibinfo {pages} {4722} (\bibinfo {year} {2021}{\natexlab{a}})},\ \Eprint {https://arxiv.org/abs/2106.11136} {arXiv:2106.11136 [astro-ph.CO]} \BibitemShut {NoStop}%
\bibitem [{\citenamefont {{Khadka}}\ \emph {et~al.}(2022)\citenamefont {{Khadka}}, \citenamefont {{Mart{\'\i}nez-Aldama}}, \citenamefont {{Zaja{\v{c}}ek}}, \citenamefont {{Czerny}},\ and\ \citenamefont {{Ratra}}}]{Khadkaetal2021c}%
  \BibitemOpen
  \bibfield  {author} {\bibinfo {author} {\bibfnamefont {N.}~\bibnamefont {{Khadka}}}, \bibinfo {author} {\bibfnamefont {M.~L.}\ \bibnamefont {{Mart{\'\i}nez-Aldama}}}, \bibinfo {author} {\bibfnamefont {M.}~\bibnamefont {{Zaja{\v{c}}ek}}}, \bibinfo {author} {\bibfnamefont {B.}~\bibnamefont {{Czerny}}},\ and\ \bibinfo {author} {\bibfnamefont {B.}~\bibnamefont {{Ratra}}},\ }\href {https://doi.org/10.1093/mnras/stac914} {\bibfield  {journal} {\bibinfo  {journal} {Mon. Not. R. Astron. Soc.}\ }\textbf {\bibinfo {volume} {513}},\ \bibinfo {pages} {1985} (\bibinfo {year} {2022})},\ \Eprint {https://arxiv.org/abs/2112.00052} {arXiv:2112.00052 [astro-ph.CO]} \BibitemShut {NoStop}%
\bibitem [{\citenamefont {Khadka}\ \emph {et~al.}(2022)\citenamefont {Khadka}, \citenamefont {Zaja\v{c}ek}, \citenamefont {Panda}, \citenamefont {Mart\'\i{}nez-Aldama},\ and\ \citenamefont {Ratra}}]{Khadka:2022ooh}%
  \BibitemOpen
  \bibfield  {author} {\bibinfo {author} {\bibfnamefont {N.}~\bibnamefont {Khadka}}, \bibinfo {author} {\bibfnamefont {M.}~\bibnamefont {Zaja\v{c}ek}}, \bibinfo {author} {\bibfnamefont {S.}~\bibnamefont {Panda}}, \bibinfo {author} {\bibfnamefont {M.~L.}\ \bibnamefont {Mart\'\i{}nez-Aldama}},\ and\ \bibinfo {author} {\bibfnamefont {B.}~\bibnamefont {Ratra}},\ }\href {https://doi.org/10.1093/mnras/stac1940} {\bibfield  {journal} {\bibinfo  {journal} {Mon. Not. Roy. Astron. Soc.}\ }\textbf {\bibinfo {volume} {515}},\ \bibinfo {pages} {3729} (\bibinfo {year} {2022})},\ \Eprint {https://arxiv.org/abs/2205.05813} {arXiv:2205.05813 [astro-ph.CO]} \BibitemShut {NoStop}%
\bibitem [{\citenamefont {Cao}\ \emph {et~al.}(2022)\citenamefont {Cao}, \citenamefont {Zaja\v{c}ek}, \citenamefont {Panda}, \citenamefont {Mart\'\i{}nez-Aldama}, \citenamefont {Czerny},\ and\ \citenamefont {Ratra}}]{Cao:2022pdv}%
  \BibitemOpen
  \bibfield  {author} {\bibinfo {author} {\bibfnamefont {S.}~\bibnamefont {Cao}}, \bibinfo {author} {\bibfnamefont {M.}~\bibnamefont {Zaja\v{c}ek}}, \bibinfo {author} {\bibfnamefont {S.}~\bibnamefont {Panda}}, \bibinfo {author} {\bibfnamefont {M.~L.}\ \bibnamefont {Mart\'\i{}nez-Aldama}}, \bibinfo {author} {\bibfnamefont {B.}~\bibnamefont {Czerny}},\ and\ \bibinfo {author} {\bibfnamefont {B.}~\bibnamefont {Ratra}},\ }\href {https://doi.org/10.1093/mnras/stac2325} {\bibfield  {journal} {\bibinfo  {journal} {Mon. Not. Roy. Astron. Soc.}\ }\textbf {\bibinfo {volume} {516}},\ \bibinfo {pages} {1721} (\bibinfo {year} {2022})},\ \Eprint {https://arxiv.org/abs/2205.15552} {arXiv:2205.15552 [astro-ph.CO]} \BibitemShut {NoStop}%
\bibitem [{\citenamefont {Czerny}\ \emph {et~al.}(2023)\citenamefont {Czerny} \emph {et~al.}}]{Czerny:2022xfj}%
  \BibitemOpen
  \bibfield  {author} {\bibinfo {author} {\bibfnamefont {B.}~\bibnamefont {Czerny}} \emph {et~al.},\ }\href {https://doi.org/10.1007/s10509-023-04165-7} {\bibfield  {journal} {\bibinfo  {journal} {Astrophys. Space Sci.}\ }\textbf {\bibinfo {volume} {368}},\ \bibinfo {pages} {8} (\bibinfo {year} {2023})},\ \Eprint {https://arxiv.org/abs/2209.06563} {arXiv:2209.06563 [astro-ph.GA]} \BibitemShut {NoStop}%
\bibitem [{\citenamefont {{Cao}}\ \emph {et~al.}(2024)\citenamefont {{Cao}}, \citenamefont {{Zaja{\v{c}}ek}}, \citenamefont {{Czerny}}, \citenamefont {{Panda}},\ and\ \citenamefont {{Ratra}}}]{Caoetal2024}%
  \BibitemOpen
  \bibfield  {author} {\bibinfo {author} {\bibfnamefont {S.}~\bibnamefont {{Cao}}}, \bibinfo {author} {\bibfnamefont {M.}~\bibnamefont {{Zaja{\v{c}}ek}}}, \bibinfo {author} {\bibfnamefont {B.}~\bibnamefont {{Czerny}}}, \bibinfo {author} {\bibfnamefont {S.}~\bibnamefont {{Panda}}},\ and\ \bibinfo {author} {\bibfnamefont {B.}~\bibnamefont {{Ratra}}},\ }\href {https://doi.org/10.1093/mnras/stae433} {\bibfield  {journal} {\bibinfo  {journal} {\mnras}\ }\textbf {\bibinfo {volume} {528}},\ \bibinfo {pages} {6444} (\bibinfo {year} {2024})},\ \Eprint {https://arxiv.org/abs/2309.16516} {arXiv:2309.16516 [astro-ph.CO]} \BibitemShut {NoStop}%
\bibitem [{\citenamefont {{Wang}}\ \emph {et~al.}(2016)\citenamefont {{Wang}}, \citenamefont {{Wang}}, \citenamefont {{Cheng}},\ and\ \citenamefont {{Dai}}}]{Wang_2016}%
  \BibitemOpen
  \bibfield  {author} {\bibinfo {author} {\bibfnamefont {J.~S.}\ \bibnamefont {{Wang}}}, \bibinfo {author} {\bibfnamefont {F.~Y.}\ \bibnamefont {{Wang}}}, \bibinfo {author} {\bibfnamefont {K.~S.}\ \bibnamefont {{Cheng}}},\ and\ \bibinfo {author} {\bibfnamefont {Z.~G.}\ \bibnamefont {{Dai}}},\ }\href {https://doi.org/10.1051/0004-6361/201526485} {\bibfield  {journal} {\bibinfo  {journal} {Astron. Astrophys.}\ }\textbf {\bibinfo {volume} {585}},\ \bibinfo {eid} {A68} (\bibinfo {year} {2016})},\ \Eprint {https://arxiv.org/abs/1509.08558} {arXiv:1509.08558 [astro-ph.HE]} \BibitemShut {NoStop}%
\bibitem [{\citenamefont {{Fana Dirirsa}}\ \emph {et~al.}(2019)\citenamefont {{Fana Dirirsa}}, \citenamefont {{Razzaque}}, \citenamefont {{Piron}}, \citenamefont {{Arimoto}}, \citenamefont {{Axelsson}}, \citenamefont {{Kocevski}}, \citenamefont {{Longo}}, \citenamefont {{Ohno}},\ and\ \citenamefont {{Zhu}}}]{Dirirsa2019}%
  \BibitemOpen
  \bibfield  {author} {\bibinfo {author} {\bibfnamefont {F.}~\bibnamefont {{Fana Dirirsa}}}, \bibinfo {author} {\bibfnamefont {S.}~\bibnamefont {{Razzaque}}}, \bibinfo {author} {\bibfnamefont {F.}~\bibnamefont {{Piron}}}, \bibinfo {author} {\bibfnamefont {M.}~\bibnamefont {{Arimoto}}}, \bibinfo {author} {\bibfnamefont {M.}~\bibnamefont {{Axelsson}}}, \bibinfo {author} {\bibfnamefont {D.}~\bibnamefont {{Kocevski}}}, \bibinfo {author} {\bibfnamefont {F.}~\bibnamefont {{Longo}}}, \bibinfo {author} {\bibfnamefont {M.}~\bibnamefont {{Ohno}}},\ and\ \bibinfo {author} {\bibfnamefont {S.}~\bibnamefont {{Zhu}}},\ }\href {https://doi.org/10.3847/1538-4357/ab4e11} {\bibfield  {journal} {\bibinfo  {journal} {Astrophys. J.}\ }\textbf {\bibinfo {volume} {887}},\ \bibinfo {eid} {13} (\bibinfo {year} {2019})},\ \Eprint {https://arxiv.org/abs/1910.07009} {arXiv:1910.07009 [astro-ph.HE]} \BibitemShut {NoStop}%
\bibitem [{\citenamefont {{Khadka}}\ and\ \citenamefont {{Ratra}}(2020{\natexlab{a}})}]{KhadkaRatra2020c}%
  \BibitemOpen
  \bibfield  {author} {\bibinfo {author} {\bibfnamefont {N.}~\bibnamefont {{Khadka}}}\ and\ \bibinfo {author} {\bibfnamefont {B.}~\bibnamefont {{Ratra}}},\ }\href {https://doi.org/10.1093/mnras/staa2779} {\bibfield  {journal} {\bibinfo  {journal} {Mon. Not. R. Astron. Soc.}\ }\textbf {\bibinfo {volume} {499}},\ \bibinfo {pages} {391} (\bibinfo {year} {2020}{\natexlab{a}})},\ \Eprint {https://arxiv.org/abs/2007.13907} {arXiv:2007.13907 [astro-ph.CO]} \BibitemShut {NoStop}%
\bibitem [{\citenamefont {{Dainotti}}\ \emph {et~al.}(2020)\citenamefont {{Dainotti}}, \citenamefont {{Lenart}}, \citenamefont {{Sarracino}}, \citenamefont {{Nagataki}}, \citenamefont {{Capozziello}},\ and\ \citenamefont {{Fraija}}}]{Dainottetal2020}%
  \BibitemOpen
  \bibfield  {author} {\bibinfo {author} {\bibfnamefont {M.~G.}\ \bibnamefont {{Dainotti}}}, \bibinfo {author} {\bibfnamefont {A.~{\L}.}\ \bibnamefont {{Lenart}}}, \bibinfo {author} {\bibfnamefont {G.}~\bibnamefont {{Sarracino}}}, \bibinfo {author} {\bibfnamefont {S.}~\bibnamefont {{Nagataki}}}, \bibinfo {author} {\bibfnamefont {S.}~\bibnamefont {{Capozziello}}},\ and\ \bibinfo {author} {\bibfnamefont {N.}~\bibnamefont {{Fraija}}},\ }\href {https://doi.org/10.3847/1538-4357/abbe8a} {\bibfield  {journal} {\bibinfo  {journal} {Astrophys. J.}\ }\textbf {\bibinfo {volume} {904}},\ \bibinfo {eid} {97} (\bibinfo {year} {2020})},\ \Eprint {https://arxiv.org/abs/2010.02092} {arXiv:2010.02092 [astro-ph.HE]} \BibitemShut {NoStop}%
\bibitem [{\citenamefont {{Hu}}\ \emph {et~al.}(2021)\citenamefont {{Hu}}, \citenamefont {{Wang}},\ and\ \citenamefont {{Dai}}}]{Huetal_2021}%
  \BibitemOpen
  \bibfield  {author} {\bibinfo {author} {\bibfnamefont {J.~P.}\ \bibnamefont {{Hu}}}, \bibinfo {author} {\bibfnamefont {F.~Y.}\ \bibnamefont {{Wang}}},\ and\ \bibinfo {author} {\bibfnamefont {Z.~G.}\ \bibnamefont {{Dai}}},\ }\href {https://doi.org/10.1093/mnras/stab2180} {\bibfield  {journal} {\bibinfo  {journal} {Mon. Not. R. Astron. Soc.}\ }\textbf {\bibinfo {volume} {507}},\ \bibinfo {pages} {730} (\bibinfo {year} {2021})},\ \Eprint {https://arxiv.org/abs/2107.12718} {arXiv:2107.12718 [astro-ph.CO]} \BibitemShut {NoStop}%
\bibitem [{\citenamefont {{Dai}}\ \emph {et~al.}(2021)\citenamefont {{Dai}}, \citenamefont {{Zheng}}, \citenamefont {{Li}}, \citenamefont {{Gao}},\ and\ \citenamefont {{Zhu}}}]{Daietal_2021}%
  \BibitemOpen
  \bibfield  {author} {\bibinfo {author} {\bibfnamefont {Y.}~\bibnamefont {{Dai}}}, \bibinfo {author} {\bibfnamefont {X.-G.}\ \bibnamefont {{Zheng}}}, \bibinfo {author} {\bibfnamefont {Z.-X.}\ \bibnamefont {{Li}}}, \bibinfo {author} {\bibfnamefont {H.}~\bibnamefont {{Gao}}},\ and\ \bibinfo {author} {\bibfnamefont {Z.-H.}\ \bibnamefont {{Zhu}}},\ }\href {https://doi.org/10.1051/0004-6361/202140895} {\bibfield  {journal} {\bibinfo  {journal} {Astron. Astrophys.}\ }\textbf {\bibinfo {volume} {651}},\ \bibinfo {eid} {L8} (\bibinfo {year} {2021})},\ \Eprint {https://arxiv.org/abs/2111.05544} {arXiv:2111.05544 [astro-ph.HE]} \BibitemShut {NoStop}%
\bibitem [{\citenamefont {{Demianski}}\ \emph {et~al.}(2021)\citenamefont {{Demianski}}, \citenamefont {{Piedipalumbo}}, \citenamefont {{Sawant}},\ and\ \citenamefont {{Amati}}}]{Demianskietal_2021}%
  \BibitemOpen
  \bibfield  {author} {\bibinfo {author} {\bibfnamefont {M.}~\bibnamefont {{Demianski}}}, \bibinfo {author} {\bibfnamefont {E.}~\bibnamefont {{Piedipalumbo}}}, \bibinfo {author} {\bibfnamefont {D.}~\bibnamefont {{Sawant}}},\ and\ \bibinfo {author} {\bibfnamefont {L.}~\bibnamefont {{Amati}}},\ }\href {https://doi.org/10.1093/mnras/stab1669} {\bibfield  {journal} {\bibinfo  {journal} {\mnras}\ }\textbf {\bibinfo {volume} {506}},\ \bibinfo {pages} {903} (\bibinfo {year} {2021})}\BibitemShut {NoStop}%
\bibitem [{\citenamefont {{Khadka}}\ \emph {et~al.}(2021{\natexlab{b}})\citenamefont {{Khadka}}, \citenamefont {{Luongo}}, \citenamefont {{Muccino}},\ and\ \citenamefont {{Ratra}}}]{Khadkaetal_2021b}%
  \BibitemOpen
  \bibfield  {author} {\bibinfo {author} {\bibfnamefont {N.}~\bibnamefont {{Khadka}}}, \bibinfo {author} {\bibfnamefont {O.}~\bibnamefont {{Luongo}}}, \bibinfo {author} {\bibfnamefont {M.}~\bibnamefont {{Muccino}}},\ and\ \bibinfo {author} {\bibfnamefont {B.}~\bibnamefont {{Ratra}}},\ }\href {https://doi.org/10.1088/1475-7516/2021/09/042} {\bibfield  {journal} {\bibinfo  {journal} {J. Cosmol. Astropart. Phys.}\ }\textbf {\bibinfo {volume} {2021}}\bibfield  {number} {\bibinfo  {number} { (9)},\ \bibinfo {eid} {042}},\ }\Eprint {https://arxiv.org/abs/2105.12692} {arXiv:2105.12692 [astro-ph.CO]} \BibitemShut {NoStop}%
\bibitem [{\citenamefont {{Cao}}\ \emph {et~al.}(2022{\natexlab{a}})\citenamefont {{Cao}}, \citenamefont {{Dainotti}},\ and\ \citenamefont {{Ratra}}}]{CaoDainottiRatra2022b}%
  \BibitemOpen
  \bibfield  {author} {\bibinfo {author} {\bibfnamefont {S.}~\bibnamefont {{Cao}}}, \bibinfo {author} {\bibfnamefont {M.}~\bibnamefont {{Dainotti}}},\ and\ \bibinfo {author} {\bibfnamefont {B.}~\bibnamefont {{Ratra}}},\ }\href {https://doi.org/10.1093/mnras/stac2170} {\bibfield  {journal} {\bibinfo  {journal} {Mon. Not. R. Astron. Soc.}\ }\textbf {\bibinfo {volume} {516}},\ \bibinfo {pages} {1386} (\bibinfo {year} {2022}{\natexlab{a}})},\ \Eprint {https://arxiv.org/abs/2204.08710} {arXiv:2204.08710 [astro-ph.CO]} \BibitemShut {NoStop}%
\bibitem [{\citenamefont {{Dainotti}}\ \emph {et~al.}(2022{\natexlab{a}})\citenamefont {{Dainotti}}, \citenamefont {{Nielson}}, \citenamefont {{Sarracino}}, \citenamefont {{Rinaldi}}, \citenamefont {{Nagataki}}, \citenamefont {{Capozziello}}, \citenamefont {{Gnedin}},\ and\ \citenamefont {{Bargiacchi}}}]{DainottiNielson2022}%
  \BibitemOpen
  \bibfield  {author} {\bibinfo {author} {\bibfnamefont {M.~G.}\ \bibnamefont {{Dainotti}}}, \bibinfo {author} {\bibfnamefont {V.}~\bibnamefont {{Nielson}}}, \bibinfo {author} {\bibfnamefont {G.}~\bibnamefont {{Sarracino}}}, \bibinfo {author} {\bibfnamefont {E.}~\bibnamefont {{Rinaldi}}}, \bibinfo {author} {\bibfnamefont {S.}~\bibnamefont {{Nagataki}}}, \bibinfo {author} {\bibfnamefont {S.}~\bibnamefont {{Capozziello}}}, \bibinfo {author} {\bibfnamefont {O.~Y.}\ \bibnamefont {{Gnedin}}},\ and\ \bibinfo {author} {\bibfnamefont {G.}~\bibnamefont {{Bargiacchi}}},\ }\href {https://doi.org/10.1093/mnras/stac1141} {\bibfield  {journal} {\bibinfo  {journal} {Mon. Not. R. Astron. Soc.}\ }\textbf {\bibinfo {volume} {514}},\ \bibinfo {pages} {1828} (\bibinfo {year} {2022}{\natexlab{a}})},\ \Eprint {https://arxiv.org/abs/2203.15538} {arXiv:2203.15538 [astro-ph.CO]} \BibitemShut {NoStop}%
\bibitem [{\citenamefont {{Luongo}}\ and\ \citenamefont {{Muccino}}(2021)}]{LuongoMuccino2021}%
  \BibitemOpen
  \bibfield  {author} {\bibinfo {author} {\bibfnamefont {O.}~\bibnamefont {{Luongo}}}\ and\ \bibinfo {author} {\bibfnamefont {M.}~\bibnamefont {{Muccino}}},\ }\href {https://doi.org/10.3390/galaxies9040077} {\bibfield  {journal} {\bibinfo  {journal} {Galaxies}\ }\textbf {\bibinfo {volume} {9}},\ \bibinfo {pages} {77} (\bibinfo {year} {2021})},\ \Eprint {https://arxiv.org/abs/2110.14408} {arXiv:2110.14408 [astro-ph.HE]} \BibitemShut {NoStop}%
\bibitem [{\citenamefont {{Cao}}\ \emph {et~al.}(2022{\natexlab{b}})\citenamefont {{Cao}}, \citenamefont {{Khadka}},\ and\ \citenamefont {{Ratra}}}]{CaoKhadkaRatra2021}%
  \BibitemOpen
  \bibfield  {author} {\bibinfo {author} {\bibfnamefont {S.}~\bibnamefont {{Cao}}}, \bibinfo {author} {\bibfnamefont {N.}~\bibnamefont {{Khadka}}},\ and\ \bibinfo {author} {\bibfnamefont {B.}~\bibnamefont {{Ratra}}},\ }\href {https://doi.org/10.1093/mnras/stab3559} {\bibfield  {journal} {\bibinfo  {journal} {Mon. Not. R. Astron. Soc.}\ }\textbf {\bibinfo {volume} {510}},\ \bibinfo {pages} {2928} (\bibinfo {year} {2022}{\natexlab{b}})},\ \Eprint {https://arxiv.org/abs/2110.14840} {arXiv:2110.14840 [astro-ph.CO]} \BibitemShut {NoStop}%
\bibitem [{\citenamefont {{Cao}}\ \emph {et~al.}(2022{\natexlab{c}})\citenamefont {{Cao}}, \citenamefont {{Dainotti}},\ and\ \citenamefont {{Ratra}}}]{CaoDainottiRatra2022}%
  \BibitemOpen
  \bibfield  {author} {\bibinfo {author} {\bibfnamefont {S.}~\bibnamefont {{Cao}}}, \bibinfo {author} {\bibfnamefont {M.}~\bibnamefont {{Dainotti}}},\ and\ \bibinfo {author} {\bibfnamefont {B.}~\bibnamefont {{Ratra}}},\ }\href {https://doi.org/10.1093/mnras/stac517} {\bibfield  {journal} {\bibinfo  {journal} {Mon. Not. R. Astron. Soc.}\ }\textbf {\bibinfo {volume} {512}},\ \bibinfo {pages} {439} (\bibinfo {year} {2022}{\natexlab{c}})},\ \Eprint {https://arxiv.org/abs/2201.05245} {arXiv:2201.05245 [astro-ph.CO]} \BibitemShut {NoStop}%
\bibitem [{\citenamefont {{Liu}}\ \emph {et~al.}(2022)\citenamefont {{Liu}}, \citenamefont {{Chen}}, \citenamefont {{Liang}}, \citenamefont {{Yuan}}, \citenamefont {{Yu}},\ and\ \citenamefont {{Wu}}}]{Liuetal2022}%
  \BibitemOpen
  \bibfield  {author} {\bibinfo {author} {\bibfnamefont {Y.}~\bibnamefont {{Liu}}}, \bibinfo {author} {\bibfnamefont {F.}~\bibnamefont {{Chen}}}, \bibinfo {author} {\bibfnamefont {N.}~\bibnamefont {{Liang}}}, \bibinfo {author} {\bibfnamefont {Z.}~\bibnamefont {{Yuan}}}, \bibinfo {author} {\bibfnamefont {H.}~\bibnamefont {{Yu}}},\ and\ \bibinfo {author} {\bibfnamefont {P.}~\bibnamefont {{Wu}}},\ }\href {https://doi.org/10.3847/1538-4357/ac66d3} {\bibfield  {journal} {\bibinfo  {journal} {Astrophys. J.}\ }\textbf {\bibinfo {volume} {931}},\ \bibinfo {eid} {50} (\bibinfo {year} {2022})},\ \Eprint {https://arxiv.org/abs/2203.03178} {arXiv:2203.03178 [astro-ph.CO]} \BibitemShut {NoStop}%
\bibitem [{\citenamefont {{Cao}}\ \emph {et~al.}(2021{\natexlab{b}})\citenamefont {{Cao}}, \citenamefont {{Ryan}}, \citenamefont {{Khadka}},\ and\ \citenamefont {{Ratra}}}]{Caoetal_2021}%
  \BibitemOpen
  \bibfield  {author} {\bibinfo {author} {\bibfnamefont {S.}~\bibnamefont {{Cao}}}, \bibinfo {author} {\bibfnamefont {J.}~\bibnamefont {{Ryan}}}, \bibinfo {author} {\bibfnamefont {N.}~\bibnamefont {{Khadka}}},\ and\ \bibinfo {author} {\bibfnamefont {B.}~\bibnamefont {{Ratra}}},\ }\href {https://doi.org/10.1093/mnras/staa3748} {\bibfield  {journal} {\bibinfo  {journal} {Mon. Not. R. Astron. Soc.}\ }\textbf {\bibinfo {volume} {501}},\ \bibinfo {pages} {1520} (\bibinfo {year} {2021}{\natexlab{b}})},\ \Eprint {https://arxiv.org/abs/2009.12953} {arXiv:2009.12953 [astro-ph.CO]} \BibitemShut {NoStop}%
\bibitem [{\citenamefont {{Risaliti}}\ and\ \citenamefont {{Lusso}}(2015)}]{RisalitiLusso2015}%
  \BibitemOpen
  \bibfield  {author} {\bibinfo {author} {\bibfnamefont {G.}~\bibnamefont {{Risaliti}}}\ and\ \bibinfo {author} {\bibfnamefont {E.}~\bibnamefont {{Lusso}}},\ }\href {https://doi.org/10.1088/0004-637X/815/1/33} {\bibfield  {journal} {\bibinfo  {journal} {Astrophys. J.}\ }\textbf {\bibinfo {volume} {815}},\ \bibinfo {eid} {33} (\bibinfo {year} {2015})},\ \Eprint {https://arxiv.org/abs/1505.07118} {arXiv:1505.07118 [astro-ph.CO]} \BibitemShut {NoStop}%
\bibitem [{\citenamefont {{Risaliti}}\ and\ \citenamefont {{Lusso}}(2019)}]{RisalitiLusso2019}%
  \BibitemOpen
  \bibfield  {author} {\bibinfo {author} {\bibfnamefont {G.}~\bibnamefont {{Risaliti}}}\ and\ \bibinfo {author} {\bibfnamefont {E.}~\bibnamefont {{Lusso}}},\ }\href {https://doi.org/10.1038/s41550-018-0657-z} {\bibfield  {journal} {\bibinfo  {journal} {Nat. Astron.}\ }\textbf {\bibinfo {volume} {3}},\ \bibinfo {pages} {272} (\bibinfo {year} {2019})},\ \Eprint {https://arxiv.org/abs/1811.02590} {arXiv:1811.02590 [astro-ph.CO]} \BibitemShut {NoStop}%
\bibitem [{\citenamefont {{Khadka}}\ and\ \citenamefont {{Ratra}}(2020{\natexlab{b}})}]{KhadkaRatra2020a}%
  \BibitemOpen
  \bibfield  {author} {\bibinfo {author} {\bibfnamefont {N.}~\bibnamefont {{Khadka}}}\ and\ \bibinfo {author} {\bibfnamefont {B.}~\bibnamefont {{Ratra}}},\ }\href {https://doi.org/10.1093/mnras/staa101} {\bibfield  {journal} {\bibinfo  {journal} {Mon. Not. R. Astron. Soc.}\ }\textbf {\bibinfo {volume} {492}},\ \bibinfo {pages} {4456} (\bibinfo {year} {2020}{\natexlab{b}})},\ \Eprint {https://arxiv.org/abs/1909.01400} {arXiv:1909.01400 [astro-ph.CO]} \BibitemShut {NoStop}%
\bibitem [{\citenamefont {{Yang}}\ \emph {et~al.}(2020)\citenamefont {{Yang}}, \citenamefont {{Banerjee}},\ and\ \citenamefont {{{\'O} Colg{\'a}in}}}]{Yangetal2020}%
  \BibitemOpen
  \bibfield  {author} {\bibinfo {author} {\bibfnamefont {T.}~\bibnamefont {{Yang}}}, \bibinfo {author} {\bibfnamefont {A.}~\bibnamefont {{Banerjee}}},\ and\ \bibinfo {author} {\bibfnamefont {E.}~\bibnamefont {{{\'O} Colg{\'a}in}}},\ }\href {https://doi.org/10.1103/PhysRevD.102.123532} {\bibfield  {journal} {\bibinfo  {journal} {Phys. Rev. D}\ }\textbf {\bibinfo {volume} {102}},\ \bibinfo {eid} {123532} (\bibinfo {year} {2020})},\ \Eprint {https://arxiv.org/abs/1911.01681} {arXiv:1911.01681 [astro-ph.CO]} \BibitemShut {NoStop}%
\bibitem [{\citenamefont {{Khadka}}\ and\ \citenamefont {{Ratra}}(2020{\natexlab{c}})}]{KhadkaRatra2020b}%
  \BibitemOpen
  \bibfield  {author} {\bibinfo {author} {\bibfnamefont {N.}~\bibnamefont {{Khadka}}}\ and\ \bibinfo {author} {\bibfnamefont {B.}~\bibnamefont {{Ratra}}},\ }\href {https://doi.org/10.1093/mnras/staa1855} {\bibfield  {journal} {\bibinfo  {journal} {Mon. Not. R. Astron. Soc.}\ }\textbf {\bibinfo {volume} {497}},\ \bibinfo {pages} {263} (\bibinfo {year} {2020}{\natexlab{c}})},\ \Eprint {https://arxiv.org/abs/2004.09979} {arXiv:2004.09979 [astro-ph.CO]} \BibitemShut {NoStop}%
\bibitem [{\citenamefont {{Lusso}}\ \emph {et~al.}(2020)\citenamefont {{Lusso}}, \citenamefont {{Risaliti}}, \citenamefont {{Nardini}}, \citenamefont {{Bargiacchi}}, \citenamefont {{Benetti}}, \citenamefont {{Bisogni}}, \citenamefont {{Capozziello}}, \citenamefont {{Civano}}, \citenamefont {{Eggleston}}, \citenamefont {{Elvis}} \emph {et~al.}}]{Lussoetal2020}%
  \BibitemOpen
  \bibfield  {author} {\bibinfo {author} {\bibfnamefont {E.}~\bibnamefont {{Lusso}}}, \bibinfo {author} {\bibfnamefont {G.}~\bibnamefont {{Risaliti}}}, \bibinfo {author} {\bibfnamefont {E.}~\bibnamefont {{Nardini}}}, \bibinfo {author} {\bibfnamefont {G.}~\bibnamefont {{Bargiacchi}}}, \bibinfo {author} {\bibfnamefont {M.}~\bibnamefont {{Benetti}}}, \bibinfo {author} {\bibfnamefont {S.}~\bibnamefont {{Bisogni}}}, \bibinfo {author} {\bibfnamefont {S.}~\bibnamefont {{Capozziello}}}, \bibinfo {author} {\bibfnamefont {F.}~\bibnamefont {{Civano}}}, \bibinfo {author} {\bibfnamefont {L.}~\bibnamefont {{Eggleston}}}, \bibinfo {author} {\bibfnamefont {M.}~\bibnamefont {{Elvis}}}, \emph {et~al.},\ }\href {https://doi.org/10.1051/0004-6361/202038899} {\bibfield  {journal} {\bibinfo  {journal} {Astron. Astrophys.}\ }\textbf {\bibinfo {volume} {642}},\ \bibinfo {eid} {A150} (\bibinfo {year} {2020})},\ \Eprint {https://arxiv.org/abs/2008.08586} {arXiv:2008.08586 [astro-ph.GA]} \BibitemShut {NoStop}%
\bibitem [{\citenamefont {{Khadka}}\ and\ \citenamefont {{Ratra}}(2021)}]{KhadkaRatra2021}%
  \BibitemOpen
  \bibfield  {author} {\bibinfo {author} {\bibfnamefont {N.}~\bibnamefont {{Khadka}}}\ and\ \bibinfo {author} {\bibfnamefont {B.}~\bibnamefont {{Ratra}}},\ }\href {https://doi.org/10.1093/mnras/stab486} {\bibfield  {journal} {\bibinfo  {journal} {Mon. Not. R. Astron. Soc.}\ }\textbf {\bibinfo {volume} {502}},\ \bibinfo {pages} {6140} (\bibinfo {year} {2021})},\ \Eprint {https://arxiv.org/abs/2012.09291} {arXiv:2012.09291 [astro-ph.CO]} \BibitemShut {NoStop}%
\bibitem [{\citenamefont {{Khadka}}\ and\ \citenamefont {{Ratra}}(2022)}]{KhadkaRatra2022}%
  \BibitemOpen
  \bibfield  {author} {\bibinfo {author} {\bibfnamefont {N.}~\bibnamefont {{Khadka}}}\ and\ \bibinfo {author} {\bibfnamefont {B.}~\bibnamefont {{Ratra}}},\ }\href {https://doi.org/10.1093/mnras/stab3678} {\bibfield  {journal} {\bibinfo  {journal} {Mon. Not. R. Astron. Soc.}\ }\textbf {\bibinfo {volume} {510}},\ \bibinfo {pages} {2753} (\bibinfo {year} {2022})},\ \Eprint {https://arxiv.org/abs/2107.07600} {arXiv:2107.07600 [astro-ph.CO]} \BibitemShut {NoStop}%
\bibitem [{\citenamefont {{Rezaei}}\ \emph {et~al.}(2022)\citenamefont {{Rezaei}}, \citenamefont {{Sol{\`a} Peracaula}},\ and\ \citenamefont {{Malekjani}}}]{Rezaeietal2022}%
  \BibitemOpen
  \bibfield  {author} {\bibinfo {author} {\bibfnamefont {M.}~\bibnamefont {{Rezaei}}}, \bibinfo {author} {\bibfnamefont {J.}~\bibnamefont {{Sol{\`a} Peracaula}}},\ and\ \bibinfo {author} {\bibfnamefont {M.}~\bibnamefont {{Malekjani}}},\ }\href {https://doi.org/10.1093/mnras/stab3117} {\bibfield  {journal} {\bibinfo  {journal} {Mon. Not. R. Astron. Soc.}\ }\textbf {\bibinfo {volume} {509}},\ \bibinfo {pages} {2593} (\bibinfo {year} {2022})},\ \Eprint {https://arxiv.org/abs/2108.06255} {arXiv:2108.06255 [astro-ph.CO]} \BibitemShut {NoStop}%
\bibitem [{\citenamefont {{Luongo}}\ \emph {et~al.}(2022)\citenamefont {{Luongo}}, \citenamefont {{Muccino}}, \citenamefont {{Colg{\'a}in}}, \citenamefont {{Sheikh-Jabbari}},\ and\ \citenamefont {{Yin}}}]{Luongoetal2021}%
  \BibitemOpen
  \bibfield  {author} {\bibinfo {author} {\bibfnamefont {O.}~\bibnamefont {{Luongo}}}, \bibinfo {author} {\bibfnamefont {M.}~\bibnamefont {{Muccino}}}, \bibinfo {author} {\bibfnamefont {E.~{\'O}.}\ \bibnamefont {{Colg{\'a}in}}}, \bibinfo {author} {\bibfnamefont {M.~M.}\ \bibnamefont {{Sheikh-Jabbari}}},\ and\ \bibinfo {author} {\bibfnamefont {L.}~\bibnamefont {{Yin}}},\ }\href {https://doi.org/10.1103/PhysRevD.105.103510} {\bibfield  {journal} {\bibinfo  {journal} {Phys. Rev. D}\ }\textbf {\bibinfo {volume} {105}},\ \bibinfo {eid} {103510} (\bibinfo {year} {2022})},\ \Eprint {https://arxiv.org/abs/2108.13228} {arXiv:2108.13228 [astro-ph.CO]} \BibitemShut {NoStop}%
\bibitem [{\citenamefont {{Dainotti}}\ \emph {et~al.}(2022{\natexlab{b}})\citenamefont {{Dainotti}}, \citenamefont {{Bargiacchi}}, \citenamefont {{Lenart}}, \citenamefont {{Capozziello}}, \citenamefont {{{\'O} Colg{\'a}in}}, \citenamefont {{Solomon}}, \citenamefont {{Stojkovic}},\ and\ \citenamefont {{Sheikh-Jabbari}}}]{DainottiBardiacchi2022}%
  \BibitemOpen
  \bibfield  {author} {\bibinfo {author} {\bibfnamefont {M.~G.}\ \bibnamefont {{Dainotti}}}, \bibinfo {author} {\bibfnamefont {G.}~\bibnamefont {{Bargiacchi}}}, \bibinfo {author} {\bibfnamefont {A.~{\L}.}\ \bibnamefont {{Lenart}}}, \bibinfo {author} {\bibfnamefont {S.}~\bibnamefont {{Capozziello}}}, \bibinfo {author} {\bibfnamefont {E.}~\bibnamefont {{{\'O} Colg{\'a}in}}}, \bibinfo {author} {\bibfnamefont {R.}~\bibnamefont {{Solomon}}}, \bibinfo {author} {\bibfnamefont {D.}~\bibnamefont {{Stojkovic}}},\ and\ \bibinfo {author} {\bibfnamefont {M.~M.}\ \bibnamefont {{Sheikh-Jabbari}}},\ }\href {https://doi.org/10.3847/1538-4357/ac6593} {\bibfield  {journal} {\bibinfo  {journal} {Astrophys. J.}\ }\textbf {\bibinfo {volume} {931}},\ \bibinfo {eid} {106} (\bibinfo {year} {2022}{\natexlab{b}})},\ \Eprint {https://arxiv.org/abs/2203.12914} {arXiv:2203.12914 [astro-ph.HE]} \BibitemShut {NoStop}%
\bibitem [{\citenamefont {Petrosian}\ \emph {et~al.}(2022)\citenamefont {Petrosian}, \citenamefont {Singal},\ and\ \citenamefont {Mutchnick}}]{Petrosian:2022tlp}%
  \BibitemOpen
  \bibfield  {author} {\bibinfo {author} {\bibfnamefont {V.}~\bibnamefont {Petrosian}}, \bibinfo {author} {\bibfnamefont {J.}~\bibnamefont {Singal}},\ and\ \bibinfo {author} {\bibfnamefont {S.}~\bibnamefont {Mutchnick}},\ }\href {https://doi.org/10.3847/2041-8213/ac85ac} {\bibfield  {journal} {\bibinfo  {journal} {Astrophys. J. Lett.}\ }\textbf {\bibinfo {volume} {935}},\ \bibinfo {pages} {L19} (\bibinfo {year} {2022})},\ \Eprint {https://arxiv.org/abs/2205.07981} {arXiv:2205.07981 [astro-ph.CO]} \BibitemShut {NoStop}%
\bibitem [{\citenamefont {{Khadka}}\ \emph {et~al.}(2023)\citenamefont {{Khadka}}, \citenamefont {{Zaja{\v{c}}ek}}, \citenamefont {{Prince}}, \citenamefont {{Panda}}, \citenamefont {{Czerny}}, \citenamefont {{Mart{\'\i}nez-Aldama}}, \citenamefont {{Jaiswal}},\ and\ \citenamefont {{Ratra}}}]{Khadka:2022aeg}%
  \BibitemOpen
  \bibfield  {author} {\bibinfo {author} {\bibfnamefont {N.}~\bibnamefont {{Khadka}}}, \bibinfo {author} {\bibfnamefont {M.}~\bibnamefont {{Zaja{\v{c}}ek}}}, \bibinfo {author} {\bibfnamefont {R.}~\bibnamefont {{Prince}}}, \bibinfo {author} {\bibfnamefont {S.}~\bibnamefont {{Panda}}}, \bibinfo {author} {\bibfnamefont {B.}~\bibnamefont {{Czerny}}}, \bibinfo {author} {\bibfnamefont {M.~L.}\ \bibnamefont {{Mart{\'\i}nez-Aldama}}}, \bibinfo {author} {\bibfnamefont {V.~K.}\ \bibnamefont {{Jaiswal}}},\ and\ \bibinfo {author} {\bibfnamefont {B.}~\bibnamefont {{Ratra}}},\ }\href {https://doi.org/10.1093/mnras/stad1040} {\bibfield  {journal} {\bibinfo  {journal} {\mnras}\ }\textbf {\bibinfo {volume} {522}},\ \bibinfo {pages} {1247} (\bibinfo {year} {2023})},\ \Eprint {https://arxiv.org/abs/2212.10483} {arXiv:2212.10483 [astro-ph.CO]} \BibitemShut {NoStop}%
\bibitem [{\citenamefont {{Zaja{\v{c}}ek}}\ \emph {et~al.}(2024)\citenamefont {{Zaja{\v{c}}ek}}, \citenamefont {{Czerny}}, \citenamefont {{Khadka}}, \citenamefont {{Mart{\'\i}nez-Aldama}}, \citenamefont {{Prince}}, \citenamefont {{Panda}},\ and\ \citenamefont {{Ratra}}}]{Zajaceketal2024}%
  \BibitemOpen
  \bibfield  {author} {\bibinfo {author} {\bibfnamefont {M.}~\bibnamefont {{Zaja{\v{c}}ek}}}, \bibinfo {author} {\bibfnamefont {B.}~\bibnamefont {{Czerny}}}, \bibinfo {author} {\bibfnamefont {N.}~\bibnamefont {{Khadka}}}, \bibinfo {author} {\bibfnamefont {M.~L.}\ \bibnamefont {{Mart{\'\i}nez-Aldama}}}, \bibinfo {author} {\bibfnamefont {R.}~\bibnamefont {{Prince}}}, \bibinfo {author} {\bibfnamefont {S.}~\bibnamefont {{Panda}}},\ and\ \bibinfo {author} {\bibfnamefont {B.}~\bibnamefont {{Ratra}}},\ }\href {https://doi.org/10.3847/1538-4357/ad11dc} {\bibfield  {journal} {\bibinfo  {journal} {\apj}\ }\textbf {\bibinfo {volume} {961}},\ \bibinfo {eid} {229} (\bibinfo {year} {2024})},\ \Eprint {https://arxiv.org/abs/2305.08179} {arXiv:2305.08179 [astro-ph.GA]} \BibitemShut {NoStop}%
\bibitem [{\citenamefont {{Peebles}}(1984)}]{peeb84}%
  \BibitemOpen
  \bibfield  {author} {\bibinfo {author} {\bibfnamefont {P.~J.~E.}\ \bibnamefont {{Peebles}}},\ }\href {https://doi.org/10.1086/162425} {\bibfield  {journal} {\bibinfo  {journal} {Astrophys. J.}\ }\textbf {\bibinfo {volume} {284}},\ \bibinfo {pages} {439} (\bibinfo {year} {1984})}\BibitemShut {NoStop}%
\bibitem [{\citenamefont {{Perivolaropoulos}}\ and\ \citenamefont {{Skara}}(2022)}]{PerivolaropoulosSkara2021}%
  \BibitemOpen
  \bibfield  {author} {\bibinfo {author} {\bibfnamefont {L.}~\bibnamefont {{Perivolaropoulos}}}\ and\ \bibinfo {author} {\bibfnamefont {F.}~\bibnamefont {{Skara}}},\ }\href {https://doi.org/10.1016/j.newar.2022.101659} {\bibfield  {journal} {\bibinfo  {journal} {New Astronomy Reviews}\ }\textbf {\bibinfo {volume} {95}},\ \bibinfo {eid} {101659} (\bibinfo {year} {2022})},\ \Eprint {https://arxiv.org/abs/2105.05208} {arXiv:2105.05208 [astro-ph.CO]} \BibitemShut {NoStop}%
\bibitem [{\citenamefont {{Moresco}}\ \emph {et~al.}(2022)\citenamefont {{Moresco}}, \citenamefont {{Amati}}, \citenamefont {{Amendola}}, \citenamefont {{Birrer}}, \citenamefont {{Blakeslee}}, \citenamefont {{Cantiello}}, \citenamefont {{Cimatti}}, \citenamefont {{Darling}}, \citenamefont {{Della Valle}}, \citenamefont {{Fishbach}} \emph {et~al.}}]{Morescoetal2022}%
  \BibitemOpen
  \bibfield  {author} {\bibinfo {author} {\bibfnamefont {M.}~\bibnamefont {{Moresco}}}, \bibinfo {author} {\bibfnamefont {L.}~\bibnamefont {{Amati}}}, \bibinfo {author} {\bibfnamefont {L.}~\bibnamefont {{Amendola}}}, \bibinfo {author} {\bibfnamefont {S.}~\bibnamefont {{Birrer}}}, \bibinfo {author} {\bibfnamefont {J.~P.}\ \bibnamefont {{Blakeslee}}}, \bibinfo {author} {\bibfnamefont {M.}~\bibnamefont {{Cantiello}}}, \bibinfo {author} {\bibfnamefont {A.}~\bibnamefont {{Cimatti}}}, \bibinfo {author} {\bibfnamefont {J.}~\bibnamefont {{Darling}}}, \bibinfo {author} {\bibfnamefont {M.}~\bibnamefont {{Della Valle}}}, \bibinfo {author} {\bibfnamefont {M.}~\bibnamefont {{Fishbach}}}, \emph {et~al.},\ }\href {https://doi.org/10.1007/s41114-022-00040-z} {\bibfield  {journal} {\bibinfo  {journal} {Living Reviews in Relativity}\ }\textbf {\bibinfo {volume} {25}},\ \bibinfo {eid} {6} (\bibinfo {year} {2022})},\ \Eprint {https://arxiv.org/abs/2201.07241} {arXiv:2201.07241 [astro-ph.CO]} \BibitemShut {NoStop}%
\bibitem [{\citenamefont {{Abdalla}}\ \emph {et~al.}(2022)\citenamefont {{Abdalla}}, \citenamefont {{Abell{\'a}n}}, \citenamefont {{Aboubrahim}}, \citenamefont {{Agnello}}, \citenamefont {{Akarsu}}, \citenamefont {{Akrami}}, \citenamefont {{Alestas}}, \citenamefont {{Aloni}}, \citenamefont {{Amendola}}, \citenamefont {{Anchordoqui}} \emph {et~al.}}]{Abdallaetal2022}%
  \BibitemOpen
  \bibfield  {author} {\bibinfo {author} {\bibfnamefont {E.}~\bibnamefont {{Abdalla}}}, \bibinfo {author} {\bibfnamefont {G.~F.}\ \bibnamefont {{Abell{\'a}n}}}, \bibinfo {author} {\bibfnamefont {A.}~\bibnamefont {{Aboubrahim}}}, \bibinfo {author} {\bibfnamefont {A.}~\bibnamefont {{Agnello}}}, \bibinfo {author} {\bibfnamefont {{\"O}.}~\bibnamefont {{Akarsu}}}, \bibinfo {author} {\bibfnamefont {Y.}~\bibnamefont {{Akrami}}}, \bibinfo {author} {\bibfnamefont {G.}~\bibnamefont {{Alestas}}}, \bibinfo {author} {\bibfnamefont {D.}~\bibnamefont {{Aloni}}}, \bibinfo {author} {\bibfnamefont {L.}~\bibnamefont {{Amendola}}}, \bibinfo {author} {\bibfnamefont {L.~A.}\ \bibnamefont {{Anchordoqui}}}, \emph {et~al.},\ }\href {https://doi.org/10.1016/j.jheap.2022.04.002} {\bibfield  {journal} {\bibinfo  {journal} {Journal of High Energy Astrophysics}\ }\textbf {\bibinfo {volume} {34}},\ \bibinfo {pages} {49} (\bibinfo {year} {2022})},\ \Eprint {https://arxiv.org/abs/2203.06142} {arXiv:2203.06142 [astro-ph.CO]} \BibitemShut
  {NoStop}%
\bibitem [{\citenamefont {{Hu}}\ and\ \citenamefont {{Wang}}(2023)}]{Hu:2023jqc}%
  \BibitemOpen
  \bibfield  {author} {\bibinfo {author} {\bibfnamefont {J.-P.}\ \bibnamefont {{Hu}}}\ and\ \bibinfo {author} {\bibfnamefont {F.-Y.}\ \bibnamefont {{Wang}}},\ }\href {https://doi.org/10.3390/universe9020094} {\bibfield  {journal} {\bibinfo  {journal} {Universe}\ }\textbf {\bibinfo {volume} {9}},\ \bibinfo {eid} {94} (\bibinfo {year} {2023})},\ \Eprint {https://arxiv.org/abs/2302.05709} {arXiv:2302.05709 [astro-ph.CO]} \BibitemShut {NoStop}%
\bibitem [{\citenamefont {{Ooba}}\ \emph {et~al.}(2018{\natexlab{a}})\citenamefont {{Ooba}}, \citenamefont {{Ratra}},\ and\ \citenamefont {{Sugiyama}}}]{Oobaetal2018b}%
  \BibitemOpen
  \bibfield  {author} {\bibinfo {author} {\bibfnamefont {J.}~\bibnamefont {{Ooba}}}, \bibinfo {author} {\bibfnamefont {B.}~\bibnamefont {{Ratra}}},\ and\ \bibinfo {author} {\bibfnamefont {N.}~\bibnamefont {{Sugiyama}}},\ }\href {https://doi.org/10.3847/1538-4357/aaec6f} {\bibfield  {journal} {\bibinfo  {journal} {\apj}\ }\textbf {\bibinfo {volume} {869}},\ \bibinfo {eid} {34} (\bibinfo {year} {2018}{\natexlab{a}})},\ \Eprint {https://arxiv.org/abs/1710.03271} {arXiv:1710.03271 [astro-ph.CO]} \BibitemShut {NoStop}%
\bibitem [{\citenamefont {{Park}}\ and\ \citenamefont {{Ratra}}(2019{\natexlab{a}})}]{ParkRatra2019b}%
  \BibitemOpen
  \bibfield  {author} {\bibinfo {author} {\bibfnamefont {C.-G.}\ \bibnamefont {{Park}}}\ and\ \bibinfo {author} {\bibfnamefont {B.}~\bibnamefont {{Ratra}}},\ }\href {https://doi.org/10.1007/s10509-019-3567-3} {\bibfield  {journal} {\bibinfo  {journal} {\apss}\ }\textbf {\bibinfo {volume} {364}},\ \bibinfo {eid} {82} (\bibinfo {year} {2019}{\natexlab{a}})},\ \Eprint {https://arxiv.org/abs/1803.05522} {arXiv:1803.05522 [astro-ph.CO]} \BibitemShut {NoStop}%
\bibitem [{\citenamefont {{Di Valentino}}\ \emph {et~al.}(2021)\citenamefont {{Di Valentino}}, \citenamefont {{Melchiorri}},\ and\ \citenamefont {{Silk}}}]{DiValentinoetal2021a}%
  \BibitemOpen
  \bibfield  {author} {\bibinfo {author} {\bibfnamefont {E.}~\bibnamefont {{Di Valentino}}}, \bibinfo {author} {\bibfnamefont {A.}~\bibnamefont {{Melchiorri}}},\ and\ \bibinfo {author} {\bibfnamefont {J.}~\bibnamefont {{Silk}}},\ }\href {https://doi.org/10.3847/2041-8213/abe1c4} {\bibfield  {journal} {\bibinfo  {journal} {Astrophys. J. Lett.}\ }\textbf {\bibinfo {volume} {908}},\ \bibinfo {eid} {L9} (\bibinfo {year} {2021})},\ \Eprint {https://arxiv.org/abs/2003.04935} {arXiv:2003.04935 [astro-ph.CO]} \BibitemShut {NoStop}%
\bibitem [{\citenamefont {{Arjona}}\ and\ \citenamefont {{Nesseris}}(2021)}]{ArjonaNesseris2021}%
  \BibitemOpen
  \bibfield  {author} {\bibinfo {author} {\bibfnamefont {R.}~\bibnamefont {{Arjona}}}\ and\ \bibinfo {author} {\bibfnamefont {S.}~\bibnamefont {{Nesseris}}},\ }\href {https://doi.org/10.1103/PhysRevD.103.103539} {\bibfield  {journal} {\bibinfo  {journal} {Phys. Rev. D}\ }\textbf {\bibinfo {volume} {103}},\ \bibinfo {eid} {103539} (\bibinfo {year} {2021})},\ \Eprint {https://arxiv.org/abs/2103.06789} {arXiv:2103.06789 [astro-ph.CO]} \BibitemShut {NoStop}%
\bibitem [{\citenamefont {{Dhawan}}\ \emph {et~al.}(2021)\citenamefont {{Dhawan}}, \citenamefont {{Alsing}},\ and\ \citenamefont {{Vagnozzi}}}]{Dhawanetal2021}%
  \BibitemOpen
  \bibfield  {author} {\bibinfo {author} {\bibfnamefont {S.}~\bibnamefont {{Dhawan}}}, \bibinfo {author} {\bibfnamefont {J.}~\bibnamefont {{Alsing}}},\ and\ \bibinfo {author} {\bibfnamefont {S.}~\bibnamefont {{Vagnozzi}}},\ }\href {https://doi.org/10.1093/mnrasl/slab058} {\bibfield  {journal} {\bibinfo  {journal} {Mon. Not. R. Astron. Soc. Lett.}\ }\textbf {\bibinfo {volume} {506}},\ \bibinfo {pages} {L1} (\bibinfo {year} {2021})},\ \Eprint {https://arxiv.org/abs/2104.02485} {arXiv:2104.02485 [astro-ph.CO]} \BibitemShut {NoStop}%
\bibitem [{\citenamefont {{Renzi}}\ \emph {et~al.}(2022)\citenamefont {{Renzi}}, \citenamefont {{Hogg}},\ and\ \citenamefont {{Giar{\`e}}}}]{Renzietal2022}%
  \BibitemOpen
  \bibfield  {author} {\bibinfo {author} {\bibfnamefont {F.}~\bibnamefont {{Renzi}}}, \bibinfo {author} {\bibfnamefont {N.~B.}\ \bibnamefont {{Hogg}}},\ and\ \bibinfo {author} {\bibfnamefont {W.}~\bibnamefont {{Giar{\`e}}}},\ }\href {https://doi.org/10.1093/mnras/stac1030} {\bibfield  {journal} {\bibinfo  {journal} {\mnras}\ }\textbf {\bibinfo {volume} {513}},\ \bibinfo {pages} {4004} (\bibinfo {year} {2022})},\ \Eprint {https://arxiv.org/abs/2112.05701} {arXiv:2112.05701 [astro-ph.CO]} \BibitemShut {NoStop}%
\bibitem [{\citenamefont {{Geng}}\ \emph {et~al.}(2022)\citenamefont {{Geng}}, \citenamefont {{Hsu}},\ and\ \citenamefont {{Lu}}}]{Gengetal2022}%
  \BibitemOpen
  \bibfield  {author} {\bibinfo {author} {\bibfnamefont {C.-Q.}\ \bibnamefont {{Geng}}}, \bibinfo {author} {\bibfnamefont {Y.-T.}\ \bibnamefont {{Hsu}}},\ and\ \bibinfo {author} {\bibfnamefont {J.-R.}\ \bibnamefont {{Lu}}},\ }\href {https://doi.org/10.3847/1538-4357/ac4495} {\bibfield  {journal} {\bibinfo  {journal} {Astrophys. J.}\ }\textbf {\bibinfo {volume} {926}},\ \bibinfo {eid} {74} (\bibinfo {year} {2022})},\ \Eprint {https://arxiv.org/abs/2112.10041} {arXiv:2112.10041 [astro-ph.CO]} \BibitemShut {NoStop}%
\bibitem [{\citenamefont {{Mukherjee}}\ and\ \citenamefont {{Banerjee}}(2022)}]{MukherjeeBanerjee2022}%
  \BibitemOpen
  \bibfield  {author} {\bibinfo {author} {\bibfnamefont {P.}~\bibnamefont {{Mukherjee}}}\ and\ \bibinfo {author} {\bibfnamefont {N.}~\bibnamefont {{Banerjee}}},\ }\href {https://doi.org/10.1103/PhysRevD.105.063516} {\bibfield  {journal} {\bibinfo  {journal} {Phys. Rev. D}\ }\textbf {\bibinfo {volume} {105}},\ \bibinfo {eid} {063516} (\bibinfo {year} {2022})},\ \Eprint {https://arxiv.org/abs/2202.07886} {arXiv:2202.07886 [astro-ph.CO]} \BibitemShut {NoStop}%
\bibitem [{\citenamefont {{Glanville}}\ \emph {et~al.}(2022)\citenamefont {{Glanville}}, \citenamefont {{Howlett}},\ and\ \citenamefont {{Davis}}}]{Glanvilleetal2022}%
  \BibitemOpen
  \bibfield  {author} {\bibinfo {author} {\bibfnamefont {A.}~\bibnamefont {{Glanville}}}, \bibinfo {author} {\bibfnamefont {C.}~\bibnamefont {{Howlett}}},\ and\ \bibinfo {author} {\bibfnamefont {T.}~\bibnamefont {{Davis}}},\ }\href {https://doi.org/10.1093/mnras/stac2891} {\bibfield  {journal} {\bibinfo  {journal} {\mnras}\ }\textbf {\bibinfo {volume} {517}},\ \bibinfo {pages} {3087} (\bibinfo {year} {2022})},\ \Eprint {https://arxiv.org/abs/2205.05892} {arXiv:2205.05892 [astro-ph.CO]} \BibitemShut {NoStop}%
\bibitem [{\citenamefont {{Wu}}\ \emph {et~al.}(2023)\citenamefont {{Wu}}, \citenamefont {{Qi}},\ and\ \citenamefont {{Zhang}}}]{Wuetal2023}%
  \BibitemOpen
  \bibfield  {author} {\bibinfo {author} {\bibfnamefont {P.-J.}\ \bibnamefont {{Wu}}}, \bibinfo {author} {\bibfnamefont {J.-Z.}\ \bibnamefont {{Qi}}},\ and\ \bibinfo {author} {\bibfnamefont {X.}~\bibnamefont {{Zhang}}},\ }\href {https://doi.org/10.1088/1674-1137/acc647} {\bibfield  {journal} {\bibinfo  {journal} {Chinese Physics C}\ }\textbf {\bibinfo {volume} {47}},\ \bibinfo {eid} {055106} (\bibinfo {year} {2023})},\ \Eprint {https://arxiv.org/abs/2209.08502} {arXiv:2209.08502 [astro-ph.CO]} \BibitemShut {NoStop}%
\bibitem [{\citenamefont {{\MakeLowercase{D}e Cruz P{\'e}rez}}\ \emph {et~al.}(2023)\citenamefont {{\MakeLowercase{D}e Cruz P{\'e}rez}}, \citenamefont {{Park}},\ and\ \citenamefont {{Ratra}}}]{deCruzPerezetal2023}%
  \BibitemOpen
  \bibfield  {author} {\bibinfo {author} {\bibfnamefont {J.}~\bibnamefont {{\MakeLowercase{D}e Cruz P{\'e}rez}}}, \bibinfo {author} {\bibfnamefont {C.-G.}\ \bibnamefont {{Park}}},\ and\ \bibinfo {author} {\bibfnamefont {B.}~\bibnamefont {{Ratra}}},\ }\href {https://doi.org/10.1103/PhysRevD.107.063522} {\bibfield  {journal} {\bibinfo  {journal} {\prd}\ }\textbf {\bibinfo {volume} {107}},\ \bibinfo {eid} {063522} (\bibinfo {year} {2023})},\ \Eprint {https://arxiv.org/abs/2211.04268} {arXiv:2211.04268 [astro-ph.CO]} \BibitemShut {NoStop}%
\bibitem [{\citenamefont {{Dahiya}}\ and\ \citenamefont {{Jain}}(2023)}]{DahiyaJain2022}%
  \BibitemOpen
  \bibfield  {author} {\bibinfo {author} {\bibfnamefont {D.}~\bibnamefont {{Dahiya}}}\ and\ \bibinfo {author} {\bibfnamefont {D.}~\bibnamefont {{Jain}}},\ }\href {https://doi.org/10.1088/1674-4527/ace17a} {\bibfield  {journal} {\bibinfo  {journal} {Research in Astronomy and Astrophysics}\ }\textbf {\bibinfo {volume} {23}},\ \bibinfo {eid} {095001} (\bibinfo {year} {2023})},\ \Eprint {https://arxiv.org/abs/2212.04751} {arXiv:2212.04751 [astro-ph.CO]} \BibitemShut {NoStop}%
\bibitem [{\citenamefont {{Stevens}}\ \emph {et~al.}(2023)\citenamefont {{Stevens}}, \citenamefont {{Khoraminezhad}},\ and\ \citenamefont {{Saito}}}]{Stevensetal2023}%
  \BibitemOpen
  \bibfield  {author} {\bibinfo {author} {\bibfnamefont {J.}~\bibnamefont {{Stevens}}}, \bibinfo {author} {\bibfnamefont {H.}~\bibnamefont {{Khoraminezhad}}},\ and\ \bibinfo {author} {\bibfnamefont {S.}~\bibnamefont {{Saito}}},\ }\href {https://doi.org/10.1088/1475-7516/2023/07/046} {\bibfield  {journal} {\bibinfo  {journal} {\jcap}\ }\textbf {\bibinfo {volume} {2023}},\ \bibinfo {eid} {046} (\bibinfo {year} {2023})},\ \Eprint {https://arxiv.org/abs/2212.09804} {arXiv:2212.09804 [astro-ph.CO]} \BibitemShut {NoStop}%
\bibitem [{\citenamefont {{Favale}}\ \emph {et~al.}(2023)\citenamefont {{Favale}}, \citenamefont {{G{\'o}mez-Valent}},\ and\ \citenamefont {{Migliaccio}}}]{Favaleetal2023}%
  \BibitemOpen
  \bibfield  {author} {\bibinfo {author} {\bibfnamefont {A.}~\bibnamefont {{Favale}}}, \bibinfo {author} {\bibfnamefont {A.}~\bibnamefont {{G{\'o}mez-Valent}}},\ and\ \bibinfo {author} {\bibfnamefont {M.}~\bibnamefont {{Migliaccio}}},\ }\href {https://doi.org/10.1093/mnras/stad1621} {\bibfield  {journal} {\bibinfo  {journal} {\mnras}\ }\textbf {\bibinfo {volume} {523}},\ \bibinfo {pages} {3406} (\bibinfo {year} {2023})},\ \Eprint {https://arxiv.org/abs/2301.09591} {arXiv:2301.09591 [astro-ph.CO]} \BibitemShut {NoStop}%
\bibitem [{\citenamefont {{Qi}}\ \emph {et~al.}(2023)\citenamefont {{Qi}}, \citenamefont {{Meng}}, \citenamefont {{Zhang}},\ and\ \citenamefont {{Zhang}}}]{Qietal2023}%
  \BibitemOpen
  \bibfield  {author} {\bibinfo {author} {\bibfnamefont {J.-Z.}\ \bibnamefont {{Qi}}}, \bibinfo {author} {\bibfnamefont {P.}~\bibnamefont {{Meng}}}, \bibinfo {author} {\bibfnamefont {J.-F.}\ \bibnamefont {{Zhang}}},\ and\ \bibinfo {author} {\bibfnamefont {X.}~\bibnamefont {{Zhang}}},\ }\href {https://doi.org/10.1103/PhysRevD.108.063522} {\bibfield  {journal} {\bibinfo  {journal} {\prd}\ }\textbf {\bibinfo {volume} {108}},\ \bibinfo {eid} {063522} (\bibinfo {year} {2023})},\ \Eprint {https://arxiv.org/abs/2302.08889} {arXiv:2302.08889 [astro-ph.CO]} \BibitemShut {NoStop}%
\bibitem [{\citenamefont {{de Cruz Perez}}\ \emph {et~al.}(2024)\citenamefont {{de Cruz Perez}}, \citenamefont {{Park}},\ and\ \citenamefont {{Ratra}}}]{deCruzPerezetal2024}%
  \BibitemOpen
  \bibfield  {author} {\bibinfo {author} {\bibfnamefont {J.}~\bibnamefont {{de Cruz Perez}}}, \bibinfo {author} {\bibfnamefont {C.-G.}\ \bibnamefont {{Park}}},\ and\ \bibinfo {author} {\bibfnamefont {B.}~\bibnamefont {{Ratra}}},\ }\href {https://doi.org/10.48550/arXiv.2404.19194} {\bibfield  {journal} {\bibinfo  {journal} {arXiv e-prints}\ ,\ \bibinfo {eid} {arXiv:2404.19194}} (\bibinfo {year} {2024})},\ \Eprint {https://arxiv.org/abs/2404.19194} {arXiv:2404.19194 [astro-ph.CO]} \BibitemShut {NoStop}%
\bibitem [{\citenamefont {{Peebles}}\ and\ \citenamefont {{Ratra}}(1988)}]{peebrat88}%
  \BibitemOpen
  \bibfield  {author} {\bibinfo {author} {\bibfnamefont {P.~J.~E.}\ \bibnamefont {{Peebles}}}\ and\ \bibinfo {author} {\bibfnamefont {B.}~\bibnamefont {{Ratra}}},\ }\href {https://doi.org/10.1086/185100} {\bibfield  {journal} {\bibinfo  {journal} {Astrophys. J. Lett.}\ }\textbf {\bibinfo {volume} {325}},\ \bibinfo {pages} {L17} (\bibinfo {year} {1988})}\BibitemShut {NoStop}%
\bibitem [{\citenamefont {{Ratra}}\ and\ \citenamefont {{Peebles}}(1988)}]{ratpeeb88}%
  \BibitemOpen
  \bibfield  {author} {\bibinfo {author} {\bibfnamefont {B.}~\bibnamefont {{Ratra}}}\ and\ \bibinfo {author} {\bibfnamefont {P.~J.~E.}\ \bibnamefont {{Peebles}}},\ }\href {https://doi.org/10.1103/PhysRevD.37.3406} {\bibfield  {journal} {\bibinfo  {journal} {Phys. Rev. D}\ }\textbf {\bibinfo {volume} {37}},\ \bibinfo {pages} {3406} (\bibinfo {year} {1988})}\BibitemShut {NoStop}%
\bibitem [{\citenamefont {{Pavlov}}\ \emph {et~al.}(2013)\citenamefont {{Pavlov}}, \citenamefont {{Westmoreland}}, \citenamefont {{Saaidi}},\ and\ \citenamefont {{Ratra}}}]{pavlov13}%
  \BibitemOpen
  \bibfield  {author} {\bibinfo {author} {\bibfnamefont {A.}~\bibnamefont {{Pavlov}}}, \bibinfo {author} {\bibfnamefont {S.}~\bibnamefont {{Westmoreland}}}, \bibinfo {author} {\bibfnamefont {K.}~\bibnamefont {{Saaidi}}},\ and\ \bibinfo {author} {\bibfnamefont {B.}~\bibnamefont {{Ratra}}},\ }\href {https://doi.org/10.1103/PhysRevD.88.123513} {\bibfield  {journal} {\bibinfo  {journal} {Phys. Rev. D}\ }\textbf {\bibinfo {volume} {88}},\ \bibinfo {eid} {123513} (\bibinfo {year} {2013})},\ \Eprint {https://arxiv.org/abs/1307.7399} {arXiv:1307.7399 [astro-ph.CO]} \BibitemShut {NoStop}%
\bibitem [{\citenamefont {{Ooba}}\ \emph {et~al.}(2018{\natexlab{b}})\citenamefont {{Ooba}}, \citenamefont {{Ratra}},\ and\ \citenamefont {{Sugiyama}}}]{ooba_etal_2018b}%
  \BibitemOpen
  \bibfield  {author} {\bibinfo {author} {\bibfnamefont {J.}~\bibnamefont {{Ooba}}}, \bibinfo {author} {\bibfnamefont {B.}~\bibnamefont {{Ratra}}},\ and\ \bibinfo {author} {\bibfnamefont {N.}~\bibnamefont {{Sugiyama}}},\ }\href {https://doi.org/10.3847/1538-4357/aadcf3} {\bibfield  {journal} {\bibinfo  {journal} {Astrophys. J.}\ }\textbf {\bibinfo {volume} {866}},\ \bibinfo {eid} {68} (\bibinfo {year} {2018}{\natexlab{b}})},\ \Eprint {https://arxiv.org/abs/1712.08617} {arXiv:1712.08617 [astro-ph.CO]} \BibitemShut {NoStop}%
\bibitem [{\citenamefont {{Ooba}}\ \emph {et~al.}(2019)\citenamefont {{Ooba}}, \citenamefont {{Ratra}},\ and\ \citenamefont {{Sugiyama}}}]{ooba_etal_2019}%
  \BibitemOpen
  \bibfield  {author} {\bibinfo {author} {\bibfnamefont {J.}~\bibnamefont {{Ooba}}}, \bibinfo {author} {\bibfnamefont {B.}~\bibnamefont {{Ratra}}},\ and\ \bibinfo {author} {\bibfnamefont {N.}~\bibnamefont {{Sugiyama}}},\ }\href {https://doi.org/10.1007/s10509-019-3663-4} {\bibfield  {journal} {\bibinfo  {journal} {Astrophys. Space Sci.}\ }\textbf {\bibinfo {volume} {364}},\ \bibinfo {eid} {176} (\bibinfo {year} {2019})},\ \Eprint {https://arxiv.org/abs/1802.05571} {arXiv:1802.05571 [astro-ph.CO]} \BibitemShut {NoStop}%
\bibitem [{\citenamefont {{Park}}\ and\ \citenamefont {{Ratra}}(2018)}]{park_ratra_2018}%
  \BibitemOpen
  \bibfield  {author} {\bibinfo {author} {\bibfnamefont {C.-G.}\ \bibnamefont {{Park}}}\ and\ \bibinfo {author} {\bibfnamefont {B.}~\bibnamefont {{Ratra}}},\ }\href {https://doi.org/10.3847/1538-4357/aae82d} {\bibfield  {journal} {\bibinfo  {journal} {Astrophys. J.}\ }\textbf {\bibinfo {volume} {868}},\ \bibinfo {eid} {83} (\bibinfo {year} {2018})},\ \Eprint {https://arxiv.org/abs/1807.07421} {arXiv:1807.07421} \BibitemShut {NoStop}%
\bibitem [{\citenamefont {{Park}}\ and\ \citenamefont {{Ratra}}(2019{\natexlab{b}})}]{park_ratra_2019b}%
  \BibitemOpen
  \bibfield  {author} {\bibinfo {author} {\bibfnamefont {C.-G.}\ \bibnamefont {{Park}}}\ and\ \bibinfo {author} {\bibfnamefont {B.}~\bibnamefont {{Ratra}}},\ }\href {https://doi.org/10.1007/s10509-019-3627-8} {\bibfield  {journal} {\bibinfo  {journal} {Astrophys. Space Sci.}\ }\textbf {\bibinfo {volume} {364}},\ \bibinfo {eid} {134} (\bibinfo {year} {2019}{\natexlab{b}})},\ \Eprint {https://arxiv.org/abs/1809.03598} {arXiv:1809.03598 [astro-ph.CO]} \BibitemShut {NoStop}%
\bibitem [{\citenamefont {{Park}}\ and\ \citenamefont {{Ratra}}(2020)}]{park_ratra_2020}%
  \BibitemOpen
  \bibfield  {author} {\bibinfo {author} {\bibfnamefont {C.-G.}\ \bibnamefont {{Park}}}\ and\ \bibinfo {author} {\bibfnamefont {B.}~\bibnamefont {{Ratra}}},\ }\href {https://doi.org/10.1103/PhysRevD.101.083508} {\bibfield  {journal} {\bibinfo  {journal} {Phys. Rev. D}\ }\textbf {\bibinfo {volume} {101}},\ \bibinfo {eid} {083508} (\bibinfo {year} {2020})},\ \Eprint {https://arxiv.org/abs/1908.08477} {arXiv:1908.08477 [astro-ph.CO]} \BibitemShut {NoStop}%
\bibitem [{\citenamefont {{Singh}}\ \emph {et~al.}(2019)\citenamefont {{Singh}}, \citenamefont {{Sangwan}},\ and\ \citenamefont {{Jassal}}}]{Singhetal2019}%
  \BibitemOpen
  \bibfield  {author} {\bibinfo {author} {\bibfnamefont {A.}~\bibnamefont {{Singh}}}, \bibinfo {author} {\bibfnamefont {A.}~\bibnamefont {{Sangwan}}},\ and\ \bibinfo {author} {\bibfnamefont {H.~K.}\ \bibnamefont {{Jassal}}},\ }\href {https://doi.org/10.1088/1475-7516/2019/04/047} {\bibfield  {journal} {\bibinfo  {journal} {J. Cosmol. Astropart. Phys.}\ }\textbf {\bibinfo {volume} {2019}}\bibfield  {number} {\bibinfo  {number} { (4)},\ \bibinfo {eid} {047}},\ }\Eprint {https://arxiv.org/abs/1811.07513} {arXiv:1811.07513 [gr-qc]} \BibitemShut {NoStop}%
\bibitem [{\citenamefont {{Ure{\~n}a-L{\'o}pez}}\ and\ \citenamefont {{Roy}}(2020)}]{UrenaLopezRoy2020}%
  \BibitemOpen
  \bibfield  {author} {\bibinfo {author} {\bibfnamefont {L.~A.}\ \bibnamefont {{Ure{\~n}a-L{\'o}pez}}}\ and\ \bibinfo {author} {\bibfnamefont {N.}~\bibnamefont {{Roy}}},\ }\href {https://doi.org/10.1103/PhysRevD.102.063510} {\bibfield  {journal} {\bibinfo  {journal} {Phys. Rev. D}\ }\textbf {\bibinfo {volume} {102}},\ \bibinfo {eid} {063510} (\bibinfo {year} {2020})},\ \Eprint {https://arxiv.org/abs/2007.08873} {arXiv:2007.08873 [astro-ph.CO]} \BibitemShut {NoStop}%
\bibitem [{\citenamefont {{Sinha}}\ and\ \citenamefont {{Banerjee}}(2021)}]{SinhaBanerjee2021}%
  \BibitemOpen
  \bibfield  {author} {\bibinfo {author} {\bibfnamefont {S.}~\bibnamefont {{Sinha}}}\ and\ \bibinfo {author} {\bibfnamefont {N.}~\bibnamefont {{Banerjee}}},\ }\href {https://doi.org/10.1088/1475-7516/2021/04/060} {\bibfield  {journal} {\bibinfo  {journal} {J. Cosmol. Astropart. Phys.}\ }\textbf {\bibinfo {volume} {2021}}\bibfield  {number} {\bibinfo  {number} { (4)},\ \bibinfo {eid} {060}},\ }\Eprint {https://arxiv.org/abs/2010.02651} {arXiv:2010.02651 [astro-ph.CO]} \BibitemShut {NoStop}%
\bibitem [{\citenamefont {{de Cruz Perez}}\ \emph {et~al.}(2021)\citenamefont {{de Cruz Perez}}, \citenamefont {{Sola Peracaula}}, \citenamefont {{Gomez-Valent}},\ and\ \citenamefont {{Moreno-Pulido}}}]{deCruzetal2021}%
  \BibitemOpen
  \bibfield  {author} {\bibinfo {author} {\bibfnamefont {J.}~\bibnamefont {{de Cruz Perez}}}, \bibinfo {author} {\bibfnamefont {J.}~\bibnamefont {{Sola Peracaula}}}, \bibinfo {author} {\bibfnamefont {A.}~\bibnamefont {{Gomez-Valent}}},\ and\ \bibinfo {author} {\bibfnamefont {C.}~\bibnamefont {{Moreno-Pulido}}},\ }\Eprint {https://arxiv.org/abs/2110.07569} {arXiv:2110.07569 [astro-ph.CO]}  (\bibinfo {year} {2021})\BibitemShut {NoStop}%
\bibitem [{\citenamefont {{Xu}}\ \emph {et~al.}(2022)\citenamefont {{Xu}}, \citenamefont {{Chen}}, \citenamefont {{Xu}},\ and\ \citenamefont {{Cao}}}]{Xuetal2022}%
  \BibitemOpen
  \bibfield  {author} {\bibinfo {author} {\bibfnamefont {T.}~\bibnamefont {{Xu}}}, \bibinfo {author} {\bibfnamefont {Y.}~\bibnamefont {{Chen}}}, \bibinfo {author} {\bibfnamefont {L.}~\bibnamefont {{Xu}}},\ and\ \bibinfo {author} {\bibfnamefont {S.}~\bibnamefont {{Cao}}},\ }\href {https://doi.org/10.1016/j.dark.2022.101023} {\bibfield  {journal} {\bibinfo  {journal} {Physics of the Dark Universe}\ }\textbf {\bibinfo {volume} {36}},\ \bibinfo {eid} {101023} (\bibinfo {year} {2022})},\ \Eprint {https://arxiv.org/abs/2109.02453} {arXiv:2109.02453 [astro-ph.CO]} \BibitemShut {NoStop}%
\bibitem [{\citenamefont {{Jesus}}\ \emph {et~al.}(2022)\citenamefont {{Jesus}}, \citenamefont {{Valentim}}, \citenamefont {{Escobal}}, \citenamefont {{Pereira}},\ and\ \citenamefont {{Benndorf}}}]{Jesusetal2022}%
  \BibitemOpen
  \bibfield  {author} {\bibinfo {author} {\bibfnamefont {J.~F.}\ \bibnamefont {{Jesus}}}, \bibinfo {author} {\bibfnamefont {R.}~\bibnamefont {{Valentim}}}, \bibinfo {author} {\bibfnamefont {A.~A.}\ \bibnamefont {{Escobal}}}, \bibinfo {author} {\bibfnamefont {S.~H.}\ \bibnamefont {{Pereira}}},\ and\ \bibinfo {author} {\bibfnamefont {D.}~\bibnamefont {{Benndorf}}},\ }\href {https://doi.org/10.1088/1475-7516/2022/11/037} {\bibfield  {journal} {\bibinfo  {journal} {J. Cosmol. Astropart. Phys.}\ }\textbf {\bibinfo {volume} {2022}}\bibfield  {number} {\bibinfo  {number} { (11)},\ \bibinfo {eid} {037}},\ }\Eprint {https://arxiv.org/abs/2112.09722} {arXiv:2112.09722 [astro-ph.CO]} \BibitemShut {NoStop}%
\bibitem [{\citenamefont {{Adil}}\ \emph {et~al.}(2023)\citenamefont {{Adil}}, \citenamefont {{Albrecht}},\ and\ \citenamefont {{Knox}}}]{Adiletal2023}%
  \BibitemOpen
  \bibfield  {author} {\bibinfo {author} {\bibfnamefont {A.}~\bibnamefont {{Adil}}}, \bibinfo {author} {\bibfnamefont {A.}~\bibnamefont {{Albrecht}}},\ and\ \bibinfo {author} {\bibfnamefont {L.}~\bibnamefont {{Knox}}},\ }\href {https://doi.org/10.1103/PhysRevD.107.063521} {\bibfield  {journal} {\bibinfo  {journal} {\prd}\ }\textbf {\bibinfo {volume} {107}},\ \bibinfo {eid} {063521} (\bibinfo {year} {2023})},\ \Eprint {https://arxiv.org/abs/2207.10235} {arXiv:2207.10235 [astro-ph.CO]} \BibitemShut {NoStop}%
\bibitem [{\citenamefont {{Dong}}\ \emph {et~al.}(2023)\citenamefont {{Dong}}, \citenamefont {{Park}}, \citenamefont {{Hong}}, \citenamefont {{Kim}}, \citenamefont {{Hwang}}, \citenamefont {{Park}},\ and\ \citenamefont {{Appleby}}}]{Dongetal2023}%
  \BibitemOpen
  \bibfield  {author} {\bibinfo {author} {\bibfnamefont {F.}~\bibnamefont {{Dong}}}, \bibinfo {author} {\bibfnamefont {C.}~\bibnamefont {{Park}}}, \bibinfo {author} {\bibfnamefont {S.~E.}\ \bibnamefont {{Hong}}}, \bibinfo {author} {\bibfnamefont {J.}~\bibnamefont {{Kim}}}, \bibinfo {author} {\bibfnamefont {H.~S.}\ \bibnamefont {{Hwang}}}, \bibinfo {author} {\bibfnamefont {H.}~\bibnamefont {{Park}}},\ and\ \bibinfo {author} {\bibfnamefont {S.}~\bibnamefont {{Appleby}}},\ }\href {https://doi.org/10.3847/1538-4357/acd185} {\bibfield  {journal} {\bibinfo  {journal} {\apj}\ }\textbf {\bibinfo {volume} {953}},\ \bibinfo {eid} {98} (\bibinfo {year} {2023})},\ \Eprint {https://arxiv.org/abs/2305.00206} {arXiv:2305.00206 [astro-ph.CO]} \BibitemShut {NoStop}%
\bibitem [{\citenamefont {{Van Raamsdonk}}\ and\ \citenamefont {{Waddell}}(2023)}]{VanRaamsdonkWaddell2023}%
  \BibitemOpen
  \bibfield  {author} {\bibinfo {author} {\bibfnamefont {M.}~\bibnamefont {{Van Raamsdonk}}}\ and\ \bibinfo {author} {\bibfnamefont {C.}~\bibnamefont {{Waddell}}},\ }\Eprint {https://arxiv.org/abs/2305.04946} {arXiv:2305.04946 [astro-ph.CO]}  (\bibinfo {year} {2023})\BibitemShut {NoStop}%
\bibitem [{\citenamefont {{Cao}}\ and\ \citenamefont {{Ratra}}(2024)}]{CaoRatra2024}%
  \BibitemOpen
  \bibfield  {author} {\bibinfo {author} {\bibfnamefont {S.}~\bibnamefont {{Cao}}}\ and\ \bibinfo {author} {\bibfnamefont {B.}~\bibnamefont {{Ratra}}},\ }\href {https://doi.org/10.48550/arXiv.2404.08697} {\bibfield  {journal} {\bibinfo  {journal} {arXiv e-prints}\ ,\ \bibinfo {eid} {arXiv:2404.08697}} (\bibinfo {year} {2024})},\ \Eprint {https://arxiv.org/abs/2404.08697} {arXiv:2404.08697 [astro-ph.CO]} \BibitemShut {NoStop}%
\bibitem [{\citenamefont {{Blas}}\ \emph {et~al.}(2011)\citenamefont {{Blas}}, \citenamefont {{Lesgourgues}},\ and\ \citenamefont {{Tram}}}]{class}%
  \BibitemOpen
  \bibfield  {author} {\bibinfo {author} {\bibfnamefont {D.}~\bibnamefont {{Blas}}}, \bibinfo {author} {\bibfnamefont {J.}~\bibnamefont {{Lesgourgues}}},\ and\ \bibinfo {author} {\bibfnamefont {T.}~\bibnamefont {{Tram}}},\ }\href {https://doi.org/10.1088/1475-7516/2011/07/034} {\bibfield  {journal} {\bibinfo  {journal} {J. Cosmol. Astropart. Phys.}\ }\textbf {\bibinfo {volume} {2011}}\bibfield  {number} {\bibinfo  {number} { (7)},\ \bibinfo {eid} {034}},\ }\Eprint {https://arxiv.org/abs/1104.2933} {arXiv:1104.2933 [astro-ph.CO]} \BibitemShut {NoStop}%
\bibitem [{\citenamefont {{Bordalo}}\ and\ \citenamefont {{Telles}}(2011)}]{BordaloTelles2011}%
  \BibitemOpen
  \bibfield  {author} {\bibinfo {author} {\bibfnamefont {V.}~\bibnamefont {{Bordalo}}}\ and\ \bibinfo {author} {\bibfnamefont {E.}~\bibnamefont {{Telles}}},\ }\href {https://doi.org/10.1088/0004-637X/735/1/52} {\bibfield  {journal} {\bibinfo  {journal} {\apj}\ }\textbf {\bibinfo {volume} {735}},\ \bibinfo {eid} {52} (\bibinfo {year} {2011})},\ \Eprint {https://arxiv.org/abs/1104.4719} {arXiv:1104.4719 [astro-ph.CO]} \BibitemShut {NoStop}%
\bibitem [{\citenamefont {{Fern{\'a}ndez Arenas}}\ \emph {et~al.}(2018)\citenamefont {{Fern{\'a}ndez Arenas}}, \citenamefont {{Terlevich}}, \citenamefont {{Terlevich}}, \citenamefont {{Melnick}}, \citenamefont {{Ch{\'a}vez}}, \citenamefont {{Bresolin}}, \citenamefont {{Telles}}, \citenamefont {{Plionis}},\ and\ \citenamefont {{Basilakos}}}]{FernandezArenas}%
  \BibitemOpen
  \bibfield  {author} {\bibinfo {author} {\bibfnamefont {D.}~\bibnamefont {{Fern{\'a}ndez Arenas}}}, \bibinfo {author} {\bibfnamefont {E.}~\bibnamefont {{Terlevich}}}, \bibinfo {author} {\bibfnamefont {R.}~\bibnamefont {{Terlevich}}}, \bibinfo {author} {\bibfnamefont {J.}~\bibnamefont {{Melnick}}}, \bibinfo {author} {\bibfnamefont {R.}~\bibnamefont {{Ch{\'a}vez}}}, \bibinfo {author} {\bibfnamefont {F.}~\bibnamefont {{Bresolin}}}, \bibinfo {author} {\bibfnamefont {E.}~\bibnamefont {{Telles}}}, \bibinfo {author} {\bibfnamefont {M.}~\bibnamefont {{Plionis}}},\ and\ \bibinfo {author} {\bibfnamefont {S.}~\bibnamefont {{Basilakos}}},\ }\href {https://doi.org/10.1093/mnras/stx2710} {\bibfield  {journal} {\bibinfo  {journal} {\mnras}\ }\textbf {\bibinfo {volume} {474}},\ \bibinfo {pages} {1250} (\bibinfo {year} {2018})},\ \Eprint {https://arxiv.org/abs/1710.05951} {arXiv:1710.05951 [astro-ph.CO]} \BibitemShut {NoStop}%
\bibitem [{\citenamefont {{Gordon}}\ \emph {et~al.}(2003)\citenamefont {{Gordon}}, \citenamefont {{Clayton}}, \citenamefont {{Misselt}}, \citenamefont {{Landolt}},\ and\ \citenamefont {{Wolff}}}]{Gordon_2003}%
  \BibitemOpen
  \bibfield  {author} {\bibinfo {author} {\bibfnamefont {K.~D.}\ \bibnamefont {{Gordon}}}, \bibinfo {author} {\bibfnamefont {G.~C.}\ \bibnamefont {{Clayton}}}, \bibinfo {author} {\bibfnamefont {K.~A.}\ \bibnamefont {{Misselt}}}, \bibinfo {author} {\bibfnamefont {A.~U.}\ \bibnamefont {{Landolt}}},\ and\ \bibinfo {author} {\bibfnamefont {M.~J.}\ \bibnamefont {{Wolff}}},\ }\href {https://doi.org/10.1086/376774} {\bibfield  {journal} {\bibinfo  {journal} {Astrophys. J.}\ }\textbf {\bibinfo {volume} {594}},\ \bibinfo {pages} {279} (\bibinfo {year} {2003})},\ \Eprint {https://arxiv.org/abs/astro-ph/0305257} {arXiv:astro-ph/0305257 [astro-ph]} \BibitemShut {NoStop}%
\bibitem [{\citenamefont {{Cao}}\ and\ \citenamefont {{Ratra}}(2023)}]{CaoRatra2023}%
  \BibitemOpen
  \bibfield  {author} {\bibinfo {author} {\bibfnamefont {S.}~\bibnamefont {{Cao}}}\ and\ \bibinfo {author} {\bibfnamefont {B.}~\bibnamefont {{Ratra}}},\ }\href {https://doi.org/10.1103/PhysRevD.107.103521} {\bibfield  {journal} {\bibinfo  {journal} {\prd}\ }\textbf {\bibinfo {volume} {107}},\ \bibinfo {eid} {103521} (\bibinfo {year} {2023})},\ \Eprint {https://arxiv.org/abs/2302.14203} {arXiv:2302.14203 [astro-ph.CO]} \BibitemShut {NoStop}%
\bibitem [{\citenamefont {{Audren}}\ \emph {et~al.}(2013)\citenamefont {{Audren}}, \citenamefont {{Lesgourgues}}, \citenamefont {{Benabed}},\ and\ \citenamefont {{Prunet}}}]{Audrenetal2013}%
  \BibitemOpen
  \bibfield  {author} {\bibinfo {author} {\bibfnamefont {B.}~\bibnamefont {{Audren}}}, \bibinfo {author} {\bibfnamefont {J.}~\bibnamefont {{Lesgourgues}}}, \bibinfo {author} {\bibfnamefont {K.}~\bibnamefont {{Benabed}}},\ and\ \bibinfo {author} {\bibfnamefont {S.}~\bibnamefont {{Prunet}}},\ }\href {https://doi.org/10.1088/1475-7516/2013/02/001} {\bibfield  {journal} {\bibinfo  {journal} {J. Cosmol. Astropart. Phys.}\ }\textbf {\bibinfo {volume} {2013}}\bibfield  {number} {\bibinfo  {number} { (2)},\ \bibinfo {eid} {001}},\ }\Eprint {https://arxiv.org/abs/1210.7183} {arXiv:1210.7183 [astro-ph.CO]} \BibitemShut {NoStop}%
\bibitem [{\citenamefont {{Brinckmann}}\ and\ \citenamefont {{Lesgourgues}}(2019)}]{Brinckmann2019}%
  \BibitemOpen
  \bibfield  {author} {\bibinfo {author} {\bibfnamefont {T.}~\bibnamefont {{Brinckmann}}}\ and\ \bibinfo {author} {\bibfnamefont {J.}~\bibnamefont {{Lesgourgues}}},\ }\href {https://doi.org/10.1016/j.dark.2018.100260} {\bibfield  {journal} {\bibinfo  {journal} {Phys. Dark Univ.}\ }\textbf {\bibinfo {volume} {24}},\ \bibinfo {eid} {100260} (\bibinfo {year} {2019})},\ \Eprint {https://arxiv.org/abs/1804.07261} {arXiv:1804.07261 [astro-ph.CO]} \BibitemShut {NoStop}%
\bibitem [{\citenamefont {{Lewis}}(2019)}]{Lewis_2019}%
  \BibitemOpen
  \bibfield  {author} {\bibinfo {author} {\bibfnamefont {A.}~\bibnamefont {{Lewis}}},\ }\Eprint {https://arxiv.org/abs/1910.13970} {arXiv:1910.13970 [astro-ph.IM]}  (\bibinfo {year} {2019})\BibitemShut {NoStop}%
\bibitem [{\citenamefont {Cao}\ and\ \citenamefont {Ratra}(2022)}]{CaoRatra2022}%
  \BibitemOpen
  \bibfield  {author} {\bibinfo {author} {\bibfnamefont {S.}~\bibnamefont {Cao}}\ and\ \bibinfo {author} {\bibfnamefont {B.}~\bibnamefont {Ratra}},\ }\href {https://doi.org/10.1093/mnras/stac1184} {\bibfield  {journal} {\bibinfo  {journal} {Mon. Not. Roy. Astron. Soc.}\ }\textbf {\bibinfo {volume} {513}},\ \bibinfo {pages} {5686} (\bibinfo {year} {2022})},\ \Eprint {https://arxiv.org/abs/2203.10825} {arXiv:2203.10825 [astro-ph.CO]} \BibitemShut {NoStop}%
\end{thebibliography}%

\end{document}